\documentclass[prx,aps,10pt,twocolumn]{revtex4-2}

\usepackage{graphicx}
\usepackage{overpic}
\usepackage{dcolumn}
\usepackage{bm}
\usepackage{amsmath}
\usepackage{amsthm}
\usepackage{amsfonts}
\usepackage{xcolor}
\usepackage{natbib}
\usepackage{enumitem}
\usepackage{tcolorbox}
\usepackage{makecell}
\usepackage{booktabs}
\usepackage{tabularx}
\usepackage{pifont}
\usepackage{balance}
\usepackage{longtable}

\newcommand{\cmark}{\ding{51}}
\newcommand{\xmark}{\ding{55}}

\newcommand{\uniformimage}[2]{
  \makecell{
    \raisebox{-.5\height}{\includegraphics[height=3cm, keepaspectratio]{#1}} \\
    \scriptsize #2
  }
}
\newcommand{\smalluniformimage}[2]{
  \makecell{
    \raisebox{-.5\height}{\includegraphics[width=2.5cm, keepaspectratio]{#1}} \\
    \scriptsize #2
  }
}

\newtheorem{definition}{Definition.}
\newtheorem{theorem}{Theorem.}
\newtheorem{lemma}{Lemma.}

\tcbset{highlighted image/.style={
  colback=gray!5,  
  boxrule=0pt,
  frame hidden,
  sharp corners,
  enhanced jigsaw,
  interior style={fill},
  boxsep=0pt,
  left=0pt, right=0pt, top=0pt, bottom=0pt
}}

\tcbset{highlighted image white/.style={
  colback=white,
  boxrule=0pt, frame hidden,
  interior style={fill},
  boxsep=0pt,
  left=0pt, right=0pt, top=0pt, bottom=0pt
}}

\begin{document}

\preprint{APS/123-QED}

\title{Angular $k$-uniformity and the Hyperinvariance of Holographic Codes}

\author{Cheng, Wanli}
 \affiliation{Faculty of Computing and Data Sciences, Boston University, Boston, MA, 02215, USA.}
 \email{cheng.wanli@outlook.com}
 
\date{\today}
\begin{abstract}
Holographic quantum error-correcting codes, often realized through tensor network architectures, have emerged as compelling toy models for exploring bulk–boundary duality in AdS/CFT. By encoding bulk information into highly entangled boundary degrees of freedom, they capture key features of holography such as subregion duality, operator reconstruction, and complementary recovery. Among them, \emph{hyperinvariant tensor networks}—characterized by the inclusion of edge tensors and the enforcement of multi-tensor isometries—offer a promising platform for realizing features such as state dependence and nontrivial boundary correlations. However, existing constructions are largely confined to two-dimensional regular tilings, and the structural principles underlying hyperinvariance remain poorly understood, especially in higher dimensions. To address this, we introduce a geometric criterion called \emph{angular $k$-uniformity}, which refines standard $k$-uniformity and its planar variants by requiring isometric behavior within angular sectors of a tensor’s rotationally symmetric layout. This condition enables the systematic identification and construction of hyperinvariant holographic codes on regular hyperbolic honeycombs in arbitrary dimension, and extends naturally to heterogeneous networks and qLEGO architectures beyond regular tilings. Altogether, angular $k$-uniformity provides a versatile, geometry-aware framework for analyzing and designing holographic tensor networks and codes with hyperinvariant features such as nontrivial boundary correlations and state dependent complementary recovery.

\end{abstract}

\maketitle


\section{Introduction\label{Sec.1}}

The holographic principle, initially motivated by black hole thermodynamics and formalized in the AdS/CFT correspondence~\cite{Holography,RT-formula}, asserts that a gravitational theory in $(d+1)$-dimensional anti–de Sitter (AdS) space is fully encoded in a $d$-dimensional conformal field theory (CFT) living on its boundary. This profound bulk–boundary duality has inspired a wide range of models that explore how geometry and quantum information intertwine.

Among these, tensor network realizations of holography provide tractable, discrete frameworks that reproduce essential features of AdS/CFT, including the Ryu–Takayanagi formula for entanglement entropy and bulk-to-boundary operator reconstruction with redundancy~\cite{HaPPY}. In particular, \emph{hyperinvariant tensor networks} (HTNs)—which incorporate edge tensors and enforce multi-tensor isometries—stand out for their ability to exhibit nontrivial boundary correlations, support state-dependent reconstruction, and mimic features of subregion duality and complementary recovery~\cite{HTN,HI-MERA,HIC-qubit,HIC-ququart}.

\begin{figure*}[t]
  \centering
  \includegraphics[width=\textwidth]{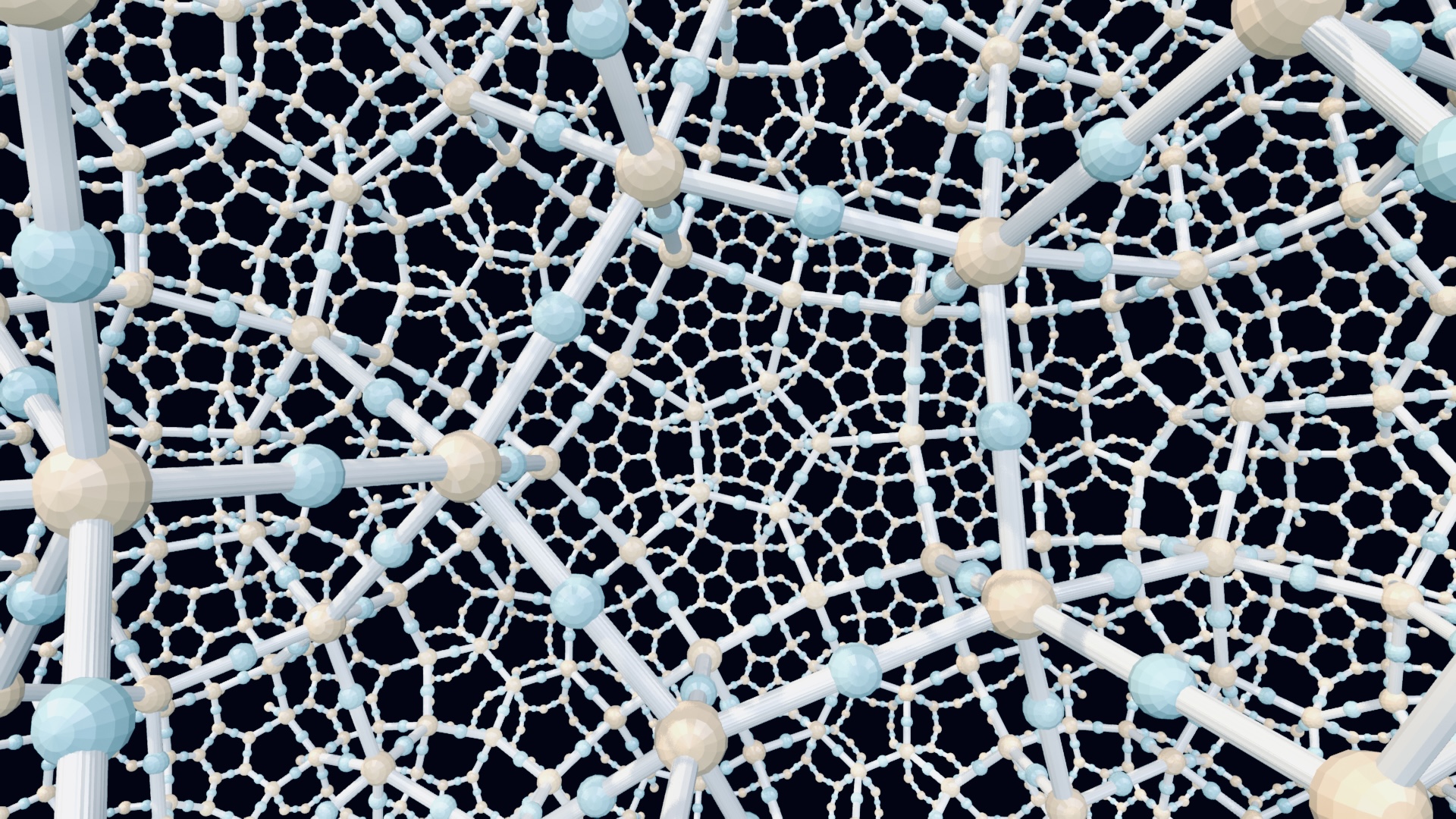}
  \caption{\textbf{Hyperinvariant Tensor Network/Code on the \{5,3,4\} Honeycomb.} 
  As in the 2D case, the network consists of two types of tensors: vertex tensors (yellow) and edge tensors (cyan). Logical indices are optional and have been omitted for clarity.}
  \label{fig:3D-HTN}
\end{figure*}

Despite these advances, most existing constructions are limited to two-dimensional hyperbolic tilings, such as $\{5,4\}$ and $\{7,3\}$. These settings, while conceptually illuminating, lack the geometric richness and algebraic complexity of higher-dimensional bulk–boundary dualities. 
Motivated by both the foundational role of higher dimensions in AdS/CFT—where many physically relevant dualities involve AdS$_{d+1}$ with $d \geq 3$—and the dimension-sensitive behavior of topological codes, such as Clifford hierarchy of transversal gates and self-correction thresholds in toric codes and fracton models, we seek to extend hyperinvariant codes into higher dimensions.

A naive generalization of 2D constructions—such as transplanting planar isometric conditions onto 3D or 4D lattices—often fails to yield codes with nontrivial boundary correlation functions, undermining the original goal of hyperinvariant tensor network design. This reveals a key gap in our current understanding: the absence of a precise, geometry-aware criterion for constructing hyperinvariant holographic codes in arbitrary dimensions.

In this work, we address this gap by introducing the concept of \emph{angular $k$-uniformity}—a geometric refinement of $k$-uniformity and its planar variants, defined in terms of isometric behavior within angular sectors of a tensor's rotationally symmetric layout. This notion enables a systematic framework for constructing and classifying hyperinvariant holographic codes on regular hyperbolic honeycombs in arbitrary dimensions. Angular $k$-uniformity serves as both a design principle and a diagnostic tool, capturing the hyperinvariant properties of a holographic tensor network.

We demonstrate the versatility of this framework through explicit tensor and CSS code constructions on uniform polytopes, and analyze their properties including multi-tensor isometries, correlation functions, complementary recovery, and uberholographic behavior. We further show that angular $k$-uniformity naturally generalizes to \emph{heterogeneous networks} and \emph{qLEGO constructions}, extending beyond regular tilings and enabling new classes of holographic codes with hyperinvariant features.

The remainder of this paper is organized as follows:  
Section~\ref{Sec.2} provides a visual overview of representative hyperinvariant codes across spatial dimensions.  
Section~\ref{Sec.3} reviews necessary background on hyperbolic lattices, tensor network isometries, and holographic coding principles.  
Section~\ref{Sec.4} presents our main results, including the definition of angular $k$-uniformity (Sec.~\ref{Sec.4.1}), the hierarchy of hyperinvariance (Sec.~\ref{Sec.4.2}), complementary recovery and fatal erasure analysis (Sec.~\ref{Sec.4.3}), the phenomenon of uberholography (Sec.~\ref{Sec.4.4}), explicit code constructions (Sec.~\ref{Sec.4.5}), scaling of code properties (Sec.~\ref{Sec.4.6}), and generalization to heterogeneous and qLEGO networks (Sec.~\ref{Sec.4.7}).  
Section~\ref{Sec.5} concludes with open problems and future directions.  
Appendix~\ref{App.A} summarizes angular $k$-uniformity realizations on all locally finite regular hyperbolic honeycombs.  
Appendix~\ref{App.B} provides technical details for vertex tensor constructions by symmetry type.

\section{Overview\label{Sec.2}}

To provide a unified overview of the properties of hyperinvariant tensor networks and codes across dimensions, we present representative examples in Table~\ref{tab:overview}.

This summary illustrates how the spatial dimensionality and angular $k$-uniformity jointly determine the key features of hyperinvariant tensor networks and codes. These include the multi-tensor isometry conditions, the (non-)triviality of two-point correlation functions, the geometry of residual regions under complementary recovery, and the behavior of fatal erasure errors. While each of these concepts will be defined and explored in detail in Sec.~\ref{Sec.4}, Table~\ref{tab:overview} already reveals that generalizing hyperinvariant tensor networks and Evenbly (hyperinvariant) codes beyond two dimensions requires increasingly complex isometric structures and leads to qualitatively distinct behaviors.

A full classification of all locally finite hyperbolic tilings and honeycombs across dimensions is deferred to Appendix~\ref{App.A}.

{\renewcommand{\arraystretch}{1.5}

\begin{table*}[ht]

\centering

\begin{tabularx}{0.9\textwidth}{@{} 
c
c
c
c >{\centering\arraybackslash}
c
c
c
@{}}

\toprule

  \textbf{Dimension}
& \textbf{Lattice}
& \makecell{\textbf{Angular}\\ \textbf{Uniformity}}
& \textbf{Multi-Tensor Block}
& \makecell{\textbf{Non-trivial}\\ \textbf{Correlation}}
& \makecell{\textbf{Residual Region}}
& \makecell{\textbf{Fatal Erasure}}
\\

\midrule
    
  \makecell{2\\ \\3\\ \\4} 
& \makecell{$\{5,4\}$\\ \\$\{5,3,4\}$\\ \\$\{5,3,3,4\}$} 
& \makecell{$k=1$\\ \\$k=2$\\ \\$k=3$} 
& \uniformimage{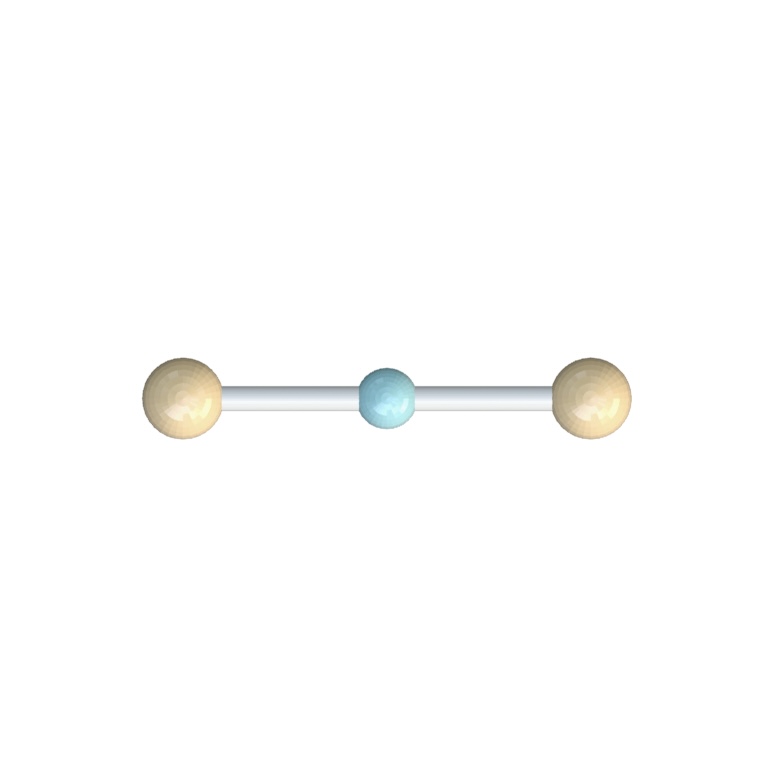}{One Edge} 
& \makecell{\cmark\\ \\ \xmark\\ \\ \xmark} 
& 1D Curve 
&\makecell{Zigzag\\ \\Line\\ \\Line}
\\
  
\midrule

  \makecell{3\\ \\4} 
& \makecell{$\{5,3,4\}$\\ \\$\{5,3,3,4\}$} 
& \makecell{$k=1$\\ \\$k=2$} 
& \uniformimage{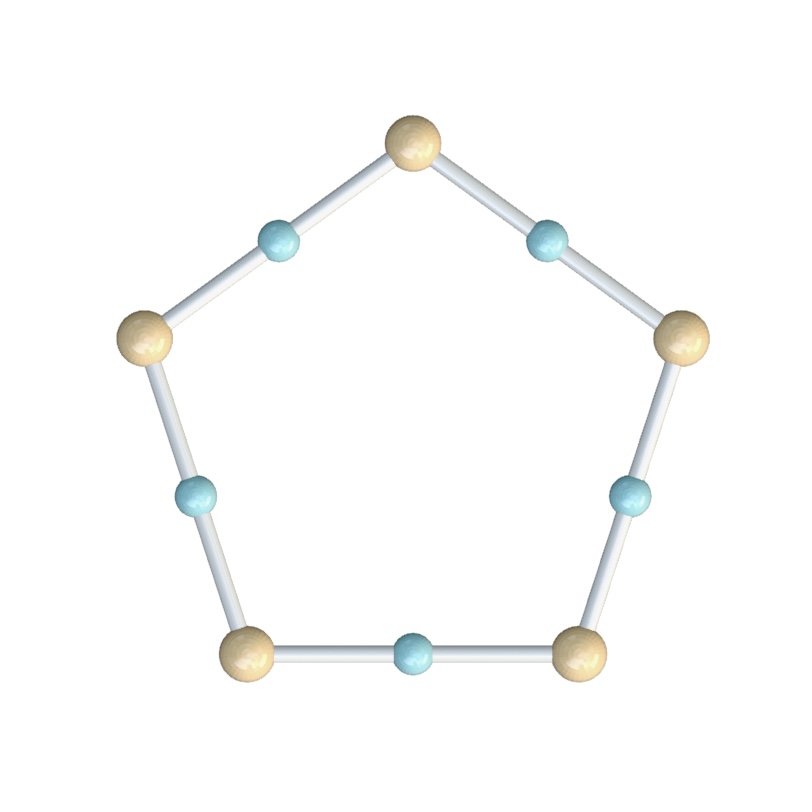}{One Pentagon} 
& \makecell{\cmark\\ \\ \xmark} 
& 2D Surface 
&\makecell{Zigzag\\ \\Line}
\\
  
\midrule

  4
& $\{5,3,3,4\}$ 
& $k=1$ 
& \uniformimage{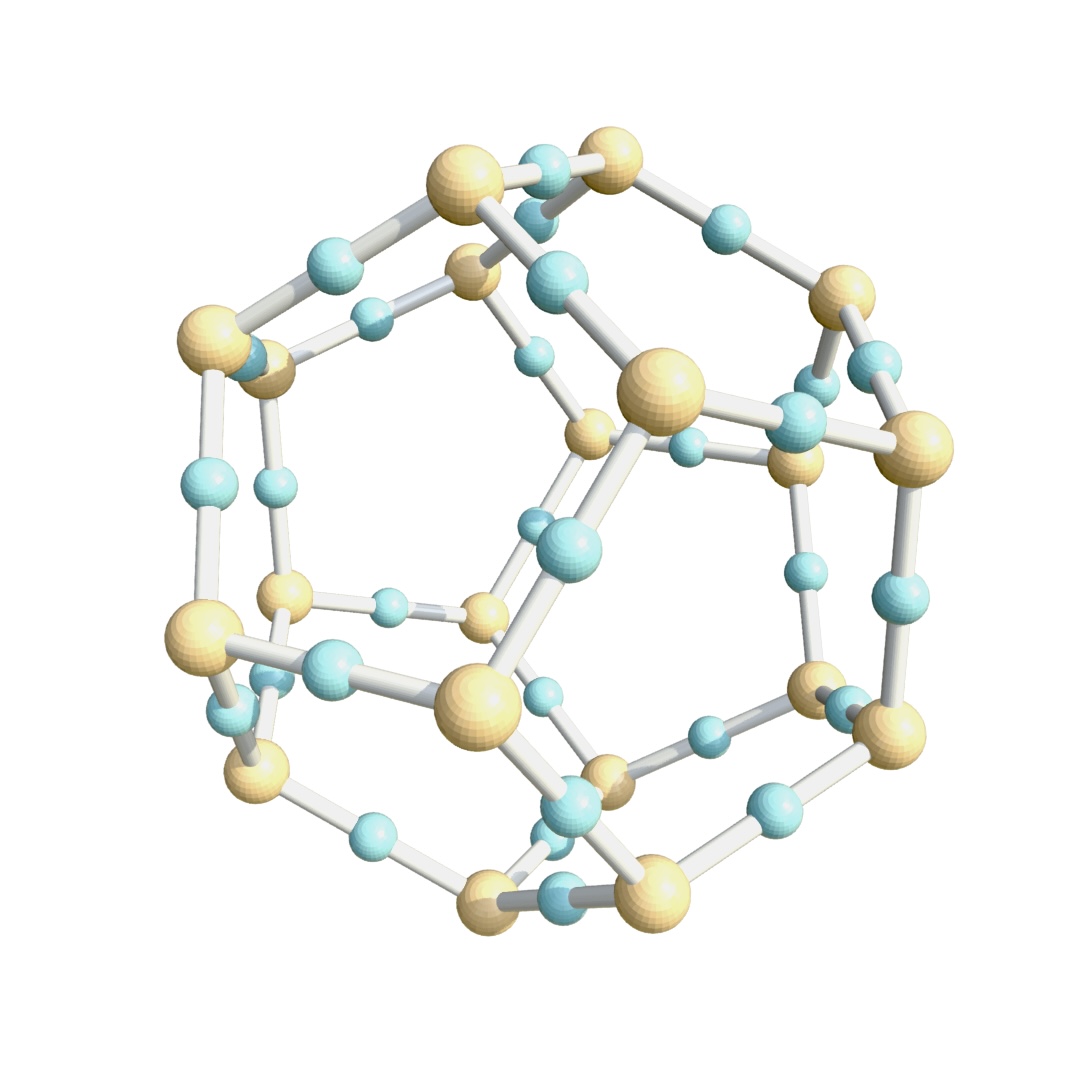}{One Dodecahedron} 
& \cmark
& 3D Volume
& Zigzag
\\
  
\bottomrule

\end{tabularx}

\caption{
Summary of key structural features of HTNs and HICs on representative hyperbolic honeycombs. Each configuration is defined by a specific angular $k$-uniformity level and associated multi-tensor block. Yellow nodes indicate vertex tensors with $k$ input legs; cyan nodes indicate edge tensors.
}

\label{tab:overview}

\end{table*}
}

\section{Background\label{Sec.3}}

\subsection{Hyperbolic Honeycombs\label{Sec.3.1}}

In the AdS$^{(d+1)}$/CFT$_d$ correspondence, a spatial slice of the $(d+1)$-dimensional anti-de Sitter (AdS) spacetime is geometrically realized as a $d$-dimensional hyperbolic space with constant negative curvature. The spatial slice of the $d$-dimensional boundary CFT resides on the $(d-1)$-dimensional ideal boundary of this hyperbolic space. To discretize such spaces in a manner compatible with tensor network constructions, it is natural to consider highly symmetric tilings.

Coxeter showed that there exist infinitely many regular hyperbolic tilings in two dimensions, characterized by the \emph{Schläfli symbol} $\{p, q\}$, which denotes a regular tiling by $p$-gons with $q$ meeting at each vertex. In three dimensions, there are four compact regular hyperbolic honeycombs: $\{3,5,3\}$, $\{4,3,5\}$, $\{5,3,4\}$, and $\{5,3,5\}$, along with five infinite families of regular quasi-compact honeycombs: $\{3,3,r\}_{r>5}$, $\{3,4,r\}_{r>3}$, $\{3,5,r\}_{r>3}$, $\{4,3,r\}_{r>5}$, and $\{5,3,r\}_{r>4}$, where “quasi-compact” refers to non-compact tilings with finite vertex degree.

In four dimensions, there are five compact regular hyperbolic honeycombs: $\{3,3,3,5\}$, $\{4,3,3,5\}$, $\{5,3,3,3\}$, $\{5,3,3,4\}$, and $\{5,3,3,5\}$, and six infinite families of quasi-compact hyperbolic honeycombs: $\{3,3,3,s\}$, $\{3,3,4,s\}$, $\{3,3,5,s\}$, $\{3,4,3,s\}$, $\{4,3,3,s\}$, and $\{5,3,3,s\}$.

Beyond four dimensions, no compact regular hyperbolic honeycombs exist. However, regular quasi-compact honeycombs can still be constructed using three infinite families of polytopes: the $d$-cube, $d$-simplex, and $d$-orthoplex families.

The \emph{Schläfli symbol} $\{p, q, r, s, \dots\}$ is defined recursively. In this notation:
\begin{itemize}
  \item $\{p\}$ denotes a regular $p$-gon (a tiling of $\mathbb{S}^1$),
  \item $\{p,q\}$ denotes a 2D tiling in $\mathbb{S}^2$, $\mathbb{E}^2$, or $\mathbb{H}^2$ with $q$ $p$-gons meeting at each vertex,
  \item $\{p,q,r\}$ represents a 3D tiling (of $\mathbb{S}^3$, $\mathbb{E}^3$, or $\mathbb{H}^3$) where $r$ cells of type $\{p,q\}$ meet around each edge,
  \item and so on, with higher-dimensional generalizations following analogously.
\end{itemize}

\subsection{Uniformity of Quantum States\label{Sec.3.2}}

In the foundational construction of the HaPPY code \cite{HaPPY}, \emph{absolutely maximally entangled} (AME) states—also known as \emph{perfect tensors}—were used as the fundamental building blocks~\cite{AME_1,AME_2}. These highly entangled states enable key holographic features such as \emph{complementary recovery} and support for \emph{uberholography}, but they also tend to trivialize boundary correlation functions due to their maximal entanglement. Subsequent works extended this framework by exploring alternative classes of highly entangled multipartite quantum states, including \emph{block-perfect tensors}~\cite{PME,Planar_k_Uniformity,Block-perfect_Steane}, which relax the stringent entanglement conditions of AME states while preserving essential properties for robust encoding and local reconstruction, thereby enabling nontrivial boundary correlation functions.

A particularly useful generalization in this context is the class of \emph{$k$-uniform states}, which extend the concept of maximal bipartite entanglement to multipartite systems. A pure $n$-qudit quantum state $|\psi\rangle$ is called \emph{$k$-uniform} if every reduction to any $k$ qudits yields the maximally mixed state:
\[
\mathrm{Tr}_{\overline{S}} \left[ |\psi\rangle\langle\psi| \right] = \frac{\mathbb{I}_{d^k}}{d^k}, \quad \text{for all } S \subset \{1, \dots, n\},\ |S| = k,
\]
where $\overline{S}$ denotes the complement of the subset $S$. These states are maximally entangled across all $k$-qudit subsystems and saturate the entanglement bound for all such reductions. The special case $k = \lfloor n/2 \rfloor$ corresponds to an AME state, or \emph{perfect tensor}, which is maximally entangled across all bipartitions. The existence of $k$-uniform states depends nontrivially on the parameters $n$, $k$, and the local dimension $d$.

To address the limitations of strict uniformity, the notion has been generalized to \emph{planar $k$-uniform states}, where the qudits are arranged on a circle (interpreted as the 1D boundary of a 2D bulk), and the maximal entanglement condition is imposed only on \emph{connected} regions of size $k$. When $k = \lfloor n/2 \rfloor$, such a state is referred to as a \emph{block-perfect tensor}. These relaxed conditions retain many of the desirable features for holographic encoding while allowing for the use of more physically realizable tensors. In particular, in \cite{HIC-qubit} planar 2-uniform tensors are used to construct Evenbly (hyperinvariant) codes that support nontrivial boundary correlation functions.

However, these notions remain mostly limited to planar or low-dimensional geometries. To address higher-dimensional and rotationally symmetric settings, we extend this concept to \emph{angular $k$-uniformity} in Sec.~\ref{Sec.4.1}.

\subsection{Holographic Tensor Networks and Holographic Codes\label{Sec.3.3}}

Tensor networks were originally developed to study quantum many-body systems with area-law entanglement. Motivated by the \emph{Ryu–Takayanagi} (RT) formula~\cite{RT-formula}, which relates entanglement entropy in holographic theories to minimal surfaces in the bulk, tensor networks were naturally introduced into the study of holography. The first such attempt was proposed by Swingle, who employed the \emph{multiscale entanglement renormalization ansatz} (MERA)—an ansatz initially designed to approximate ground states of quantum-critical spin chains—to model a spatial slice of $(2+1)$-dimensional AdS spacetime. However, the geometry of MERA does not exactly match that of a constant-curvature time slice of AdS$_{2+1}$, which is described by the hyperbolic plane.

To better reflect the geometry of AdS space, the HaPPY code—named after Harlow, Preskill, Pastawski, Yoshida, and their collaborators—was introduced using perfect tensors placed on regular hyperbolic tilings. In this construction, a perfect tensor is placed at each face of the tiling, with one leg left uncontracted to encode bulk logical information and the remaining legs contracted with neighboring tensors. The original proposal employed a 6-qubit absolutely maximally entangled (AME) state as the seed tensor. Logical bulk operators can then be reconstructed via a greedy algorithm, establishing an explicit mapping between bulk and boundary degrees of freedom. Variants of holographic codes have also been proposed using partially maximally entangled (PME) states, graph states, and other classes of seed tensors.

Perfect-tensor-based holographic codes typically exhibit a property known as \emph{uberholography}, in which bulk operators can be reconstructed from a fractal subset of the boundary. This behavior is a manifestation of the broader framework of \emph{operator algebra quantum error correction} (OAQEC), where a logical sub-algebra associated with a bulk region is recoverable from nonlocal boundary operators.

However, perfect holographic codes such as the HaPPY code fail to capture many aspects of AdS/CFT duality at finite $N$, where gravitational corrections become relevant. These include nontrivial entanglement spectra of reduced states, state-dependent bulk reconstruction, and deviations from the RT formula. To incorporate these features, it is necessary to modify either the seed tensors or the network architecture of the code. One approach that integrates both strategies is the \emph{hyperinvariant tensor network}~\cite{HTN,HIC-qubit,HIC-ququart,HI-MERA}, which generalizes the HaPPY construction to broader classes of isometries, and richer entanglement structures.

\subsection{Hyperinvariant Tensor Networks and Codes\label{Sec.3.4}}

\begin{figure*}[t]
  \centering

  \begin{minipage}[b]{0.23\textwidth}
    \includegraphics[width=\textwidth,page=1]{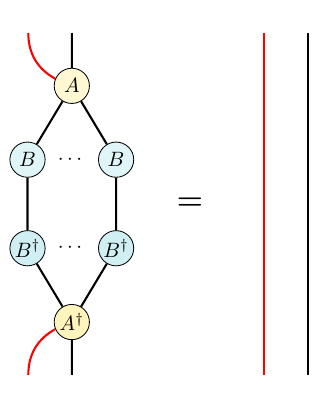}
    (a) Single-Tensor Isometry
  \end{minipage}
  \hfill
  \begin{minipage}[b]{0.4\textwidth}
    \includegraphics[width=\textwidth,page=2]{Graphics/TIC/TIC.pdf}
    (b) Double-Tensor Isometry
  \end{minipage}
  \hfill
  \begin{minipage}[b]{0.25\textwidth}
    \includegraphics[width=\textwidth,page=4]{Graphics/TIC/TIC.pdf}
    (c) Multi-Tensor Isometry
  \end{minipage}

  \caption{
  Tensor network representations of isometry constraints in HTNs and HICs. Red legs denote logical indices; black legs denote physical indices. When logical indices are omitted, the conditions reduce to those for HTNs.  
  (a) Single-tensor isometry involving tensors $A$ and $B$.  
  (b) Double-tensor isometry, originally introduced in Ref.~\cite{HTN}, constitutes the minimal structure required for 2D HTNs.  
  (c) General multi-tensor block (MTB) isometry for higher-dimensional hyperbolic honeycombs. Representative MTBs are listed in Table~\ref{tab:overview}.
  }
  \label{fig:TIC}
\end{figure*}

Hyperinvariant tensor networks (HTNs), originally introduced by Evenbly~\cite{HTN}, integrate key features from both MERA and the HaPPY code to model boundary states resembling critical 1D CFTs. Like the HaPPY code, HTNs are defined on regular hyperbolic tilings, but they relax the perfect tensor requirement in favor of more flexible isometric constraints. This allows HTNs to support nontrivial boundary correlation functions, similar to MERA.

More recently, Steinberg et al.~\cite{HIC-ququart,HIC-qubit} constructed explicit \emph{hyperinvariant Evenbly codes} (HICs) on $\{p,2n\}$ hyperbolic tilings with ququarts and qubits. These codes extend HTNs by incorporating logical degrees of freedom into the bulk, enabling holographic encoding.

Unlike HaPPY codes—which use a single type of perfect tensor—HTNs and HICs employ two distinct tensors, denoted $A$ and $B$, subject to the following conditions:
\begin{enumerate}
    \item $A$ and $B$ satisfy \emph{single-} and \emph{multi-tensor isometry constraints} (TICs), illustrated in Fig.~\ref{fig:TIC}.
    \item $A$ is invariant under the rotational symmetry group of its vertex figure (see Sec.~\ref{Sec.3.1}).
    \item $B\neq I$ is a symmetric unitary tensor, satisfying $B = B^T$ and $BB^\dagger = I$.
\end{enumerate}

Perfect tensor networks, such as the HaPPY code, also satisfy the above three conditions when one sets $B = I$. Accordingly, Refs.~\cite{HIC-ququart,HIC-qubit} proposed an additional criterion:

\begin{enumerate}[resume]
    \item* $A$ is not a perfect tensor.
\end{enumerate}

While this condition serves as a necessary criterion for distinguishing nontrivial constructions on general $\{p,q\}$ hyperbolic tilings with even $q > 4$, and is sufficient for the special case $\{p,4\}$, it does not fully characterize hyperinvariance in general settings. In Sec.~\ref{Sec.4}, we introduce the stronger notions of angular $k$-uniformity and multi-angular $k$-uniformity, which provide a more refined—but still partial—description of hyperinvariant tensor networks and codes.

Hyperbolic tilings and honeycombs naturally admit a foliation into concentric, self-similar layers~\cite{conformal_quasicrystal}. Each layer can be further (non-uniquely) decomposed into multi-tensor blocks and their substructures. Both encoding and decoding processes propagate acyclically through these blocks, with each layer acting as a scaling superoperator in a real-space renormalization group (RG) flow—reminiscent of MERA.

While the multi-tensor isometry conditions are essential for supporting nontrivial boundary correlations, their precise justification and structure remain underexplored in the existing literature. In the next section, we establish a geometric criterion—\emph{angular $k$-uniformity}—as the defining isometry condition for the vertex tensor $A$, and show how the interplay between $k$ and the underlying hyperbolic honeycomb determines the multi-tensor building blocks required for hyperinvariance.

\section{Results\label{Sec.4}}

\subsection{Angular $k$-Uniformity\label{Sec.4.1}}

To explore hyperinvariant tensor networks (HTNs) and hyperinvariant codes (HICs) beyond two dimensions, we begin with the three-dimensional case, focusing in particular on the $\{5,3,4\}$ honeycomb. As outlined in Sec.~\ref{Sec.3.4}, HTN and HICs on such tilings are constructed by assigning rotational invariant tensors $A$ to the vertices and symmetric unitaries $B$ to the edges. While the edge tensors $B$ retain the same properties as in two dimensions, the vertex tensors $A$ now require a higher-dimensional geometric formulation to fully characterize their isometry and symmetry constraints. This is difficult to express in the conventional tensor network diagram, where $A$ is visualized simply as a node with multiple legs.

\begin{figure*}[ht]
  \centering
  \begin{overpic}[width=\textwidth]{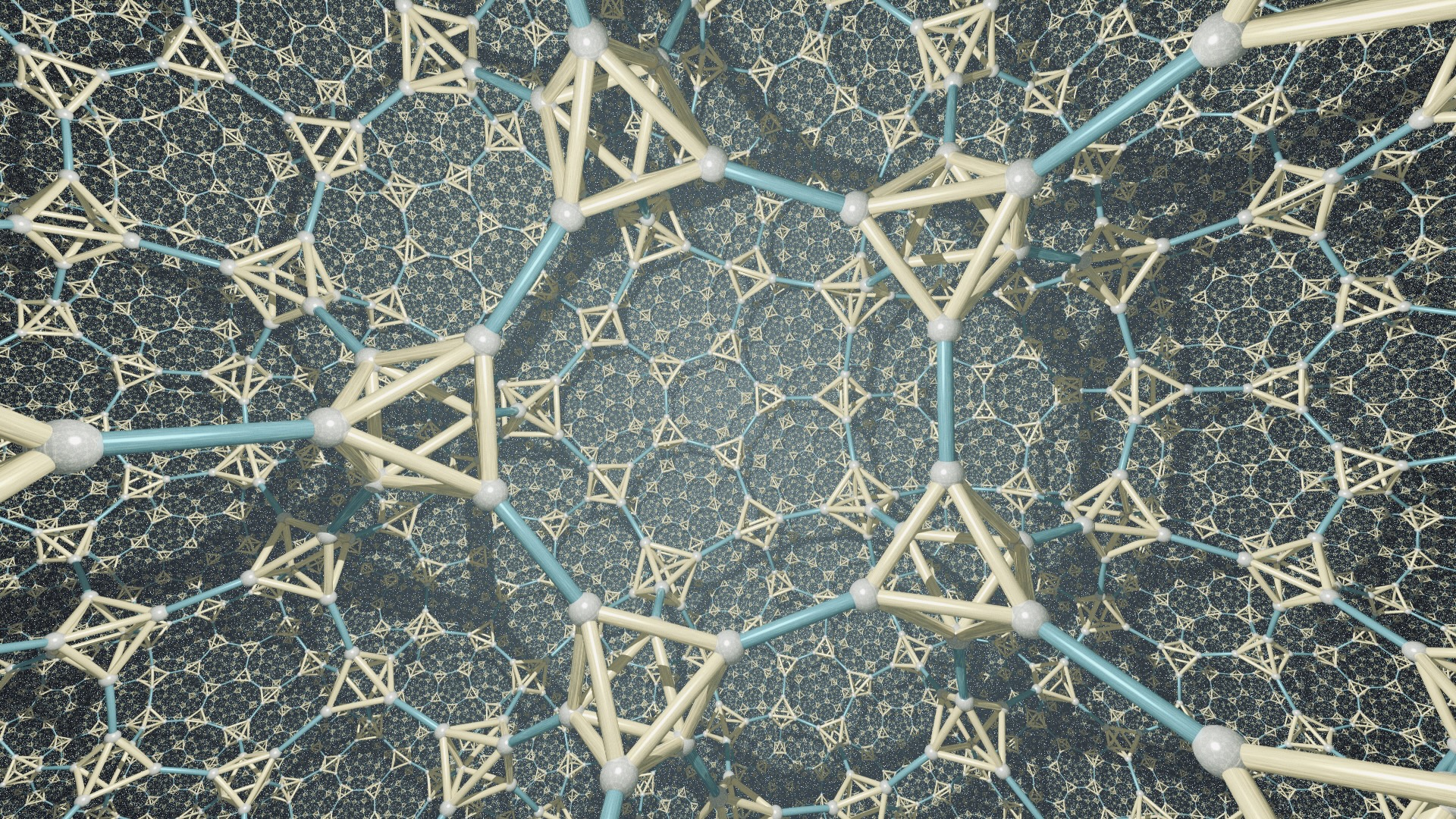}
    \put(72,16){
    \begin{tcolorbox}[colframe=black, colback=white, boxrule=1pt,sharp corners,width=0.25\textwidth]
    \includegraphics[trim=140 180 140 160, clip, width=\linewidth]{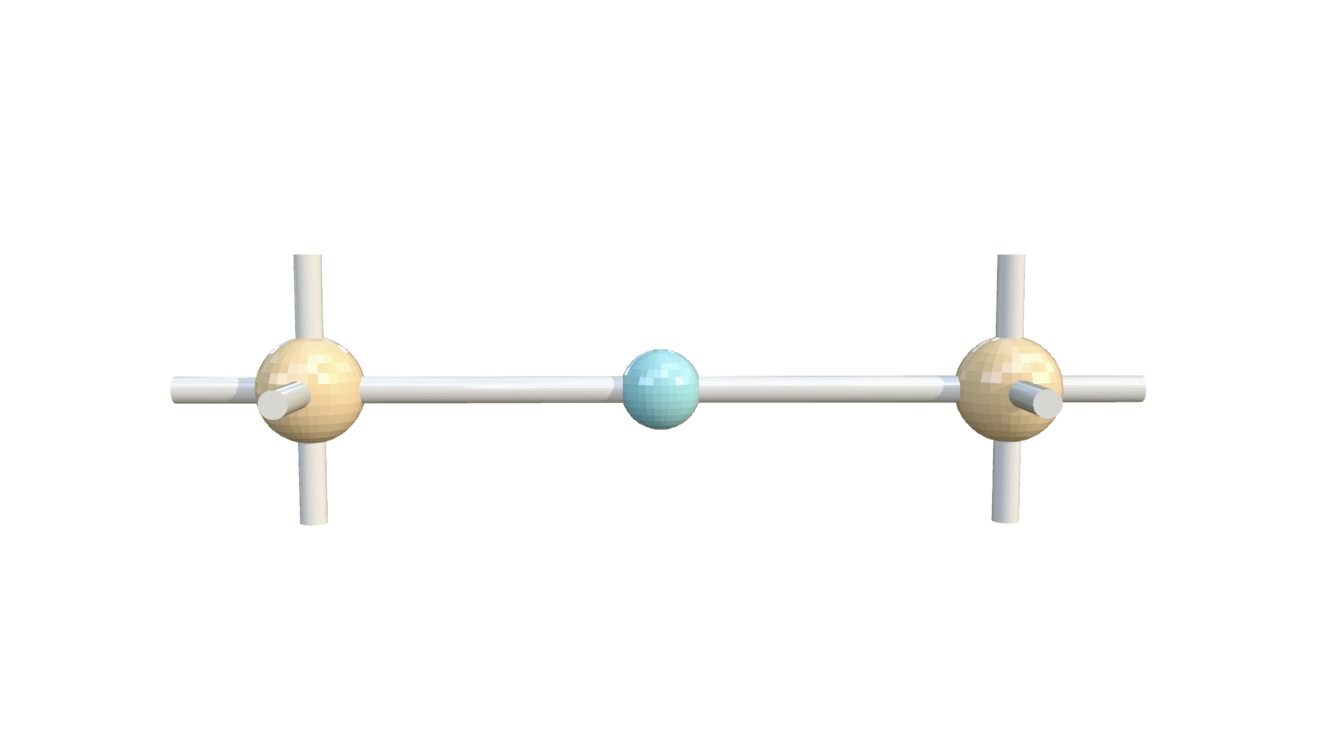}
    \small{Tensor Network Representation}
    \includegraphics[trim=140 180 140 160, clip, width=\linewidth]{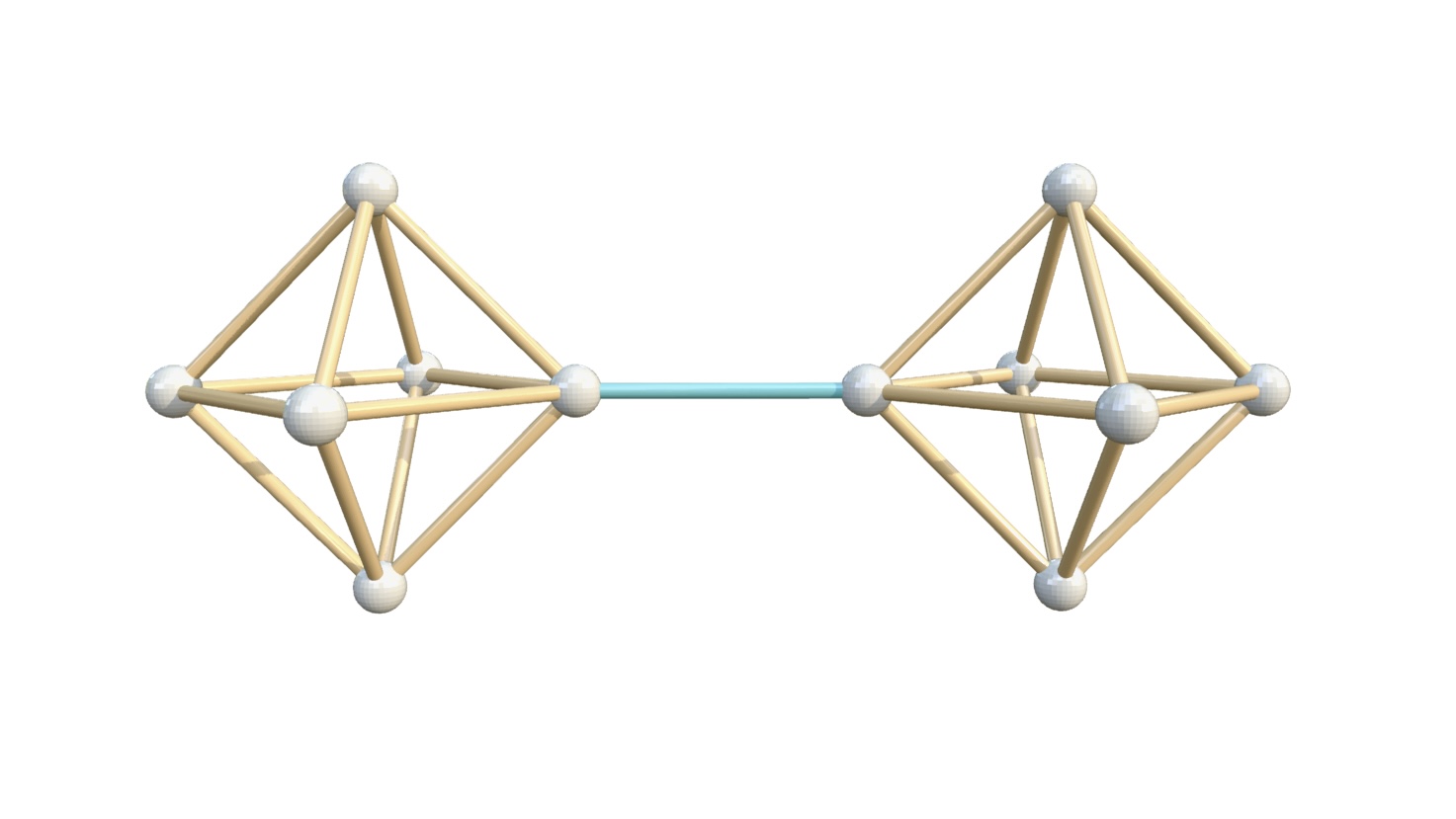}
    \small{Vertex Figure Representation}
    \end{tcolorbox}}
  \end{overpic}
    \caption{
    \textbf{Vertex Figure Representation of the \{5,3,4\} Honeycomb.} Each $A$ tensor is shown as a yellow octahedron, with its six vertices representing the tensor’s physical indices. Edge tensors $B$ are drawn as cyan edges. The coloring matches the tensor network representation in Fig.~\ref{fig:3D-HTN}.
    }
    \label{fig:vertex figure HTN}
\end{figure*}

To resolve this, we adopt the \emph{vertex figure representation}, introduced in Fig.~\ref{fig:vertex figure HTN}, which reveals the local combinatorial geometry around each vertex. In regular honeycombs, this vertex figure is itself a regular polytope.

\begin{definition}[Vertex Figure Representation]
In tensor networks embedded on a fixed lattice, especially in non-planar geometries, the standard representation often obscures local geometric relationships between indices. The \emph{vertex figure representation} remedies this by replacing each vertex in $d$-dimensional lattice with its corresponding vertex figure: a $(d{-}1)$-dimensional polytope whose vertices represent the tensor’s indices. Edges in the vertex figure connect indices that share a common facet in the underlying $d$-dimensional geometry. In this view, tensors become polytopes—e.g., a three-leg tensor becomes a triangle, a two-leg tensor an edge, and a single-leg tensor a half-edge. And tensor contractions correspond to shared vertices between polytopes.
\end{definition}

While prior works have emphasized rotational symmetry in the construction of HTNs~\cite{HTN,HIC-ququart,HIC-qubit}, the associated isometry conditions on $A$ have remained imprecisely defined. In particular, Ref.~\cite{HIC-qubit} introduced “1-isometry” and “non-perfectness” as guiding principles for balancing the number of isometric partitions (too much isometries causes triviality of boundary correlations, too few isometries prohibit holographic encoding). However, these criteria are neither necessary nor sufficient to ensure hyperinvariance—especially in higher-dimensional settings, where local geometry plays a crucial role.

To better characterize which tensors are suitable for defining hyperinvariant networks, we introduce a geometric refinement: \emph{angular $k$-uniformity}. As the name suggests, this generalizes the notions of $k$-uniformity~\cite{AME_1,AME_2} and planar $k$-uniformity~\cite{Planar_k_Uniformity,PME}, aligning isometric conditions with the angular structure of the underlying tiling.

\begin{figure}
    \includegraphics[trim=150 200 150 200, clip, width=\linewidth]{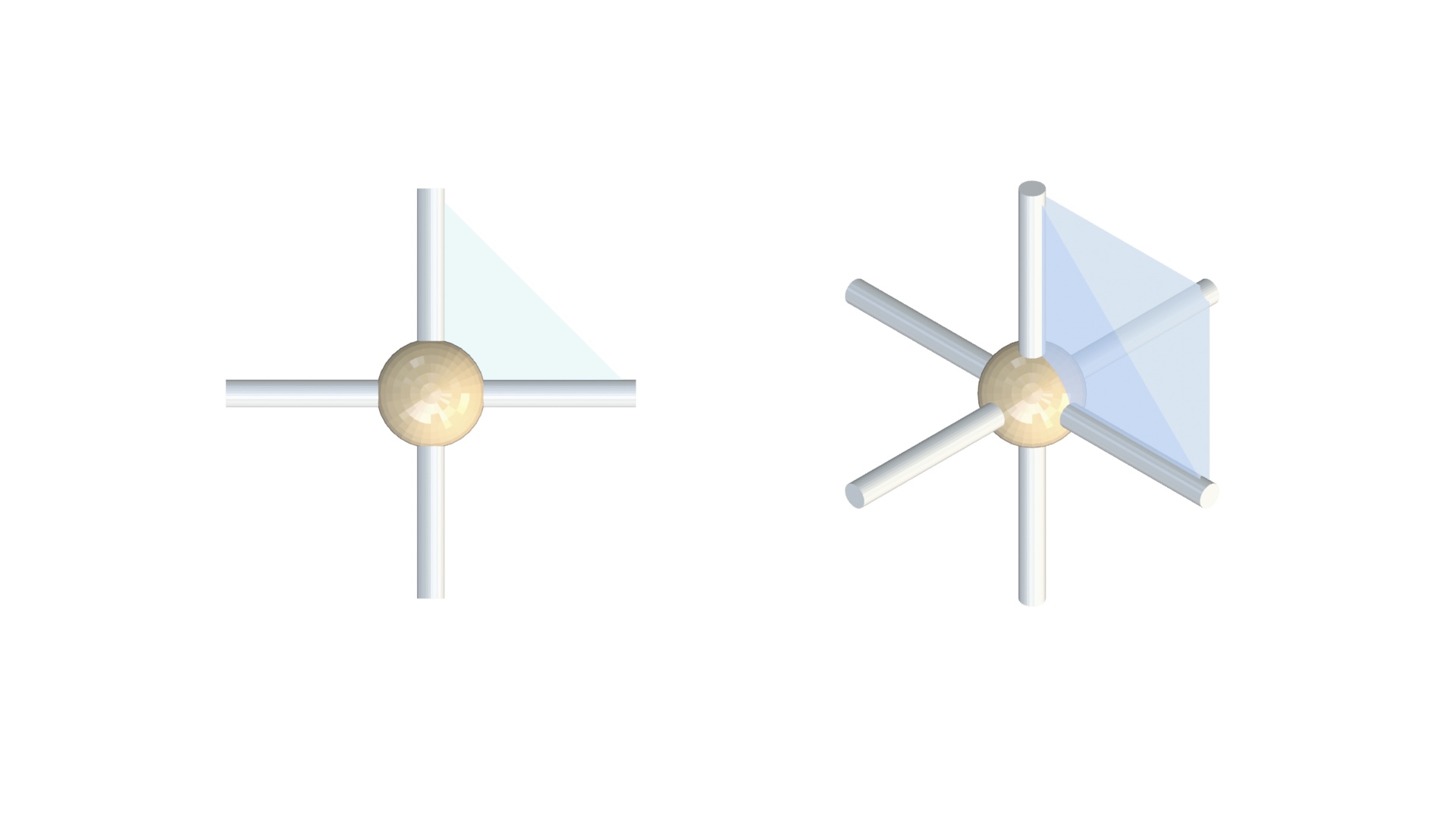}
    \includegraphics[trim=150 200 150 200, clip, width=\linewidth]{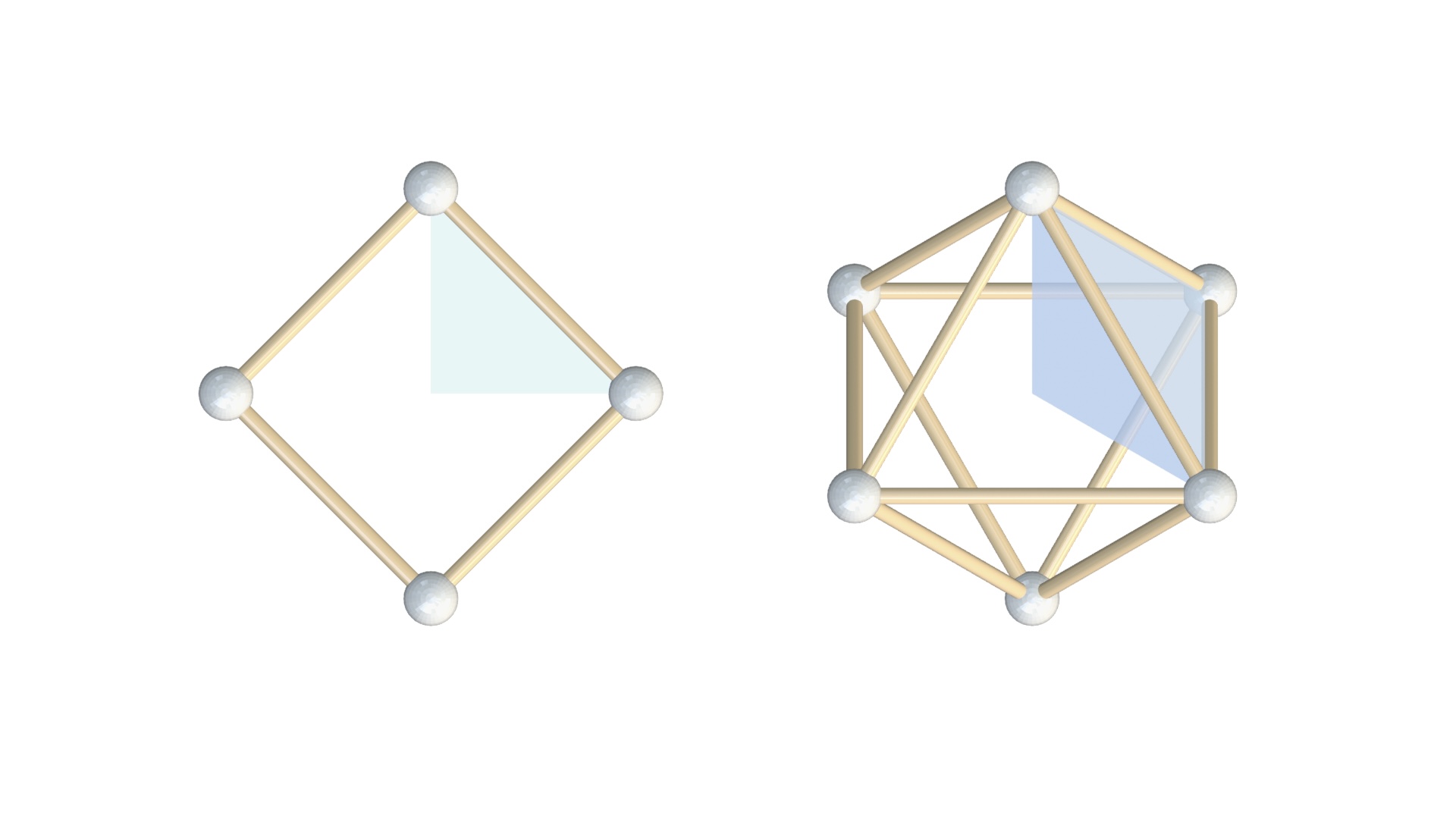}
    \caption{
    \textbf{Left:} A 2D angular sector (shaded) with four physical indices, shown in both standard tensor network and vertex figure representations. 
    \textbf{Right:} A 3D solid angle with six physical indices, comprising three overlapping 2D angular sectors. In both cases, angular locality is defined through the vertex figure.
    }
    \label{fig:solid angle}
\end{figure}

\begin{definition}[Angular Connectivity]
Let \( P \) be a \( d \)-dimensional regular polytope, and let \( S \) be a subset of its vertices. We say that \( S \) is \emph{angularly connected} if all vertices in \( S \) lie entirely within a single \((d{-}1)\)-dimensional facet of \(P\), and form a connected subgraph of the 1-skeleton of that facet.

We further say that \( S \) is \emph{strongly angularly connected} if, in addition, for every lower-dimensional subfacet (i.e., any face of dimension less than \(d{-}1\)) of that facet containing elements of \( S \), the restriction of \(S\) to that subfacet is also angularly connected within the subfacet.

Finally, two subsets of vertices \(S_1\) and \(S_2\) are said to be \emph{angularly disconnected} if they do not simultaneously lie within any common \((d{-}1)\)-dimensional facet of \(P\).
\end{definition}

\begin{definition}[Angular $k$-Uniformity]
Let \( A \) be a tensor (possibly with logical indices), whose \( n \) physical indices are arranged according to the vertex figure of a \( d \)-dimensional polytope. We say that \(A\) is \emph{angular \(k\)-uniform} if the following two conditions hold:

\begin{enumerate}[label=(\roman*)]
  \item Let \( L \) denote the set of logical indices. For every strongly angularly connected subset \( I \subset \{1, \dots, n\} \) of physical indices with \( |I| = k \), the linear map
  \[
  \bigotimes_{i \in L \cup I} \mathcal{H}_i \longrightarrow \bigotimes_{j \notin  L \cup I} \mathcal{H}_j
  \]
  is an isometry.

  \item No subset \( I' \subset \{1, \dots, n\} \) with \( |I'| > k \) satisfies the isometry condition.
\end{enumerate}
In the planar case (\(d = 2\)), where each facet reduces to an edge, this recovers the standard notion of planar $k$-uniformity for $k = 1, 2$.
\end{definition}

As illustrated in Fig.~\ref{fig:solid angle}, angular $k$-uniformity imposes an isometric constraint localized to angular sectors or solid angles in the underlying lattice geometry. This geometric perspective allows us to formulate precise conditions under which a vertex tensor supports nontrivial encoding and correlation properties.

In the next subsection, we will show that angular $k$-uniformity is not merely a formal condition, but a foundational principle underlying the hierarchy of hyperinvariance in higher-dimensional HTNs. Moreover, a more general condition—\emph{multi-angular $k$-uniformity}—will play a central role in understanding the phenomenon of uberholography, as discussed in Sec.~\ref{Sec.4.4}.

\subsection{Hierarchy of Hyperinvariance}
\label{Sec.4.2}

We use the term \emph{hyperinvariance} to refer to the necessity of enforcing multi-tensor isometries in order to ensure a well-defined encoding in a holographic tensor network (HTN). The requirement of nontrivial tensor isometries is a hallmark of hyperinvariant tensor networks and codes. In earlier works~\cite{HTN,HI-MERA,HIC-qubit,HIC-ququart}, investigations have focused exclusively on two-dimensional tilings, typically of the form $\{p,q\}$ with small $p$ and $q$. As a result, the conceptual justification for introducing such isometries was not explored in generality.

To motivate the study of hyperinvariance beyond 2D, we begin by recalling a no-go theorem from Ref.~\cite{HI-MERA}, which sets sharp conditions under which HTNs necessarily fail to exhibit nontrivial boundary correlation functions:

\begin{lemma}[2D No-Go Theorem~\cite{HI-MERA}]
\label{lem:2D-No-Go}
Consider a 2D holographic tensor network defined on a regular hyperbolic tiling, where all vertex tensors are identical and permutation-invariant. If each tensor is at least 2-isometric (i.e., 2-uniform), then the network admits only trivial boundary correlation functions.
\end{lemma}

This theorem explains why nontrivial HTNs in 2D~\cite{HTN,HIC-ququart,HIC-qubit} must be constructed using angular 1-uniform tensors. Inspired by this observation, we propose the following generalization to higher dimensions, phrased geometrically in terms of angular $k$-uniformity:

\begin{theorem}[General No-Go Theorem]
For a holographic tensor network or code defined on a non-simplicial regular hyperbolic tiling, with all vertex tensors rotationally invariant, if the tensors are maximally angular $k$-uniform—i.e., $k$ equals the full size of a $(d{-}1)$-dimensional facet of the vertex figure—then the network is not hyperinvariant and the boundary correlation functions are not universally non-trivial.
\end{theorem}

In the 2D case, this reduces to a stronger form of Lemma~\ref{lem:2D-No-Go}. For simplicial tilings, a modified version of the angular $k$-uniformity definition is required. The statement also extends naturally to semi-regular and vertex-transitive tilings as considered in \cite{HI-MERA, Heteogeneous}.

The theorem has two main implications: one concerning the emergence of hyperinvariance, and another regarding the triviality of correlation functions. We now illustrate both aspects using the $\{5,3,4\}$ honeycomb as a representative example.

\begin{figure*}
    \centering
    \includegraphics[width=0.32\linewidth]{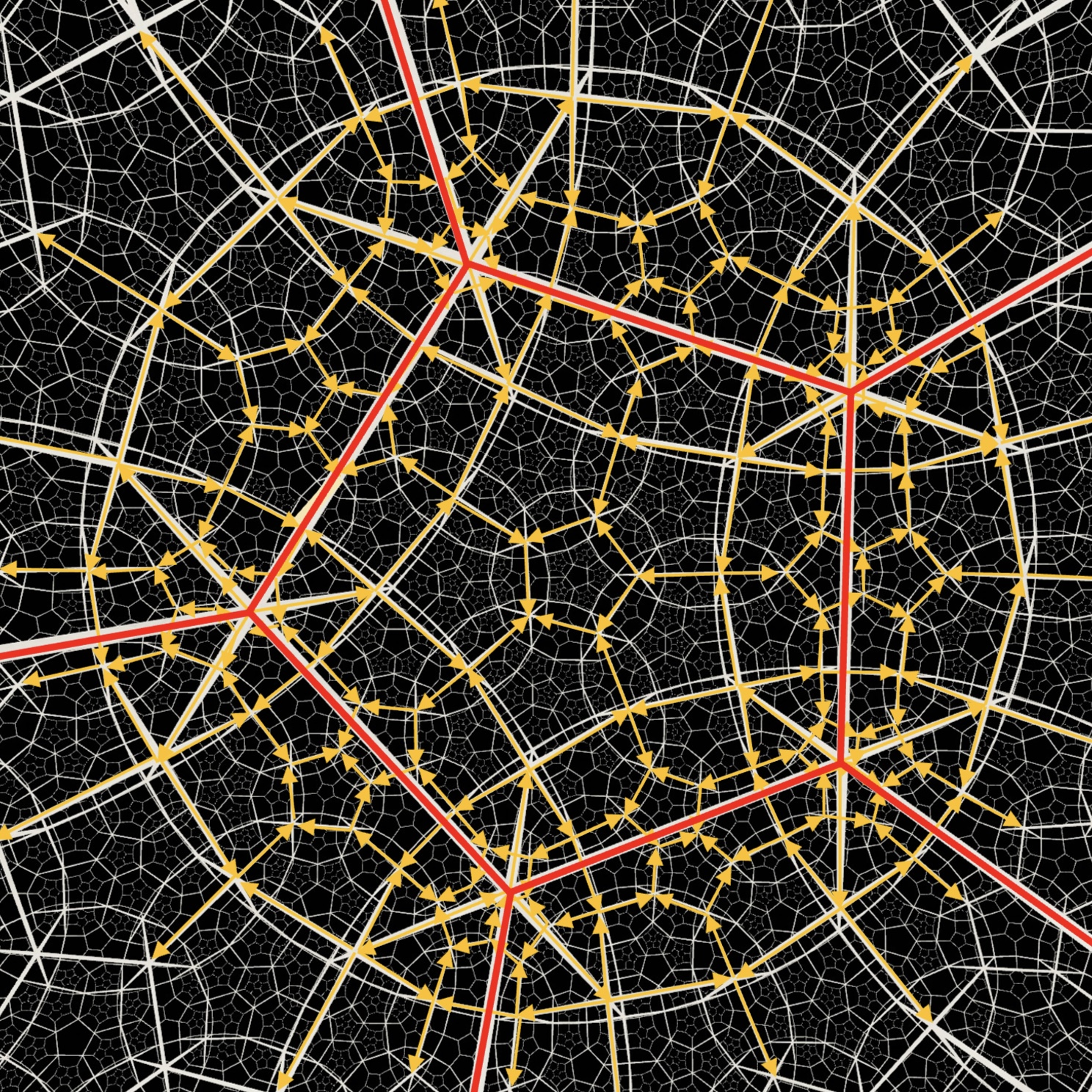}
    \includegraphics[width=0.32\linewidth]{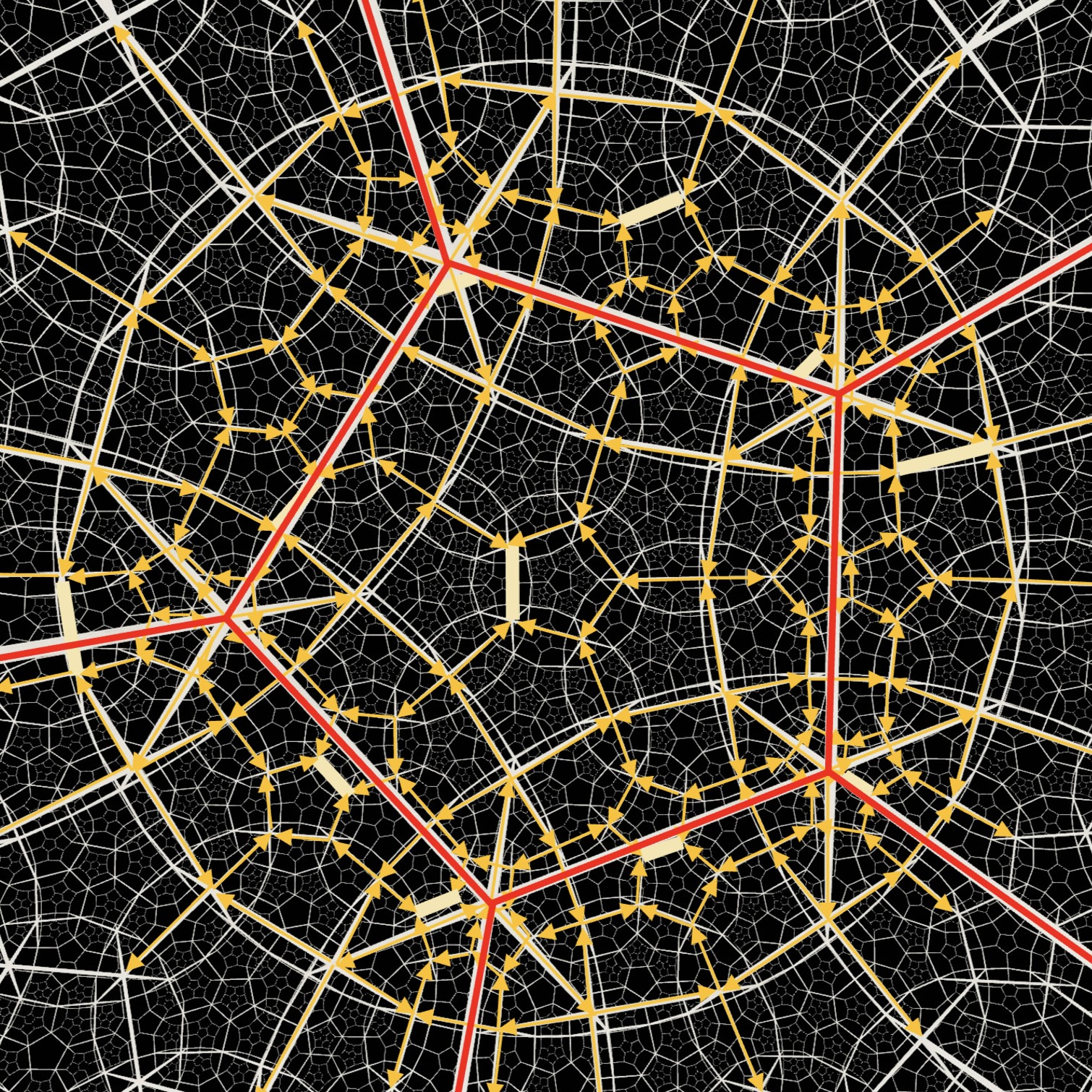}
    \includegraphics[width=0.32\linewidth]{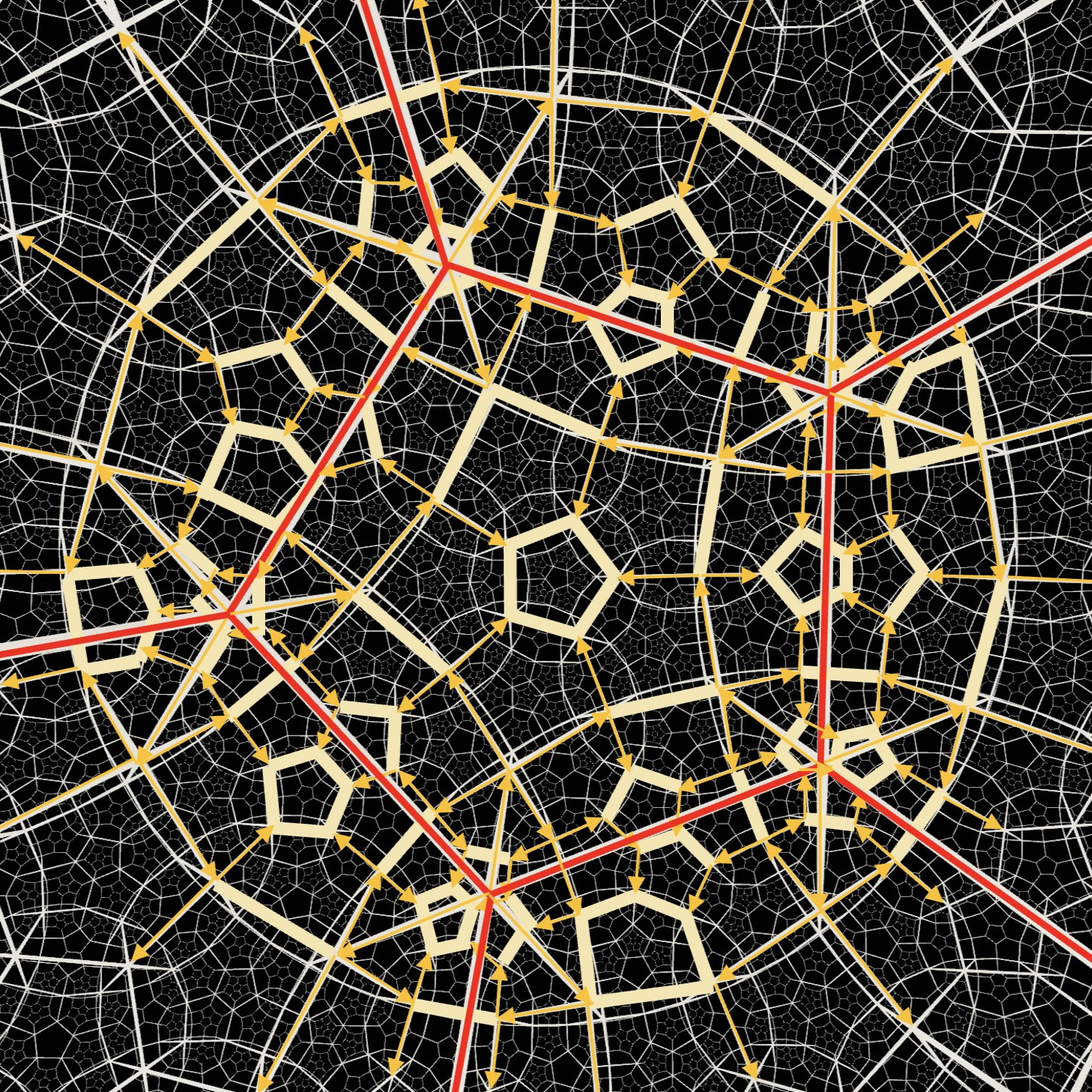}
    \caption{\textbf{Left:} A non-hyperinvariant encoding on the $\{5,3,4\}$ honeycomb, shown as a directed acyclic graph (DAG). \textbf{Middle:} Multi-tensor block decomposition of an inflation layer using angular 2-uniform vertex tensors. \textbf{Right:} Same network with angular 1-uniform vertex tensors, yielding a genuine higher-dimensional hyperinvariant structure. In both cases, orange arrows denote encoding flow, and thick yellow edges indicate multi-tensor blocks.}
    \label{fig:MTB decomposition}
\end{figure*}

For an HTN or code to be well-defined, bulk information must be isometrically encodable into the boundary. If no multi-tensor isometries are enforced, then the encoding structure corresponds to a directed acyclic graph (DAG) on the tiling—see Fig.~\ref{fig:MTB decomposition} (left). In such a DAG, each cell must contain at least one sink and one source, corresponding to solid angles at vertex positions. This necessity motivates the angular $k$-uniformity condition.

Under DAG analysis, hyperinvariance can be interpreted as the *removal* of specific directional arrows in order to cap the number of incoming edges at each vertex. For instance, in the angular $k=2$ case on $\{5,3,4\}$ (middle), a 2D-style multi-tensor isometry structure arises, while $k=1$ (right) yields a genuinely higher-dimensional configuration.

As demonstrated in these examples, enforcing multi-tensor isometries on large blocks can guarantee that isometry holds for all smaller sub-blocks. Thus, Table~\ref{tab:overview} lists only the largest representative block for each $(\text{lattice}, k)$ pair.

\begin{figure}
    \centering
    \includegraphics[width=0.48\linewidth,page=1]{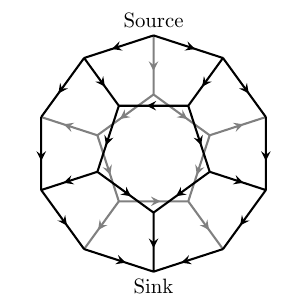}
    \includegraphics[width=0.48\linewidth,page=2]{Graphics/MTB_Decomposition/DAG_Analysis.pdf}
    \caption{\textbf{DAG Analysis.} \textbf{Left:} A DAG orientation on a dodecahedron, showing a unique sink and source. \textbf{Right:} DAG on a pentagon, embedded as a face of the dodecahedron.}
    \label{fig:DAG Analysis}
\end{figure}

\begin{figure}
    \centering
    \includegraphics[width=0.28\linewidth,page=2]{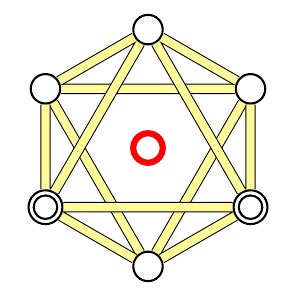}
    \includegraphics[width=0.677\linewidth,page=4]{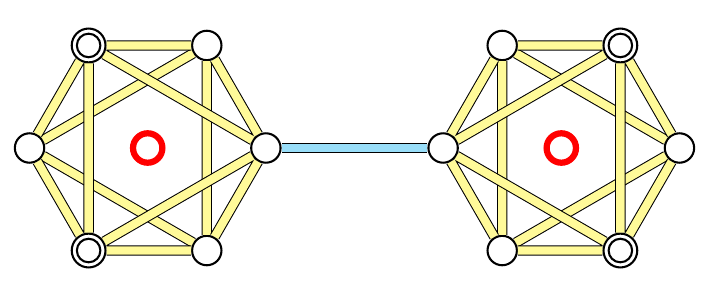}
    \caption{Vertex figure representations of $k{=}2$ multi-tensor blocks. Double-circled nodes denote input legs in the isometry condition.}
    \label{fig:order-5 cubic angular 2-uniform TIC}
\end{figure}

\begin{figure*}[htbp]
  \centering
  \begin{minipage}[t]{0.48\textwidth}
    \centering
    \includegraphics[width=0.28\textwidth,page=1]{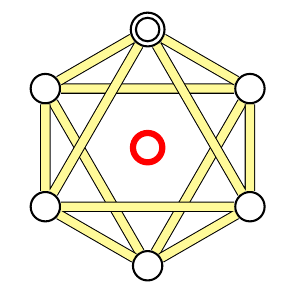}
    \hfill
    \includegraphics[width=0.677\textwidth,page=3]{Graphics/MTB_Decomposition/MTB_Decomposition.pdf}
    \includegraphics[width=\linewidth,page=5]{Graphics/MTB_Decomposition/MTB_Decomposition.pdf}
  \end{minipage}%
  \hfill
  \begin{minipage}[t]{0.48\textwidth}
  \vspace{-72pt}
    \centering
    \includegraphics[width=\linewidth,page=6]{Graphics/MTB_Decomposition/MTB_Decomposition.pdf}
  \end{minipage}

  \caption{Vertex figure representations of $k{=}1$ multi-tensor blocks. Input legs are marked with double circles.}
  \label{fig:VF-MTB-k=1}
\end{figure*}

We now turn to boundary correlation. Unlike HaPPY codes, hyperinvariant networks are capable of supporting nontrivial two-point functions, similar to MERA. However, hyperinvariance does \emph{not} guarantee such correlations. The key is whether a local operator on the boundary can propagate into the bulk.

\begin{figure}
    \centering
    \includegraphics[width=0.48\linewidth,page=3]{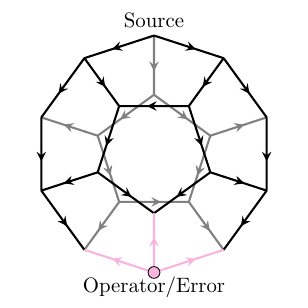}
    \includegraphics[width=0.48\linewidth,page=4]{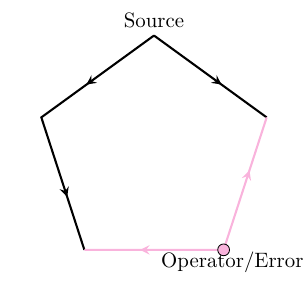}
    \caption{\textbf{Boundary operator propagation under DAG structure.} A magenta operator on a sink vertex may or may not propagate into the bulk depending on the angular $k$-uniformity.}
    \label{fig:Correlation}
\end{figure}

As argued in Refs.~\cite{HTN,HIC-ququart,HIC-qubit,HI-MERA}, HaPPY codes suppress correlations because single-site boundary operators can be corrected and erased in a very shallow region near the boundary. In the DAG perspective (Fig.~\ref{fig:Correlation}), if a sink vertex has $k_{\max}$ inputs and is acted on by an single-site operator, then all adjacent vertices have $\leq k_{\max}{-}1$ inputs and may simply treat it as an unknown input indices and leaving the operator in a local loop without propagating its influence to deeper regions of the bulk. Hence, when $k = k_{\max}$, the network suppresses bulk correlation.

In the concrete $\{5,3,4\}$ case, when $k=2$ although there is a hyperinvariant isometry condition of the same form of 2D case and the full 3D no-go condition is not met, the 2D version applies to boundary operators on a common $\{5,4\}$ plane. Thus, the presence of hyperinvariance does not imply non-triviality of correlation functions. A full summary of correlation and hyperinvariance across geometries appears in Appendix~\ref{App.A}, Table~\ref{tab:summary}.

As discussed in Refs.~\cite{HTN,HIC-ququart,HIC-qubit,HI-MERA}, HaPPY codes prohibit non-trivial correlations because single-site boundary operators can be corrected within a shallow region near the boundary. In the DAG perspective (Fig.~\ref{fig:Correlation}), if a sink vertex with $k_{\max}$ incoming edges is acted upon by a single-site operator or erasure error, then all adjacent vertices necessarily have at most $k_{\max}{-}1$ inputs. These vertices may treat the index connected to the tensor with operator applied as an unknown input and correct it, effectively trapping the operator in a local loop and preventing its influence from propagating into deeper bulk regions. Consequently, when $k = k_{\max}$, the network does not support nontrivial boundary correlations.

In the concrete $\{5,3,4\}$ example, although the $k=2$ case satisfies a hyperinvariant isometry condition analogous to the 2D scenario, the full 3D no-go condition is not violated. However, the 2D argument still applies to boundary operators supported on a common $\{5,4\}$ plane inside the honeycomb. Therefore, the presence of hyperinvariance does not necessarily imply the universal existence of nontrivial correlation functions. A comprehensive summary of correlation behavior and hyperinvariance across different geometries is provided in Appendix~\ref{App.A}, Table~\ref{tab:summary}.

\subsection{Complementary Recovery\label{Sec.4.3}}

\emph{Complementary recovery} is a fundamental error-correcting feature of AdS/CFT duality. For a bipartition $\mathcal{H} = \mathcal{H}_\mathcal{A} \otimes \mathcal{H}_{\bar{\mathcal{A}}}$ of the boundary Hilbert space (up to a cutoff), this property ensures that any local bulk operator can be reconstructed exclusively on either $\mathcal{A}$ or $\bar{\mathcal{A}}$, but not both. Geometrically, the corresponding bulk regions $a$ and $a^c$ are demarcated by the Ryu--Takayanagi (RT) surface $\gamma_\mathcal{A}$---an extremal surface (geodesics in 2D) whose area determines the leading $O(N^2)$ contribution to the boundary entanglement entropy $S_\mathcal{A} \equiv -\text{tr}(\rho_\mathcal{A} \log \rho_\mathcal{A})$, where $\rho_\mathcal{A} = \text{tr}_{\bar{\mathcal{A}}}(\rho)$. Including subleading corrections, the entropy takes the form

\begin{equation}
S_A = \frac{\text{area}(\gamma_A)}{4G} + S_a + O(G),
\end{equation}

with $S_a$ denoting the bulk entropy between $a$ and $a^c$ \cite{RT-formula}.

\begin{figure*}
    \includegraphics[width=0.48\linewidth]{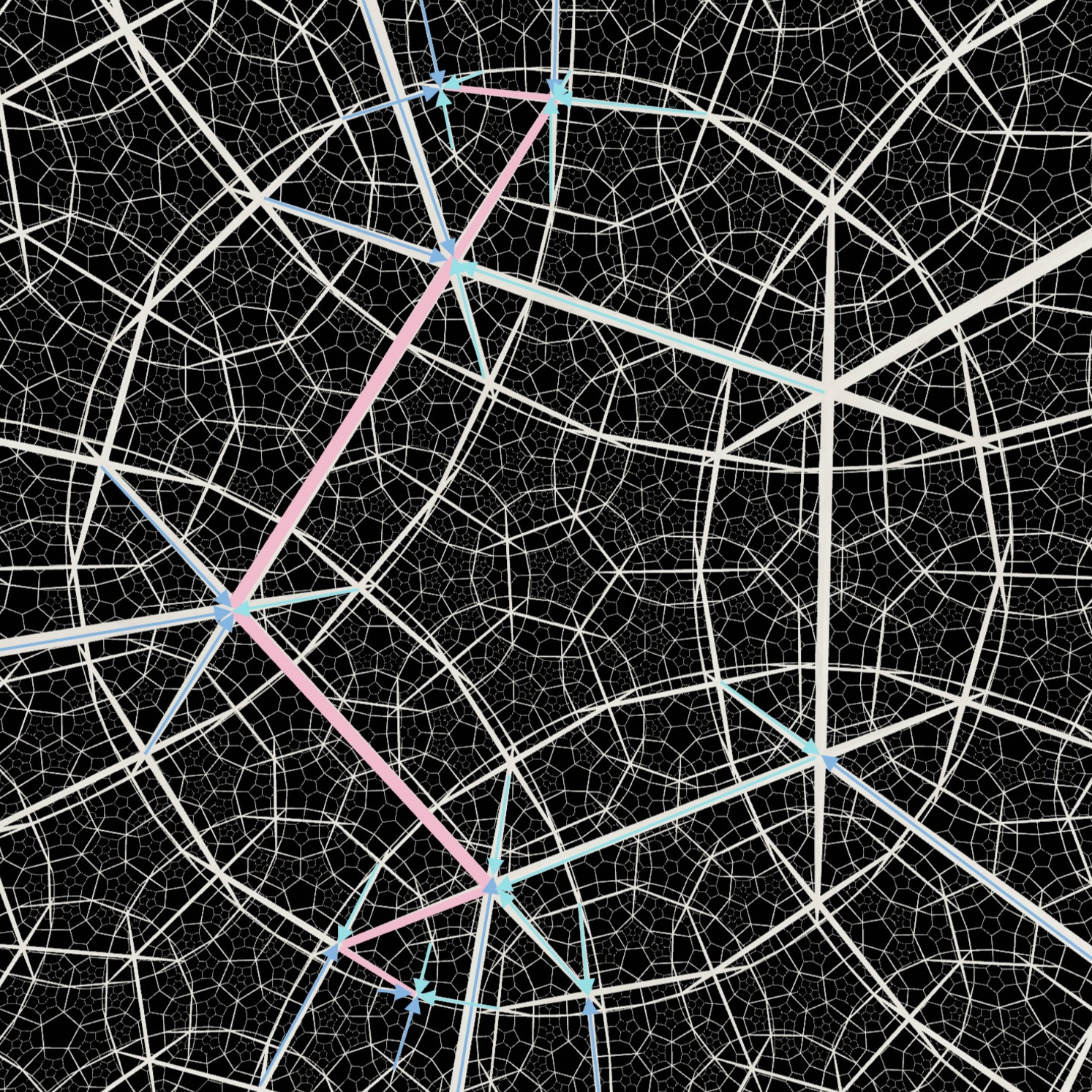}
    \includegraphics[width=0.48\linewidth]{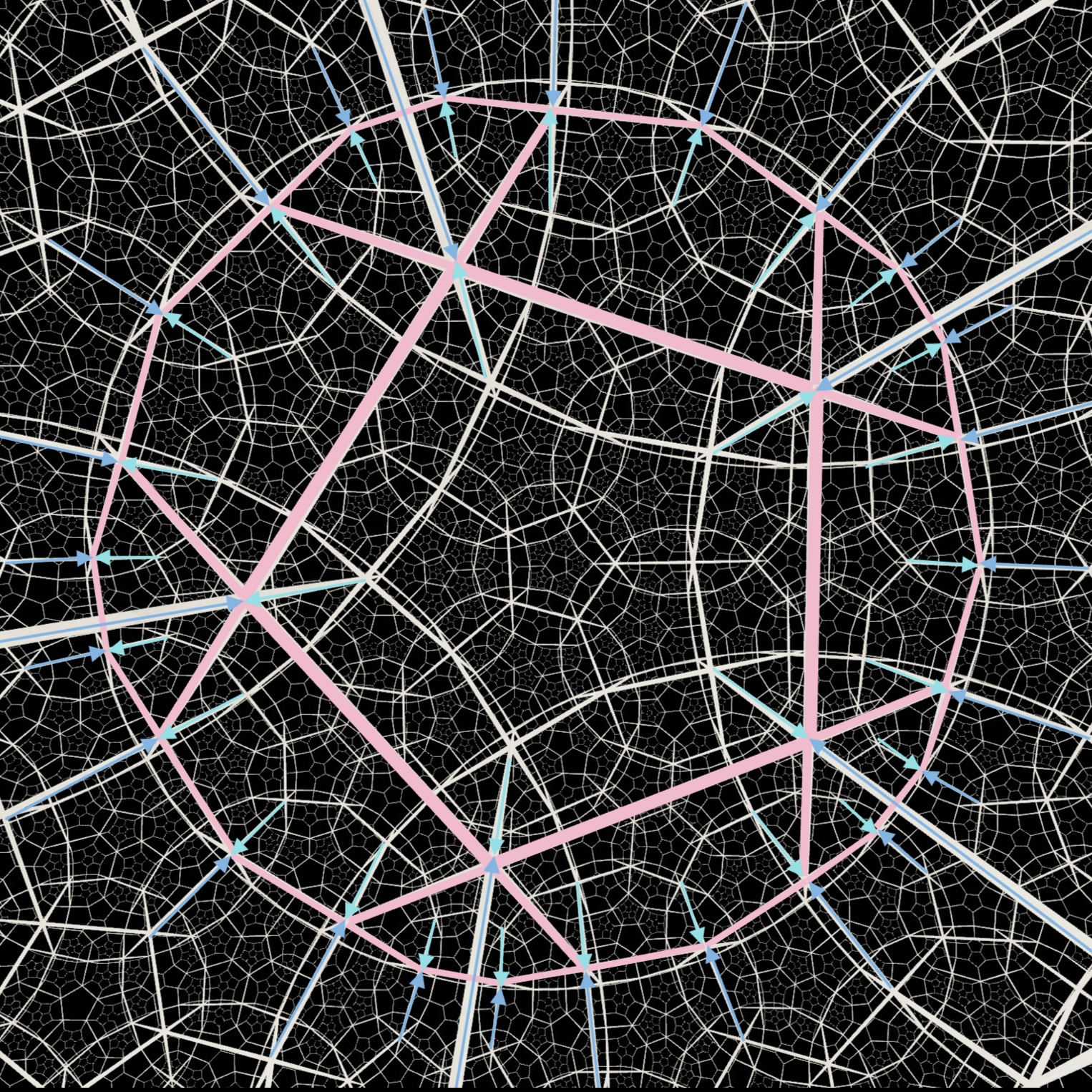}
        \caption{\textbf{left:} Residual area on \{5,3,4\} honeycomb in $k=2$ case. \textbf{Right:} Residual area on \{5,3,4\} honeycomb in $k=1$ case. The magenta region is the residual area, the sky blue and light sky blue arrows shows the decoding flow from two boundary regions.}
    \label{fig:Residual}
\end{figure*}

While holographic codes generically support complementary recovery \cite{HaPPY,Block-perfect_Steane}, the hyperinvariant Evenbly codes \cite{HIC-qubit,HIC-ququart} exhibit only \emph{approximate} complementary recovery with a residual area cannot be recovered as shown in Fig.~\ref{fig:Residual} due to their isometry constraints. In \cite{HIC-ququart}, the residual region is interpreted as the \emph{state dependent} area term in AdS/CFT under quantum corrections.

Here, we demonstrate that this behavior extends to higher-dimensional hyperinvariant codes, with the deviation from exact complementary recovery governed by two factors: (1) the degree of angular $k$-uniformity of the vertex tensors, and (2) the lattice structure of the network.

\begin{figure}
    \includegraphics[width=0.48\linewidth,page=2]{Graphics/Residual_and_Erasure/Residual.pdf}
    \includegraphics[width=0.48\linewidth,page=1]{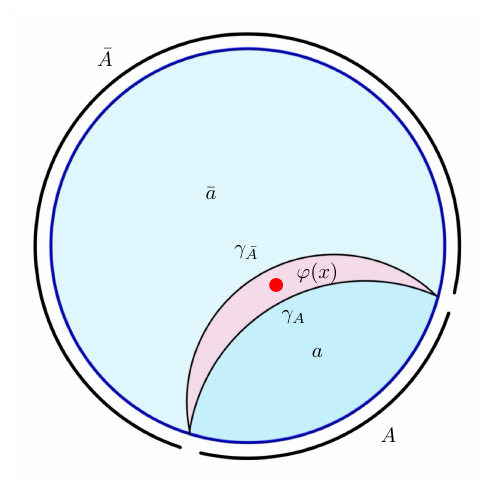}
        \caption{\textbf{left:} Exact complementary recovery typical in HaPPY-like codes. The boundary regions $A$ and its complementary $\bar{A}$ shares a same extremal surface $\gamma$, the reconstructible wedges $a$ and $\bar{a}$ corresponding to the entanglement wedges bounded by the extremal surface. Any bulk operator $\varphi(x)$ can be reconstructed either from $A$ or $\bar{A}$. \textbf{Right:} Hyperinvariant tensor networks generally shows approximate complementary }
    \label{fig:Complementary Recovery}
\end{figure}

\begin{figure}
    \centering
    \includegraphics[height=2cm,page=3]{Graphics/Residual_and_Erasure/REU.pdf}
    \includegraphics[height=2cm,page=2]{Graphics/Residual_and_Erasure/REU.pdf}
    \includegraphics[height=2cm,page=1]{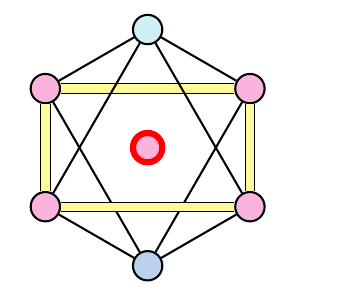}
    \includegraphics[height=2cm,page=4]{Graphics/Residual_and_Erasure/REU.pdf}
    \includegraphics[height=2cm,page=5]{Graphics/Residual_and_Erasure/REU.pdf}
    \caption{\textbf{Top:} The boundary condition of residual region in angular 1-uniform case, while all the decoding steps are prohibited.\textbf{Bottom:} Angular 1-uniform vertices in the residual region is shown on the left. Angular 2-uniform vertices in the residual region is shown on the right. The yellow edges represent the 2D extremal surface where the residual area is lying on.}
    \label{fig:Tensor in Residual}
\end{figure}

In the $\{5,3,4\}$ case, an extremal surface is a hyperbolic $\{5,4\}$ tiling in the honeycomb, like the magenta area shown on the left of Fig.~\ref{fig:Residual}. We may consider the case that $\mathcal{A}$ is the boundary of the honeycomb on one side of the extremal surface. In $k=2$ case if the entire boundary of the extremal surface is also contained in $\mathcal{A}$, then the exact complementary recovery is achieved. However when we have a boundary condition that the boundary of the $\{5,4\}$ extremal surface is also bipartite, as the case in \cite{HIC-ququart}, and one of them is contained in $\mathcal{A}$, then we will get a 1D residual area like the 2D case \cite{HIC-ququart}. However in $k=1$ case, even the boundary region of $\{5,4\}$ is entirely included in $\mathcal{A}$, there is no exact complementary recovery as shown in Fig.~\ref{fig:Tensor in Residual}. And the residual region is exactly the $\{5,4\}$ tiling, while fluctuation analogous to the 2D case is allowed \cite{HIC-ququart}.

\subsection{Uberholography and Multi-angular \(k\)-Uniformity\label{Sec.4.4}}

\emph{Uberholography} refers to the ability to reconstruct bulk information from \emph{disconnected} boundary regions—a property that distinguishes it from conventional holographic duality, where reconstruction typically relies on connected boundary segments (as in AdS/Rindler reconstruction). This phenomenon was first identified in the HaPPY code~\cite{HaPPY}, where it emerges as a direct consequence of the permutation invariance of perfect tensors. While subsequent variants using block-perfect tensors have also displayed uberholography, it has long been regarded as a non-generic feature, closely tied to the perfectness of seed tensors.

\begin{figure}
    \includegraphics[width=0.48\linewidth,page=1]{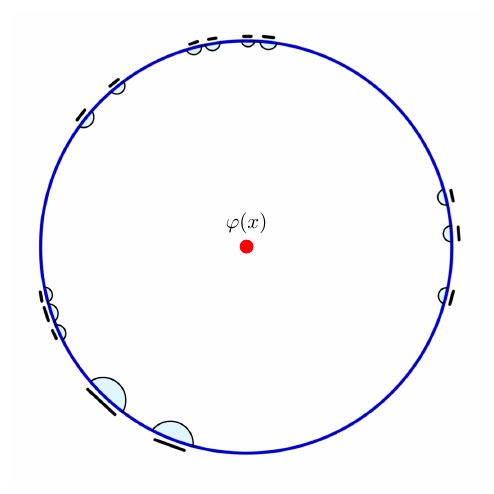}
    \includegraphics[width=0.48\linewidth,page=2]{Graphics/Uberholography/Uberholography.pdf}
    \caption{\textbf{Uberholography.} A disconnected boundary region is indicated by black marks around the boundary circle. For codes lacking uberholography (e.g., the Evenbly-type qubit codes in~\cite{HIC-qubit}), such a region supports only separate, shallow reconstruction wedges (right), and a deep bulk operator $\varphi(x)$ remains unrecoverable. In contrast, when uberholography is present (e.g., in the HaPPY code), the same boundary region yields a connected, tree-like reconstruction wedge (left), allowing $\varphi(x)$ to be recovered.}
    \label{fig:Uberholography}
\end{figure}

Notably, earlier studies of hyperinvariant tensor networks~\cite{HTN,HI-MERA,HIC-qubit,HIC-ququart}—which aim to support non-trivial boundary correlation functions—did not exhibit uberholography. This absence is unsurprising: these networks are constructed without perfect tensors, and the explicit examples in~\cite{HIC-ququart,HIC-qubit} indeed fail to exhibit uberholographic reconstruction. This aligns with the common belief that non-trivial correlations require local tensors to deviate significantly from perfectness—a requirement that seemingly conflicts with the strict isometries needed for uberholography.

Here, we demonstrate that hyperinvariance, uberholography, and non-trivial correlation functions can coexist. The key enabling condition is what we call \emph{multi-angular \(k\)-uniformity}—a generalization of angular \(k\)-uniformity in which isometric input regions may span multiple disconnected angular sectors.

\begin{definition}[Multi-Angular \(k\)-Uniformity]
Let \( A \) be a tensor satisfying condition (i) of angular \(k\)-uniformity. We say that \( A \) is \emph{multi-angular \(k\)-uniform} if the isometry condition extends to input regions composed of disjoint unions of strongly angularly connected subsets \( I_1, I_2, \dots \subset \{1,\dots,n\} \), where each \( I_\ell \) satisfies \( |I_\ell| \leq k \) and is angularly disconnected from the others.

That is, the total input region \( \mathcal{I}_{\mathrm{in}} := \bigcup_\ell I_\ell \) defines an isometry:
\[
\bigotimes_{i \in L \cup \mathcal{I}_{\mathrm{in}}} \mathcal{H}_i
\longrightarrow
\bigotimes_{j \notin L \cup \mathcal{I}_{\mathrm{in}}} \mathcal{H}_j \, ,
\]
where \( L \) denotes the set of logical indices.
\end{definition}

\begin{figure}
    \includegraphics[width=0.24\linewidth,page=6]{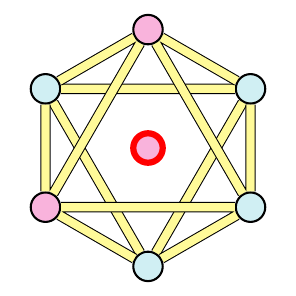}
    \includegraphics[width=0.24\linewidth,page=9]{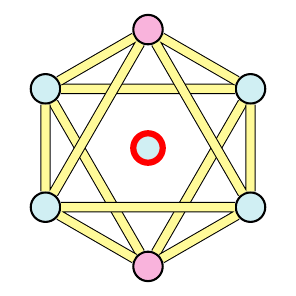}
    \includegraphics[width=0.24\linewidth,page=8]{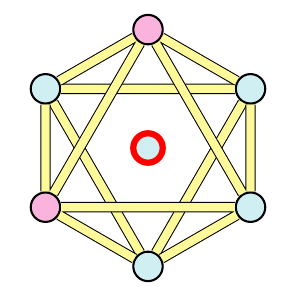}
    \includegraphics[width=0.24\linewidth,page=9]{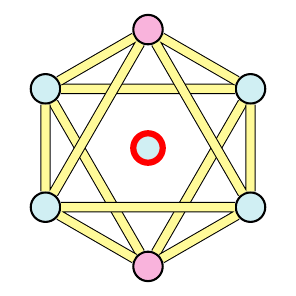}
    \caption{\textbf{Recovery behavior under erasure.} The left two panels show the recovery pattern of a multi-angular 1-uniform tensor under two physical erasures. The right two panels depict the same erasure pattern for a multi-angular 2-uniform tensor. Sky blue regions correspond to accesible/recoverable information, while magenta regions indicate irrecoverable data.}
    \label{fig:Uberholography Erasure}
\end{figure}

In the $\{5,3,4\}$ setting, the new isometries shown in Fig.~\ref{fig:Uberholography Erasure} preserve hyperinvariance due to the angular disconnectedness of the input regions. For the $k=1$ case, these additional isometries do not obstruct the propagation of boundary operators except in rare instances deep in the bulk, where two operator-supported indices may become antidotal on a vertex. Such cancellations affect only one vertex and do not affect the global feature of boundary correlation functions. When $k=2$, the network admits isometries for all input pairs, effectively rendering the local tensors perfect. As a result, non-trivial two-point correlations and two-point fatal erasure patterns are entirely prohibited-consistent with the analysis in~\cite{HI-MERA}.

This picture extends naturally to the 2D case. Recall the 2D case on the \(\{5,4\}\) tiling studied in Refs.~\cite{HIC-ququart,HIC-qubit}. With the introduction of multi-angular \(k\)-uniformity, a four-qudit perfect tensor code can be naturally reinterpreted as a multi-angular 1-uniform tensor. This perspective highlights that such tensors are capable of simultaneously exhibiting hyperinvariance and supporting non-trivial boundary correlations—two properties previously viewed as incompatible. While the explicit construction of such a perfect code and accompanying edge tensor that satisfies the required multi-tensor isometry remains an open task, this conceptual unification opens the door to new design principles for holographic tensor networks.

\subsection{Constructions of Vertex Codes \label{Sec.4.5}}

In this subsection, we will construct concrete examples that satisfies the hyperinvariance criterion for different $k$ on {5,3,4}, of which the construction methods can be extended to other hyperbolic honeycombs. A full summary will be presented in App.~\ref{App.B}.

Before we construct vertex tensor/code $A$, we recall that $A$ should be rotational invariant, hence we may first introduce an infinite family of rotational invariant CSS qubit codes can be defined on \emph{centrosymmetric} polytopes—i.e.\ each vertex \(v\) has a unique antipodal partner \(v'\), which is the key ingredient of constructing $A$ with desired symmetry and angular $k$-uniformity. We refer to it as the \emph{\(X\)–\(I\) codes} (the name echoes the celebrated \(X\)–Cube model), and the $[[4,1,2]]$ code constructed in \cite{HIC-qubit} can be considered as an $X$-$I$ code defined on a square. Actually it is the $[[4,1,2]]$ code which inspired our construction.

\begin{figure}
    \includegraphics[width=\linewidth]{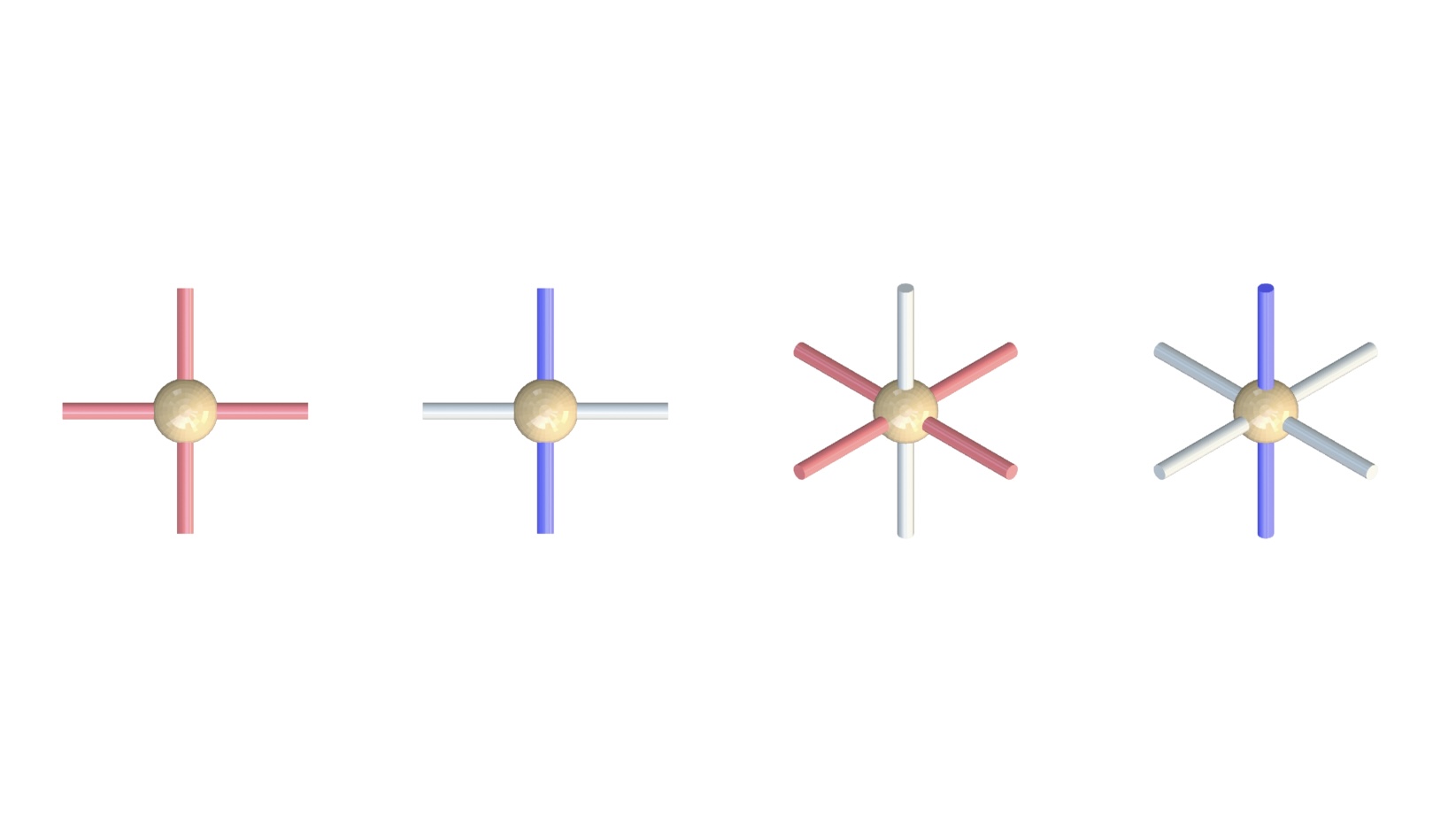}
    \caption{The stabilizers of $\Xi(4)$ and $\Xi(6)$ in the tensor network representation, which explains its name.}
    \label{fig:X-I}
\end{figure}

\begin{definition}[\(X\)–\(I\) Code]
The \(X\)–\(I\) code is the CSS code defined on $2m$ paired physical qubits with stabilizer group generated by:
\begin{itemize}
  \item \textbf{$X$-generators:} For each choice of two distinct pairs \(\{i,i'\}\) and \(\{j,j'\}\), the weight-4 operator
  \[
    \mathcal{G}_X=X_i\,X_{i'}\,X_j\,X_{j'}.
  \]
  \item \textbf{$Z$-generators:} For each antipodal pair \(\{i,i'\}\), the weight-2 operator
  \[
    \mathcal{G}_Z=Z_i\,Z_{i'}.
  \]
\end{itemize}
A convenient choice of logical operators is:
\begin{itemize}
  \item \(\bar X\) may be chosen as any weight-2 operator \(X_iX_{i'}\), or more generally as the product of \(X\)-operators over an odd number of antipodal pairs.
  \item \(\displaystyle\bar Z = \prod_{(i,i')} Z_i\), where the product runs over exactly one vertex in each antipodal pair (so \(\bar Z\) has weight \(v/2\)).
\end{itemize}
By construction, there are $m$ $Z$-generators and $m-1$ $X$-generators, which means only one logical qubit is encoded, hence is a $[[2m,1,2]]$ code. As shown in Fig.~\ref{fig:X-I}, the standard tensor network representation, the $Z$ generator looks like a ''$I$" word and $X$ generator looks like a ''$X$" word. Inspired by the $X$-Cube model with similar stabilizers, we call this family of code as $X$-$I$ code, and introduce a compact notation $\Xi(2m)$ for convenience.

By exchanging \(X\leftrightarrow Z\), one obtains an equivalent dual description in which the roles of the two stabilizer types are interchanged.
\end{definition}

\begin{figure}
    \includegraphics[width=0.24\linewidth,page=1]{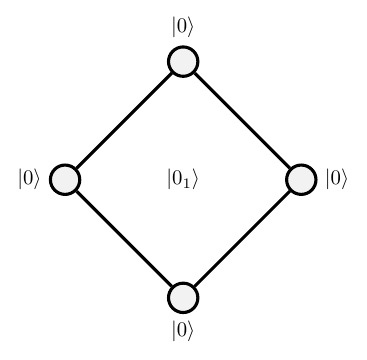}
    \includegraphics[width=0.24\linewidth,page=2]{Graphics/Truncation/X-I_4.pdf}
    \includegraphics[width=0.24\linewidth,page=3]{Graphics/Truncation/X-I_4.pdf}
    \includegraphics[width=0.24\linewidth,page=4]{Graphics/Truncation/X-I_4.pdf}
    \includegraphics[width=0.24\linewidth,page=5]{Graphics/Truncation/X-I_4.pdf}
    \includegraphics[width=0.24\linewidth,page=6]{Graphics/Truncation/X-I_4.pdf}
    \includegraphics[width=0.24\linewidth,page=7]{Graphics/Truncation/X-I_4.pdf}
    \includegraphics[width=0.24\linewidth,page=8]{Graphics/Truncation/X-I_4.pdf}

    \caption{An $\Xi$(4) code embedded in a square, which is first proposed in \cite{HIC-qubit} to construct HIC on \{5,4\} tiling.\label{fig:X-I(4)}}
\end{figure}

\begin{figure}
    \includegraphics[width=0.24\linewidth,page=1]{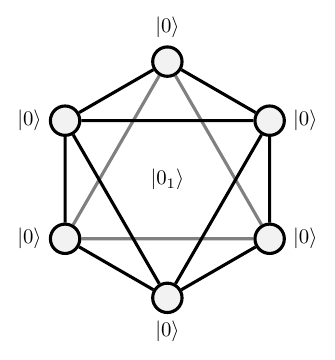}
    \includegraphics[width=0.24\linewidth,page=2]{Graphics/Truncation/X-I_6.pdf}
    \includegraphics[width=0.24\linewidth,page=3]{Graphics/Truncation/X-I_6.pdf}
    \includegraphics[width=0.24\linewidth,page=4]{Graphics/Truncation/X-I_6.pdf}
    \includegraphics[width=0.24\linewidth,page=5]{Graphics/Truncation/X-I_6.pdf}
    \includegraphics[width=0.24\linewidth,page=6]{Graphics/Truncation/X-I_6.pdf}
    \includegraphics[width=0.24\linewidth,page=7]{Graphics/Truncation/X-I_6.pdf}
    \includegraphics[width=0.24\linewidth,page=8]{Graphics/Truncation/X-I_6.pdf}
    \includegraphics[width=0.24\linewidth,page=9]{Graphics/Truncation/X-I_6.pdf}
    \includegraphics[width=0.24\linewidth,page=10]{Graphics/Truncation/X-I_6.pdf}
    \includegraphics[width=0.24\linewidth,page=11]{Graphics/Truncation/X-I_6.pdf}
    \includegraphics[width=0.24\linewidth,page=12]{Graphics/Truncation/X-I_6.pdf}

    \caption{An $\Xi$(6) code embedded in an octahedron, a hexagon or in a pair of antipodal triangles, since in all the three cases the antipodal relation between the six vertices are the same, The later two cases can also be viewed as subgraphs of the octahedron.\label{fig:X-I(6)}}
\end{figure}

The simplest $X$-$I(2m)$ code $X$-$I$(4) and  $X$-$I$(6) codes are shown in Fig.\ref{fig:X-I(4)} and Fig.\ref{fig:X-I(6)} respectively. The stabilizers can be spanned by generators of the form $\mathcal{G}_X$ and $\mathcal{G}_Z$ up to rotations. And the explicit codewords are given by
\begin{align}
    |\bar{0}\rangle_{\Xi(2m)}&=\frac{1}{\sqrt{2^{m-1}}}\sum_{i=1}^{2^{m-1}}|0_i\rangle\\
    |\bar{1}\rangle_{\Xi(2m)}&=\frac{1}{\sqrt{2^{m-1}}}\sum_{i=1}^{2^{m-1}}|1_i\rangle
\end{align}

\begin{theorem}[Full Polytope Symmetry]
Let \(P\) be a centrosymmetric polytope with full isometry group \(G\) (including all rotations and reflections).  Then the \(X\)–\(I\) code defined on \(P\) is invariant under the action of \(G\). Equivalently, for every \(g\in G\), the conjugation
\[
    S \;\mapsto\; g\,S\,g^{-1}
\]
maps each stabilizer generator and each logical operator of the \(X\)–\(I\) code to another element of the same type, and hence preserves the code space.
\end{theorem}

\begin{figure*}
    \includegraphics[trim=60 60 60 60, clip, height=3cm]{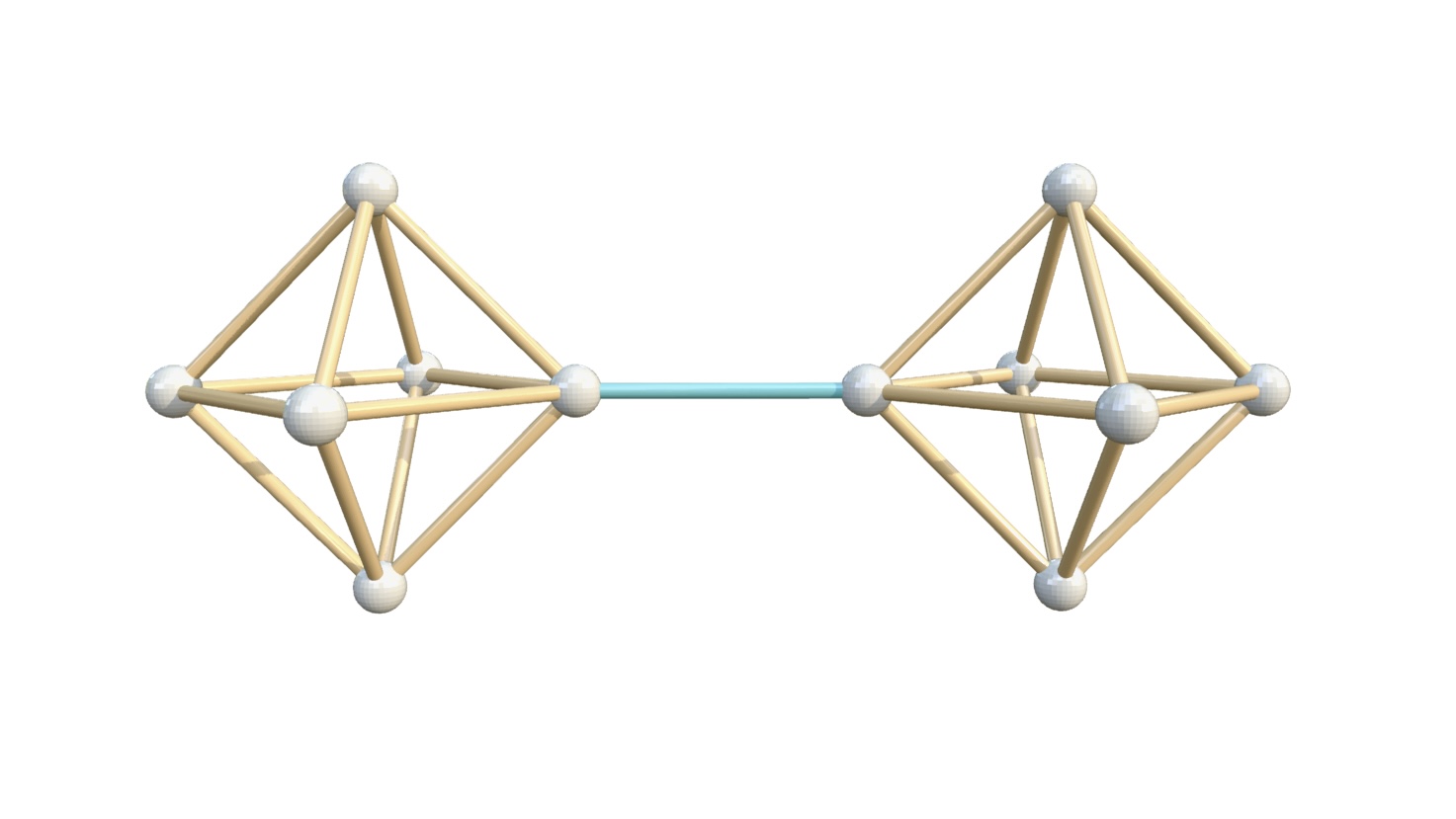}
    \includegraphics[height=3cm]{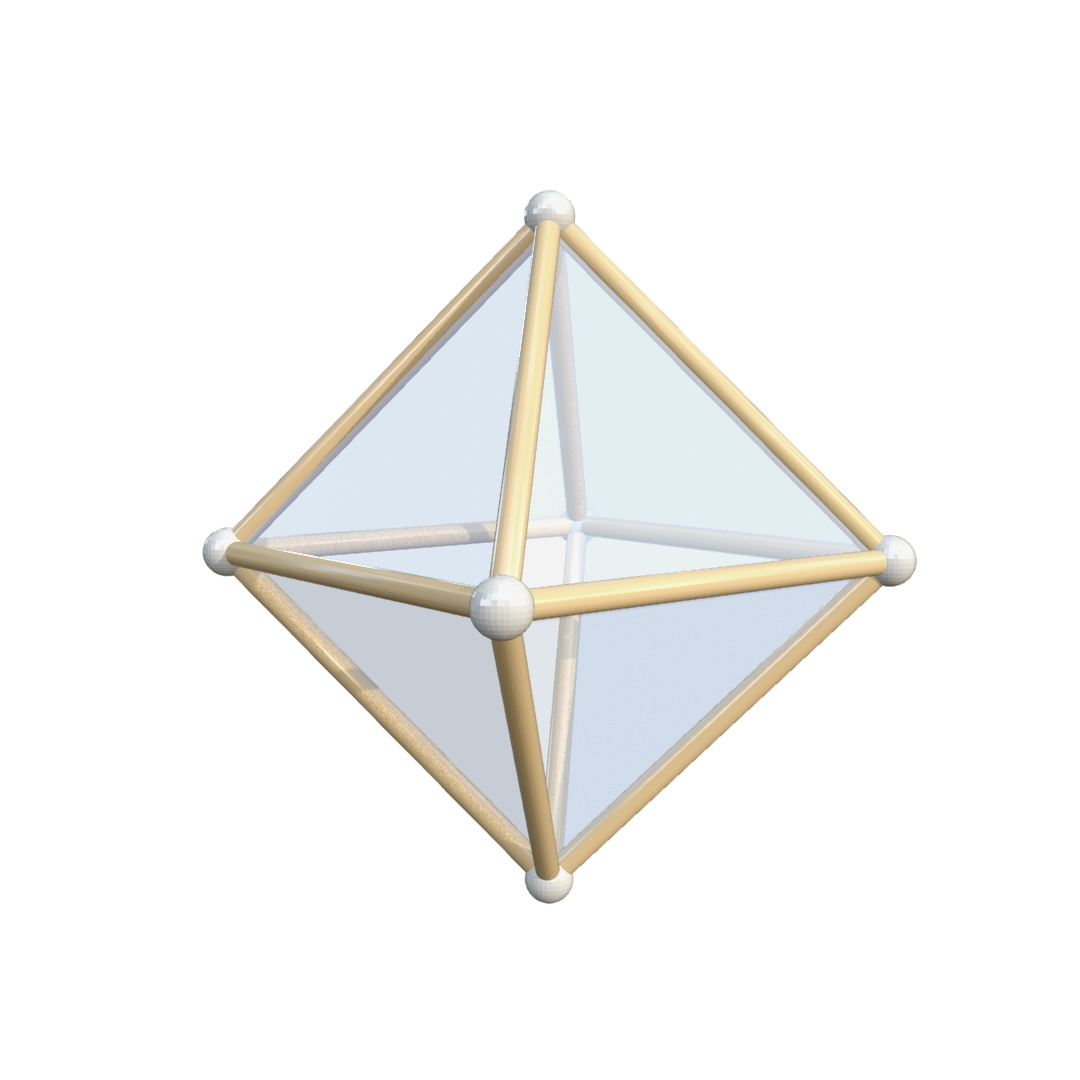}
    \includegraphics[trim=60 60 60 60, clip, height=3cm]{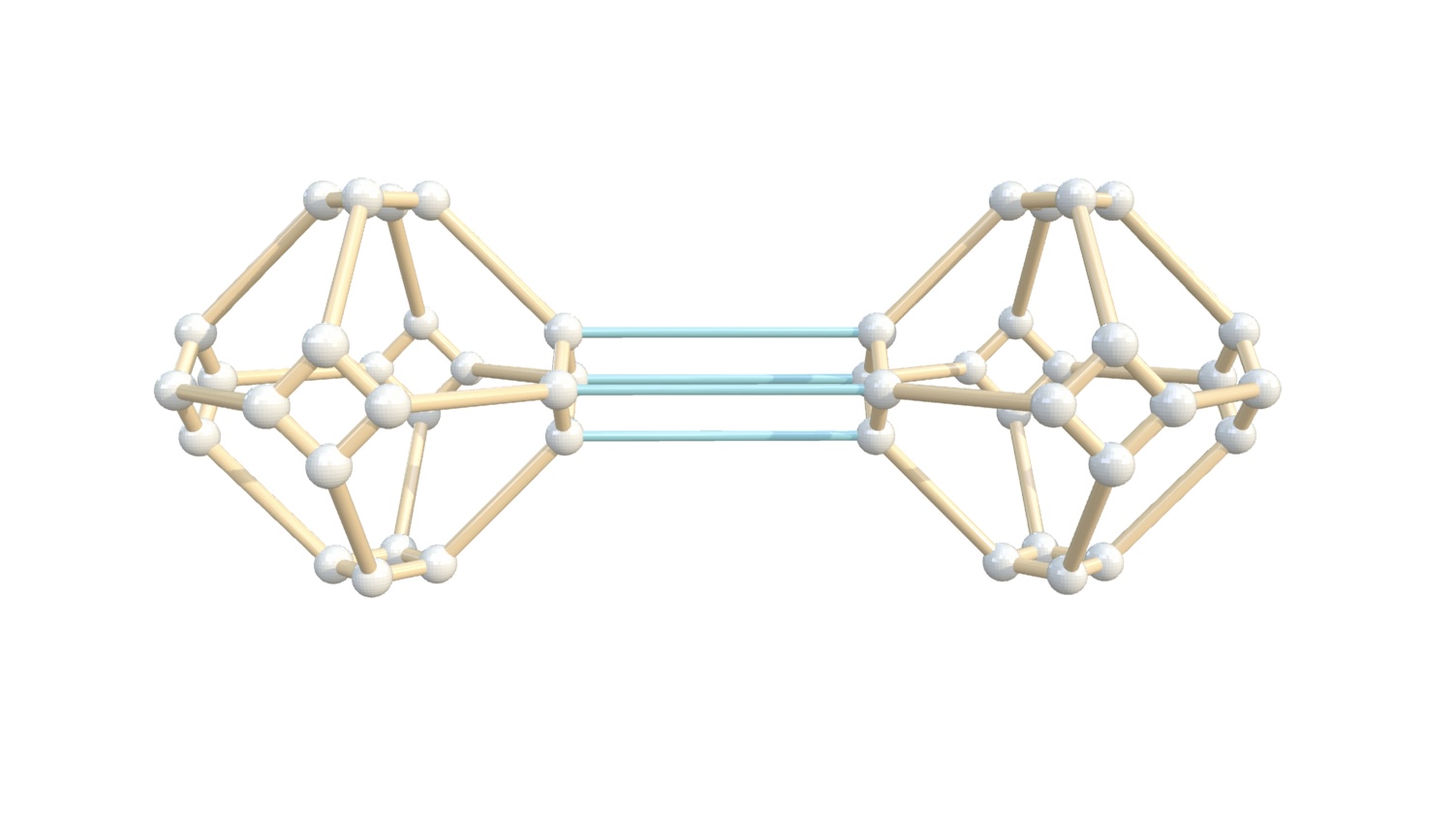}
    \includegraphics[height=3cm]{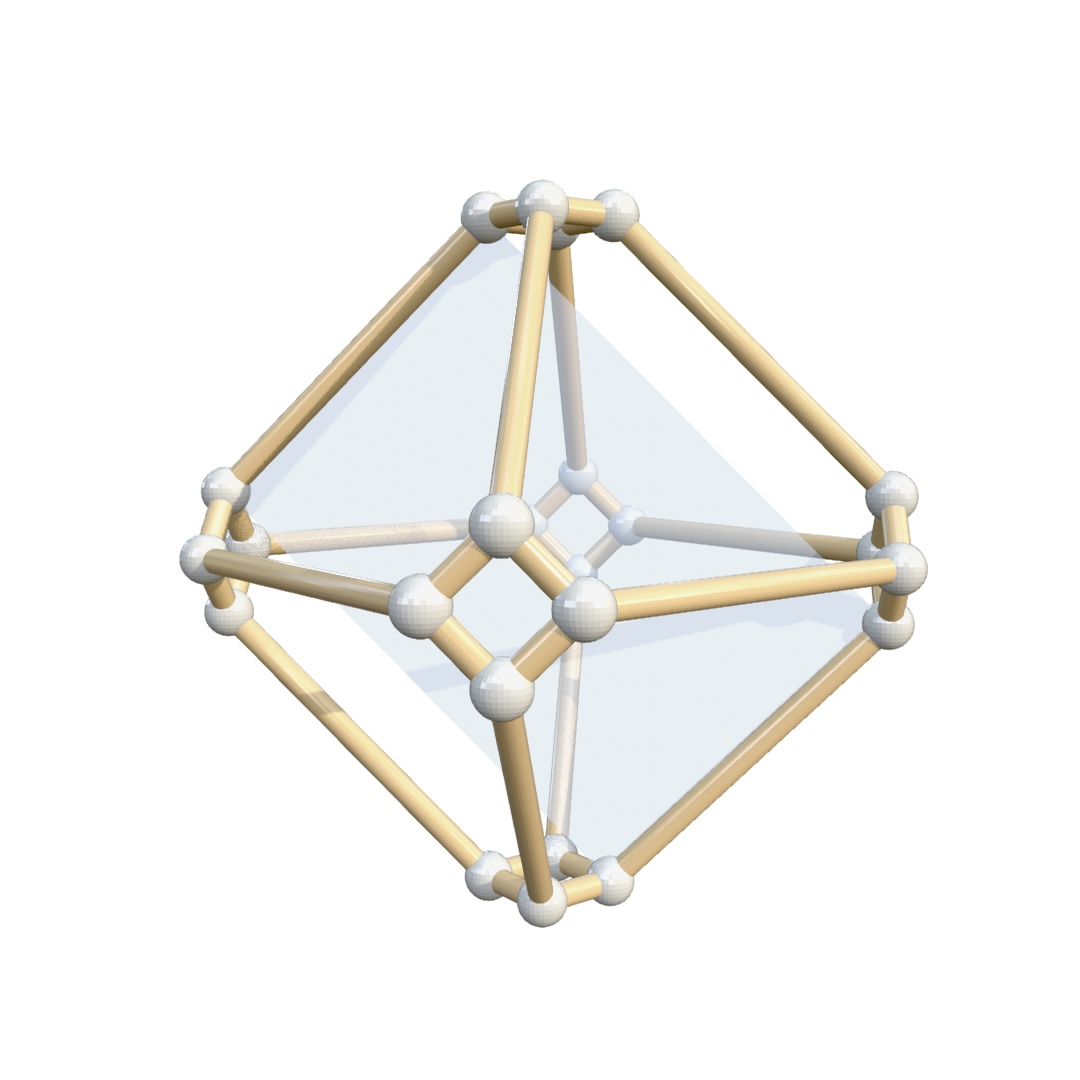}
    \includegraphics[trim=60 60 60 60, clip, height=3cm]{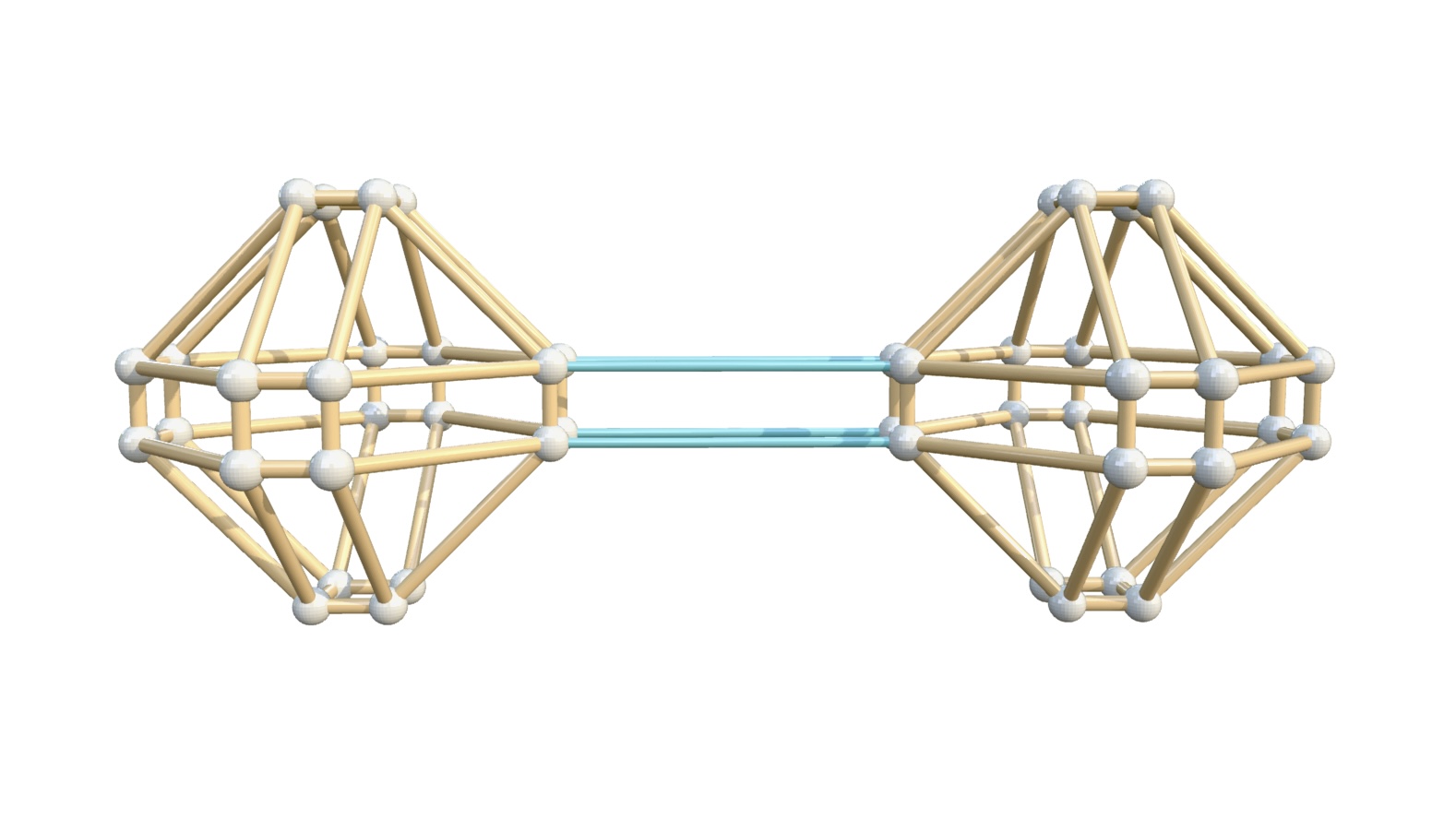}
    \includegraphics[height=3cm]{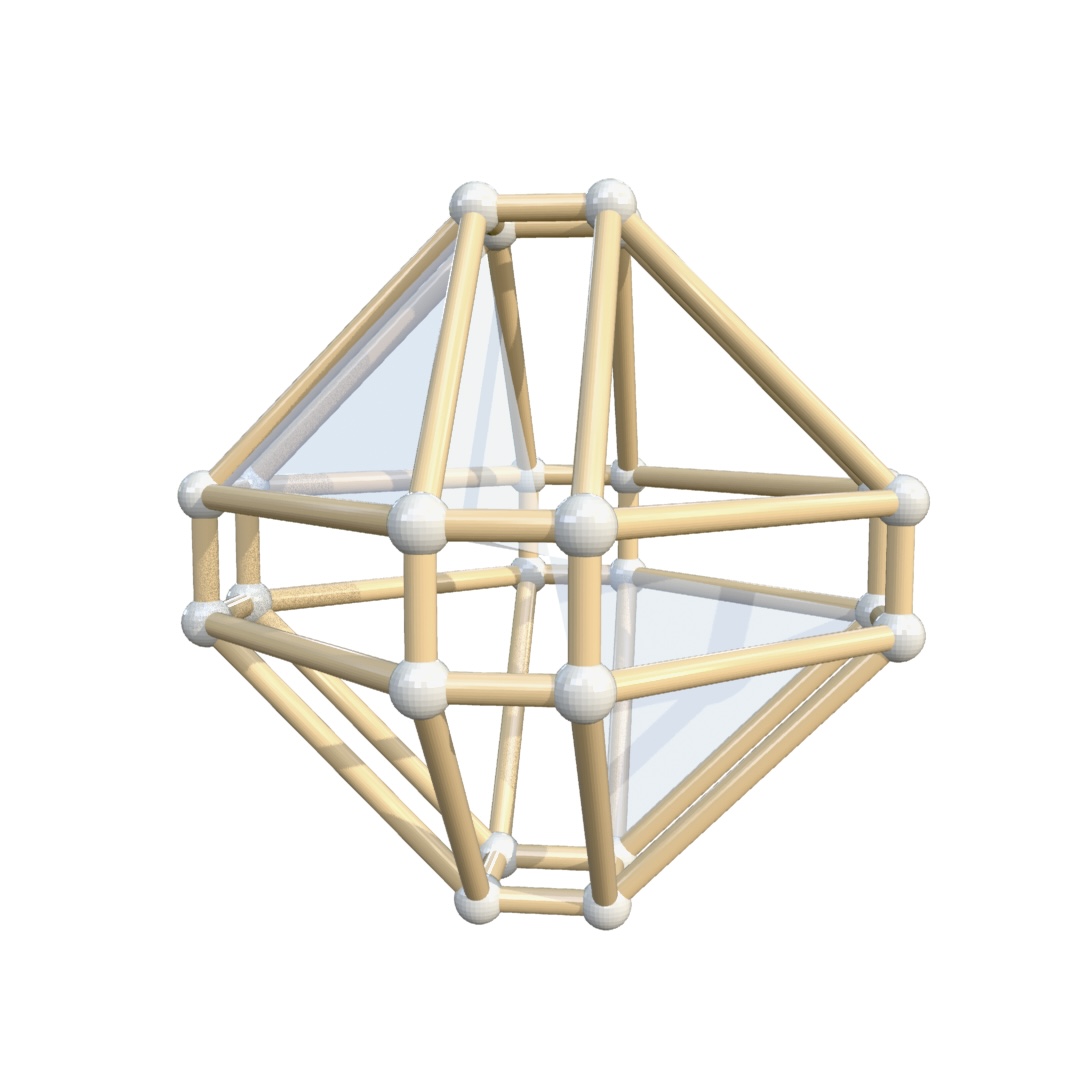}
    \includegraphics[trim=60 60 60 60, clip, height=3cm]{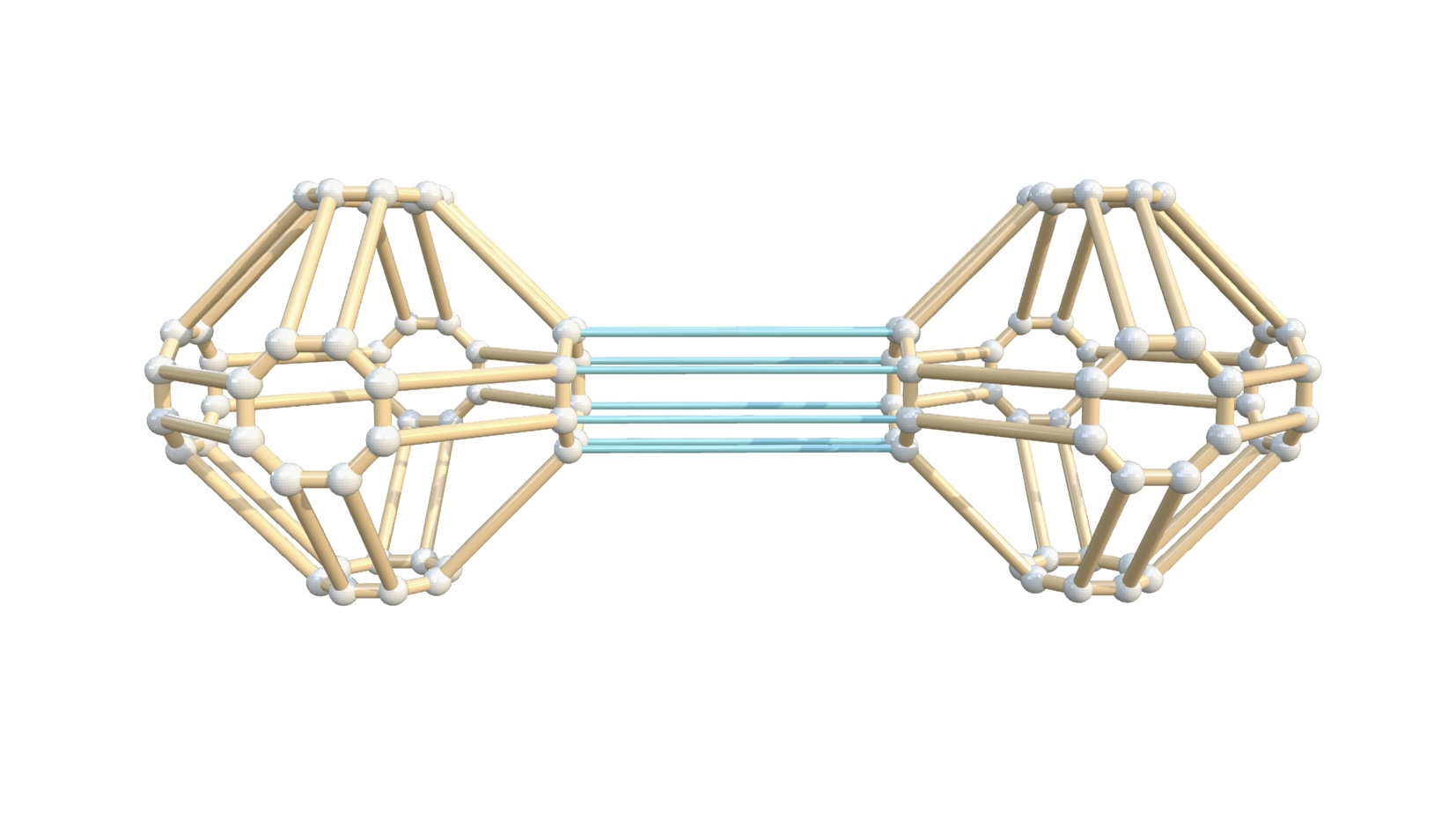}
    \includegraphics[height=3cm]{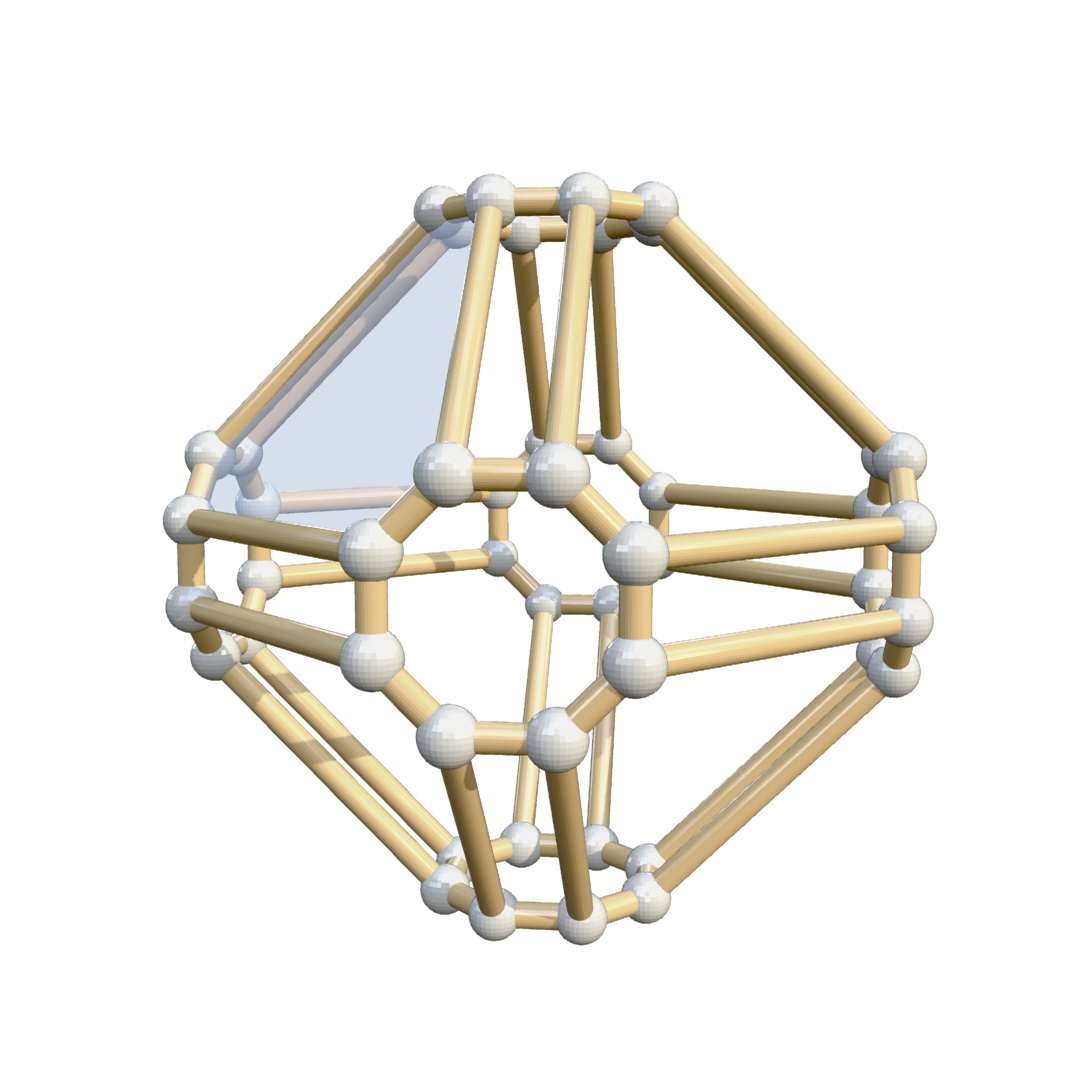}
    \caption{Four ways to construct vertex tensors for HTN/HIC on \{5,3,4\} honeycomb. \textbf{Top left:} Angular $2$-uniform code made with one $\Xi(6)$ code. \textbf{Top Right:} Angular $1$-uniform code made with six $\Xi(4)$ codes. \textbf{Bottom Left:} Angular $2$-uniform code made with four $\Xi(4)$ codes. \textbf{Bottom Right:} Multi-angular $1$-uniform code made with eight $\Xi(6)$ codes. The $X$-$I$ codes are defined on the shaded region right part of each design, and bipartition is respected to the antipodal relation. The edge tensors are simply products of Hadamard gates as shown on the left part of each design. \label{fig:Truncation Procedure}}
\end{figure*}

\begin{figure*}
    \includegraphics[width=0.24\linewidth,page=1]{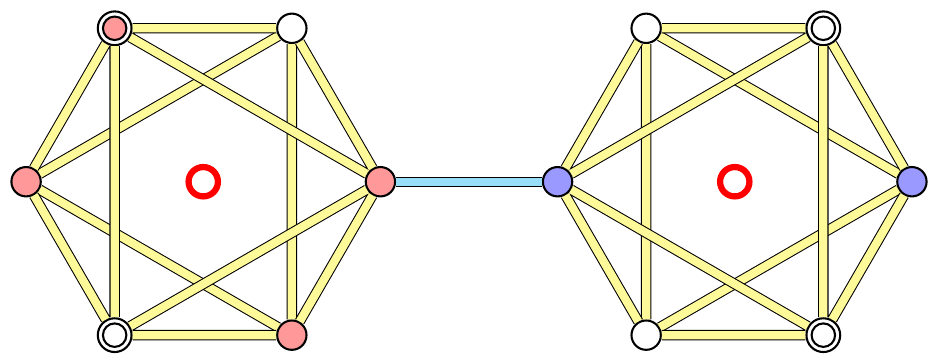}
    \includegraphics[width=0.24\linewidth,page=2]{Graphics/TIC/TIC_Proof_3,4.pdf}
    \includegraphics[width=0.24\linewidth,page=3]{Graphics/TIC/TIC_Proof_3,4.pdf}
    \includegraphics[width=0.24\linewidth,page=4]{Graphics/TIC/TIC_Proof_3,4.pdf}
    \caption{Graphical proof of the multi-tensor isometry condition via operator pushing. The red filling on nodes represent a single-qubit $X$ operator, blue filling the $Z$ operator. \label{fig:Isometry Proof A}}
\end{figure*}

\begin{figure*}
    \includegraphics[width=0.48\linewidth,page=1]{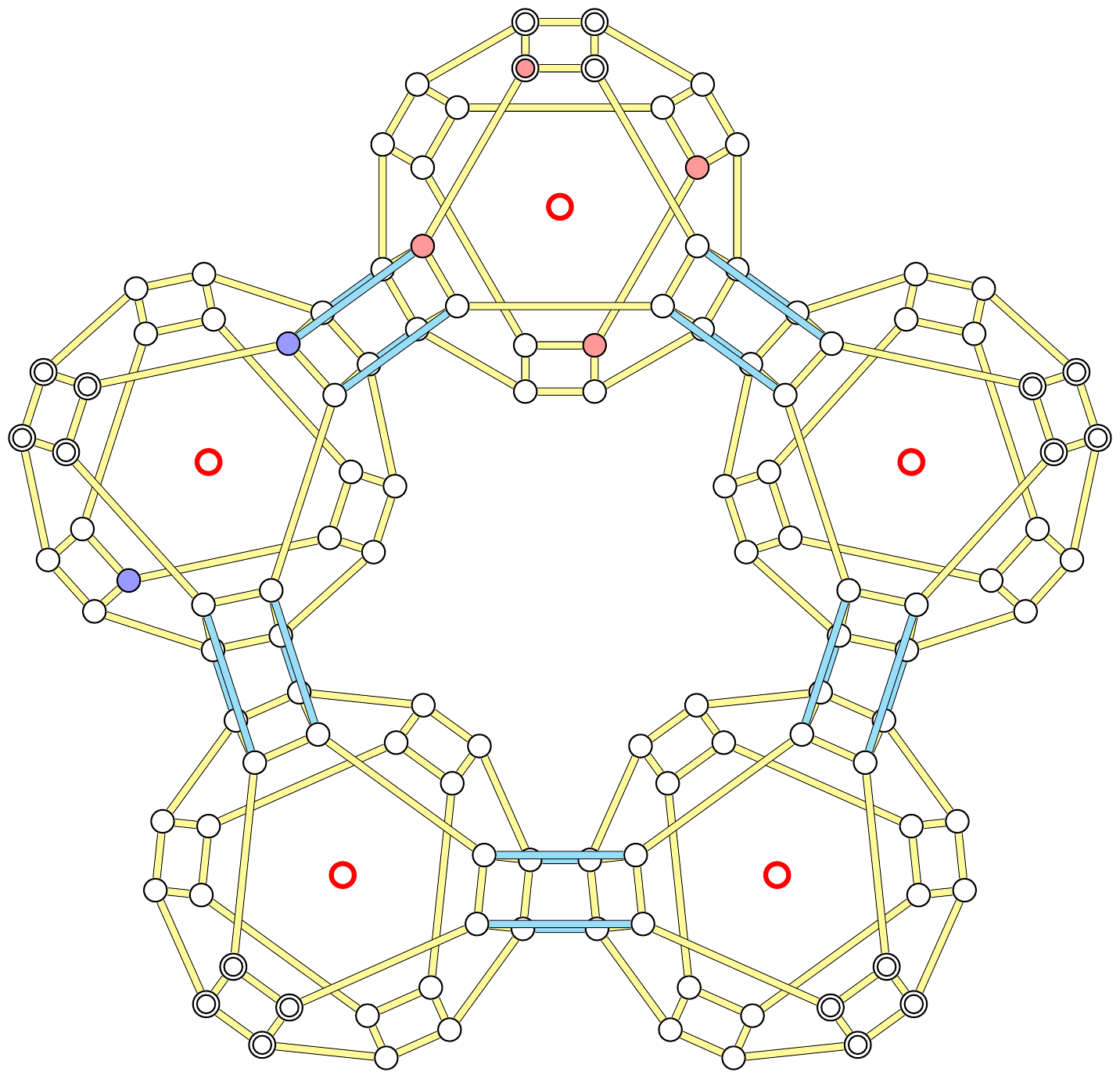}
    \includegraphics[width=0.48\linewidth,page=2]{Graphics/TIC/TIC_Proof_t_3,4.pdf}
    \includegraphics[width=0.48\linewidth,page=3]{Graphics/TIC/TIC_Proof_t_3,4.pdf}
    \includegraphics[width=0.48\linewidth,page=4]{Graphics/TIC/TIC_Proof_t_3,4.pdf}
    \label{fig:Isometry Proof B}
    \caption{The graphical proof of multi-tensor isometry condition of angular 1-uniform hyperinvariant code on $\{5,3,4\}$ honeycomb construct with GHZ of $X$-$i$ codes on t\{3,4\} truncation.}
\end{figure*}

Then, we will explicitly construct the vertex codes of HIC on $\{5,3,4\}$ honeycomb with different isometry properties as we discussed in previous sections. For the simplest case, we can simply consider an $\Xi(6)$ code with respect to the antipodal structure of a $\{3,4\}$ octahedron, as shown in Fig.~\ref{fig:X-I(4)} and top left part of Fig.~\ref{fig:Truncation Procedure}. This construction gives an angular 2-uniform vertex code, together with the Hadamard gate $H$ as the edge tensor, we will be able to construct a 3D degenerate Evenbly code. The proof of the multi-tensor isometry condition is graphically shown in Fig.~\ref{fig:Isometry Proof A} via operator pushing \cite{qLEGO,HIC-qubit,qLEGO-Exp,qLEGO-HQEC,qLEGO-XP,Heteogeneous}. For other constructions, we may use a \emph{truncation procedure} as shown in Fig.~\ref{fig:Truncation Procedure} to increase the physical bond dimension $\chi_P$ with respect to the rotational symmetry. As shown in top right of Fig.~\ref{fig:Truncation Procedure}, we can first consider code defined on truncated octahedron $t\{3,4\}$ as an $\chi_P=4$ vertex code, it is constructed by assigning each four vertices on a pair antipodal edges an $\Xi(4)$ code. One can either consider the tensor product of six $\Xi(4)$ codes which encodes six logical qubits or the GHZ entangled version that only encode one logical qubit. It is an angular 1-uniform code, and one can observe it by considering the form of $X$-stabilizers defined on it, which always cover cover two indices in an angular area. The proof of the multi-tensor isometry is graphically shown in Fig.~\ref{fig:Isometry Proof B}. Then on the bottom left we have a code defined on cantellated octahedron $rr\{4,3\}$ by assigning $\Xi(6)$ code on each antipodal triangular faces, resulting an angular 2-uniform code. And on the bottom right we have a code defined omnitruncated octahedron $tr\{4,3\}$ by assigning $\Xi(6)$ code on each hexagonal face, resulting an multi-angular 1-uniform code. The multi-tensor isometry is omitted as in the two example we shown graphically we can see that HIC made with vertex code constructed by $X$-$I$ codes always push the $Z$ input (either logical or physical) to the outgoing legs, and $X$ input may involve operator pushing through the edge tensor to neighboring tensors, but Hadamard gates on the edge transform them to physical $Z$ inputs and then pushed the operator to the outgoing legs.

\subsection{Rate and Distance \label{Sec.4.6}}

The code rate and code distance scaling is depend on the particular foliation pattern of the hyperbolic honeycombs one select. \cite{conformal_quasicrystal}. And the 2 dimensional case is already calculated in \cite{HIC-ququart}\cite{HIC-qubit}. Here we calculate the scaling of code rate and distance under both vertex inflation rule.

Given a recursive foliation pattern, the tiling/honeycombs can be decomposed into layers of different kind of vertices. We use $w_i(n)$ to denote the amount of vertices of type-$i$ (the index of the type is arbitrary) in $n^{\text{th}}$-layer. So we can use the finite dimensional vector $\mathbf{w}(n)=[w_i(n)]$ to summarize the quasiperiodic statistics of vertices in each layer and use the  inter-layer transfer matrix $T$ to characterize the recursive relation between different types of vertices such that:

\begin{equation}
    \mathbf{v}^{(n+1)}=T^{\text{Rate}}\mathbf{v}^{(n)}.
\end{equation}

In particular, the largest real eigenvalue of $T^{\text{Rate}}$, denoted as $\varphi$ is the growth rate of the tiling/honeycomb, and the corresponding normalized eigenvector $\mathbf{w}^*$ is the stationary distribution of different types of vertices in the boundary conformal quasicrystal. The asymptotic code rate can thus be represented in a closed form: 

\begin{align}
    \rho
    &\equiv\lim_{n\to\infty}\frac{N_{\text{bulk}}(n)}{N_{\text{boundary}}(n)}\\
    &=\lim_{n\to\infty}\frac{n}{kp}\frac{\mathbf{u}\cdot\mathbf{v}(n)}{\sum_{i=1}^{n}||\mathbf{v}(i)||_1}\\
    &=\frac{n}{kp}[\mathbf{u}\cdot\mathbf{v}^{*}(1-\varphi^{-1})]^{-1}
\end{align}

for $n,k$ denote the number of physical and logical qubit of each seed code $A$, $p$ the vertex degree (therefore $n/p$ is the bond dimension). $\mathbf{u}$ is the vector assigning the number of output physical indices of each type of vertices.

Specifically, the transfer matrices of the four compact honeycombs are:
\begin{equation}
    T^{\text{Rate}}_{\{5,3,4\}}=
    \begin{bmatrix}
        21 & 12 & 4 \\
        9 & 5 & 2 \\
        3 & 2 & 1
    \end{bmatrix}
\end{equation}
and the corresponding growth rates and code rates are:

\begin{align}
    \varphi_{\{5,3,4\}}&=15+4\sqrt{14} & \rho_{\{5,3,4\}}&=2
\end{align}

For the distance scaling, the methodology is similar. We can consider the operator pushing of a logical operation in the center in a recursive way. We denote the number of physical $X,Z$ operators in the $n^\text{th}$ layer of an operator pushing process as $\mathbf{w}^{(n)}=\begin{bmatrix}
    w_X^{(n)}\\
    w_Z^{(n)}
\end{bmatrix}$. In particular, the initial condition $w_X^{(1)}, w_Z^{(1)}$ are given by the specific logical operator of a single vertex code. And the recursive relation be summarized in a distance transfer matrix $T^{\text{Distance}}$ such that
\begin{equation}
    \mathbf{w}^{(n+1)}=T^{\text{Distance}}\mathbf{w}^{(n)}
\end{equation}

But different from the code rate scaling, the operator pushing not only depend on the background geometry, but also depend on the particular code structure. Therefore the the following distance scaling are only valid for the specific vertex code $\Xi(6)$ defining on $\{3,4\}$:
\begin{equation}
    T_{\{5,3,4\}}^{\text{Distance}}=
    \begin{bmatrix}
        2 & 1\\
        1 & 0
    \end{bmatrix}
\end{equation}
    
And the largest real eigenvalues are the distance scaling rates of these codes:
\begin{align}
    \lambda_{\{5,3,4\}}&=1+\sqrt{2}
\end{align}

Combine with the result of recursive growth of the hyperbolic honeycomb, the distance scaling are given by:

\begin{align}
    d_{\text{bit}}\sim \lambda^{n}= cN_{\text{boundary}}^{\log_{\varphi}\lambda}
\end{align}

for $c$ denote some constant. Numerically we have the following distance scaling:
\begin{align}
    d_{\text{bit},\{5,3,4\}}&=cN_{\text{boundary}}^{0.26}
\end{align}

which exhibits a smaller scaling exponent compared to the 2D Evenbly code on the \(\{4,5\}\) tiling~\cite{HIC-qubit}. This behavior arises because 3D hyperbolic honeycombs possess significantly higher growth rates and code rates, yet the particular \(\Xi(6)\)-based code employed here exhibits structural similarities to the 2D construction built from \(\Xi(4)\) codes. A more comprehensive analysis of distance scaling will be provided in a future version.

\subsection{Extension to Heterogeneous Architectures\label{Sec.4.7}}

While this work focuses on hyperinvariant tensor networks constructed on homogeneous regular honeycombs, the proposed framework—particularly the concepts of angular \(k\)-uniformity and multi-tensor isometry—extends naturally to more general classes of holographic architectures. These tools provide a unified method for diagnosing when a tensor network supports hyperinvariance and non-trivial correlation functions, independent of homogeneity assumptions.

As discussed in Sec.~\ref{Sec.4.2}, hyperinvariance refers to the presence of multi-tensor isometries in the holographic encoding, which can be characterized through angular \(k\)-uniformity and multi-tensor block structures. This formulation depends only on the local isometry properties of vertex tensors and the geometry of the underlying hyperbolic lattice. From this perspective, Evenbly’s original construction can be understood as a particular solution to the multi-tensor isometry condition—realized via edge tensors under a \emph{homogeneous} assumption, where a single type of vertex tensor is placed on a regular tiling.

More generally, one may consider heterogeneous solutions involving multiple types of vertex tensors. For instance, in the square tiling \(\{5,4\}\), an alternative code can be constructed by bipartitioning the lattice into vertex sets \(A\) and \(A'\), such that each \(A\)-vertex is adjacent only to \(A'\)-vertices and vice versa. Assigning \(\Xi[4]\) codes to \(A\)-vertices and their dual codes (with \(X\) and \(Z\) generators interchanged) to \(A'\)-vertices leads to a hybrid construction in which all vertices obey the same local symmetry and support the same isometry structure—achieving multi-tensor isometry without requiring explicit edge tensors.

Recent work on heterogeneous holographic codes~\cite{HI-MERA,Heteogeneous} has demonstrated that non-trivial boundary correlations and universal transversal gates can coexist in full-rate codes by combining multiple types of vertex tensors with semi-regular tilings. In this broader context, angular \(k\)-uniformity functions not only as a design principle but also as a diagnostic framework. It allows one to evaluate both regular and irregular architectures—homogeneous or heterogeneous—within a common language of local isometry constraints and their global implications. This perspective may be especially useful for generalizing qLEGO-based constructions to higher-dimensional and modular holographic codes.

\begin{figure}
    \centering
    \includegraphics[width=0.48\linewidth, page=1]{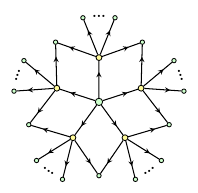}
    \includegraphics[width=0.48\linewidth, page=2]{Graphics/MTB_Decomposition/Square_Tiling_MTB.pdf}
    \caption{\textbf{Left:} Non-hyperinvariant encoding DAG on hyperbolic rhombille tilings with two types of vertices and vertex tensors, labeled as yellow and green. \textbf{Right:} Multi-tensor block decomposition of HTN on hyperbolic rhombille tiling such that both  type of vertex tensors are angular 1-uniform. The multi tensor blocks are highlighted.}
    \label{fig:rhombille tiling}
\end{figure}

One may also consider non-regular hyperbolic tilings such as the semi-regular examples discussed in~\cite{HI-MERA,Heteogeneous}, where all faces are squares but vertex degrees vary (e.g., 5 and 8, or 6 and 15). In these settings, directed acyclic graph (DAG) analysis reveals that the multi-tensor block decomposition mirrors that of regular square tilings \(\{4,q\}\). This imposes a structural constraint: at least one class of vertex tensors must be (multi-)angular 1-uniform and satisfy the multi-tensor isometry condition—such as illustrated in Fig.~\ref{fig:rhombille tiling}—in order to support non-trivial boundary correlations.

\section{Discussion \label{Sec.5}}

\subsection{Angular \(k\)-Uniformity as a Framework for Hyperinvariant Coding \label{Sec.5.1}}

In this work, we introduced \emph{angular \(k\)-uniformity} as a geometric refinement of conventional \(k\)-uniformity, enabling a systematic framework for defining and classifying hyperinvariant tensor networks and holographic codes (HICs) in higher-dimensional settings. Unlike planar \(k\)-uniformity, which is limited to 2D tilings and connected boundary regions, angular \(k\)-uniformity aligns isometric constraints with the angular structure of regular hyperbolic honeycombs. This alignment allows explicit vertex code constructions on honeycombs such as \(\{5,3,4\}\), and supports the analysis of correlation, complementary recovery, and holographic properties.

Our approach serves as a unifying diagnostic framework applicable to both \emph{homogeneous} and \emph{heterogeneous} tensor network architectures, extending beyond prior designs based on perfect tensors or restricted planar geometries. In particular, we showed that angular \(k\)-uniformity, when paired with the geometric structure of multi-tensor blocks, captures essential features of hyperinvariance and elucidates the compatibility (or tension) among nontrivial boundary correlations, uberholography, and complementary recovery.

\subsection{Limitations and Future Work \label{Sec.5.2}}

While angular \(k\)-uniformity offers a powerful framework for hyperinvariant codes, several limitations and open directions remain:

\begin{enumerate}
    \item \textbf{Incomplete concrete constructions of vertex codes.} Our constructions span all regular polytopes that serve as vertex figures in hyperbolic honeycombs, but the table remains incomplete for some \(k\) values, and multi-angularity. Systematic constructions of quantum error correcting codes satisfying (multi)-angular \(k\)-uniformity—especially with qudits or non-CSS stabilizers—remains open.

    \item \textbf{Heterogeneous Constructions.} Our framework naturally extends to heterogeneous settings, but systematic analysis of holographic codes and qLEGO architectures \cite{qLEGO,qLEGO-Exp,qLEGO-HQEC,qLEGO-XP} on semi-regular or $n$-on uniform hyperbolic honeycombs still require further development.

    \item \textbf{AdS/qCFT and variational methods.} Our results support a generalized AdS/qCFT duality where the boundary lives on a discrete conformal quasicrystal. To further explore the boundary physics—especially critical behavior—future work could incorporate variational HTNs, potentially offering practical tools for numerical studies.

    \item \textbf{Fault-tolerant and computational properties.} Although our focus was primarily on bulk reconstruction and correlation function, understanding the transversal gate sets and magic-state distillation properties—especially in connection with recent heterogeneous codes~\cite{HI-MERA,Heteogeneous}—is a promising direction for further research.

    \item \textbf{Approximate holography and subsystem codes.} Our constructions rely on exact CSS codes with full symmetry. Extending this framework to approximate quantum error correction or subsystem codes—such as the holographic Bacon–Shor model~\cite{Holographic-Bacon-Shor}—may broaden the applicability of angular \(k\)-uniformity in practical settings.ximate error correcting like \cite{Holographic-Bacon-Shor} 
\end{enumerate}

\subsection{Outlook \label{Sec.5.3}}

The geometric principles developed in this work—especially angular and multi-angular uniformity—offer a blueprint for designing modular, geometry-aware holographic codes. As tensor network methods continue to unify quantum gravity, error correction, and many-body physics, we anticipate that angularly structured holographic codes will serve as a cornerstone for both theoretical exploration and practical realization of high-dimensional holographic models.

\bibliographystyle{unsrt}
\bibliography{Main}

\appendix

\clearpage

\begin{widetext}

\section{Summary of Results\label{App.A}}

\vspace{-1em}
\noindent \textbf{TABLE II:} Summary of properties of hyperinvariant holographic tensor networks and codes on compact regular hyperbolic tiling/honeycombs. 
\vspace{-1em}

{\renewcommand{\arraystretch}{1.5}

\begin{longtable}{@{} c c c @{\hspace{13pt}} c c c @{}}

\toprule

\makecell{\textbf{Lattice}\\ \textbf{Uniformity}} & \textbf{Multi-Tensor Block(s)} & \makecell{\textbf{Corr.}\\ \textbf{Residual}} & \makecell{\textbf{Lattice}\\ \textbf{Uniformity}} & \textbf{Multi-Tensor Block(s)} & \makecell{\textbf{Corr.}\\ \textbf{Residual}}
\\

\midrule

\endfirsthead

\toprule

\makecell{\textbf{Lattice}\\ \textbf{Uniformity}} & \textbf{Multi-Tensor Block(s)} & \makecell{\textbf{Corr.}\\ \textbf{Residual}} & \makecell{\textbf{Lattice}\\ \textbf{Uniformity}} & \textbf{Multi-Tensor Block(s)} & \makecell{\textbf{Corr.}\\ \textbf{Residual}}
\\

\midrule

\endhead

\endfoot

\makecell{$\{3,q\}$\\$q\geq7$\\$k=2$} & \uniformimage{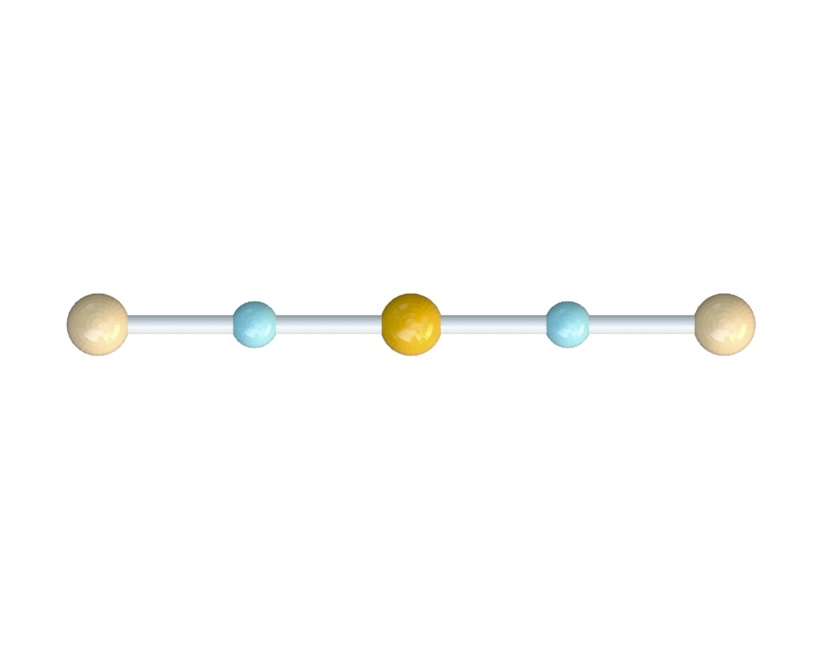}{Two Edges A} & \makecell{\cmark\\1D} &
\makecell{$\{5,3,5\}$\\$k=2$} & \uniformimage{Graphics/MTB/one_edge.jpg}{One Edge} & \makecell{\xmark\\1D}
\\

\midrule

\makecell{$\{4,q\}$\\$q\geq5$\\$k=1$} & \uniformimage{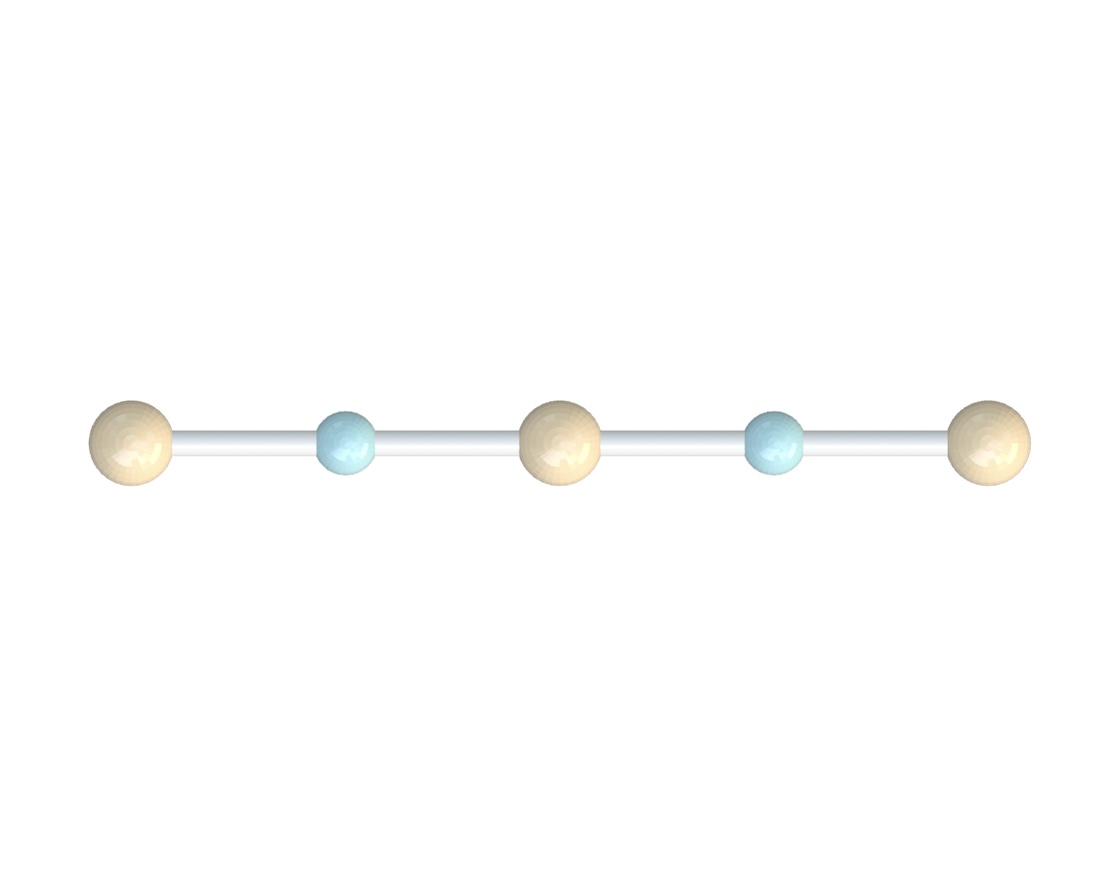}{Two Edges B} & \makecell{\cmark\\1D} &
\makecell{$\{5,3,5\}$\\$k=1$} & \uniformimage{Graphics/MTB/one_pentagon.jpg}{One Pentagon} & \makecell{\cmark\\2D}
\\

\midrule

\makecell{$\{p,3\}$\\$p\geq7$\\$k=1$\\$\chi_L/\chi_{P}\geq3$\\$(a)$} & \uniformimage{Graphics/MTB/two_edges_A.jpg}{Two Edges A} & \makecell{\cmark\\1D} &
\makecell{$\{4,3,5\}$\\$k=2$} & \uniformimage{Graphics/MTB/one_edge.jpg}{One Edge} & \makecell{\xmark\\2D}

\\

\midrule

\makecell{$\{p,q\}$ \\ $p\geq5, q\geq4$\\$k=1$} & \uniformimage{Graphics/MTB/one_edge.jpg}{One Edge} & 
\makecell{\cmark\\1D} &
\makecell{$\{4,3,5\}$\\$k=1$} & \uniformimage{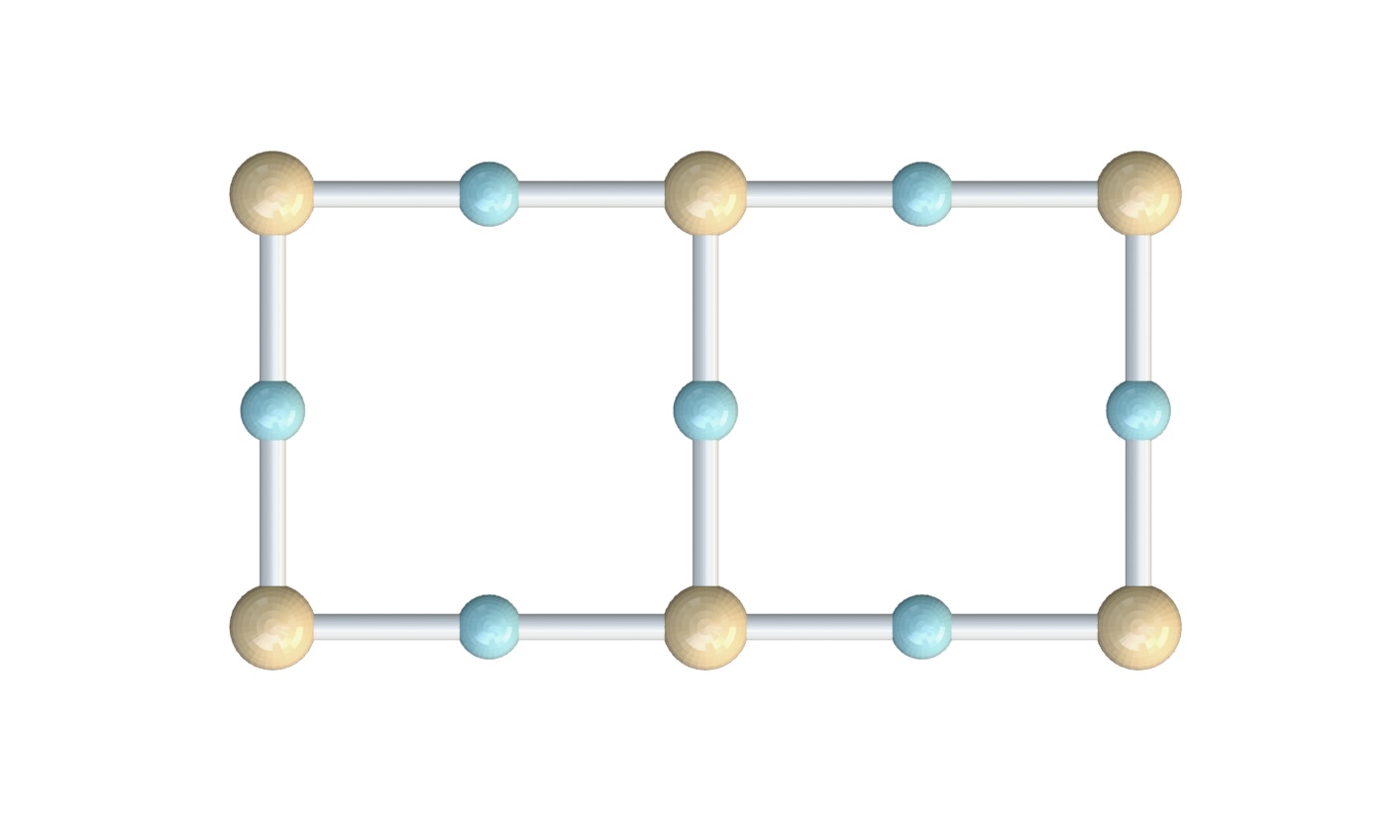}{Two Squares} & \makecell{\cmark\\2D}
\\

\midrule

\makecell{$\{5,3,4\}$\\$k=2$} & \uniformimage{Graphics/MTB/one_edge.jpg}{One Edge} & \makecell{\xmark\\1D} &
\makecell{$\{3,5,3\}$\\$k=4$} & \uniformimage{Graphics/MTB/one_edge.jpg}{One Edge} & \makecell{\xmark\\1D}
\\

\midrule
\makecell{$\{5,3,4\}$\\$k=1$} & \uniformimage{Graphics/MTB/one_pentagon.jpg}{One Pentagon} & \makecell{\cmark\\2D} &
\makecell{$\{3,5,3\}$\\$k=3$} & \uniformimage{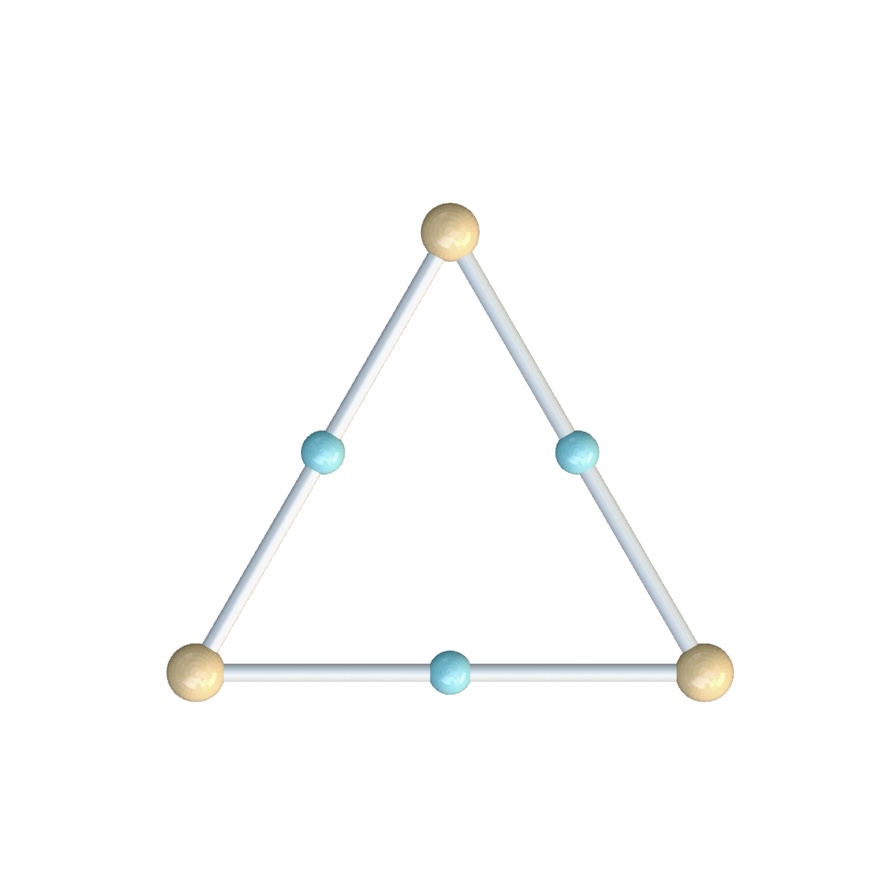}{One Triangle} & \makecell{\cmark\\2D}
\\

\midrule
\makecell{$\{3,5,3\}$\\$k=2$\\$(b)$} & \smalluniformimage{Graphics/MTB/one_pentagon.jpg}{One Pentagon}\smalluniformimage{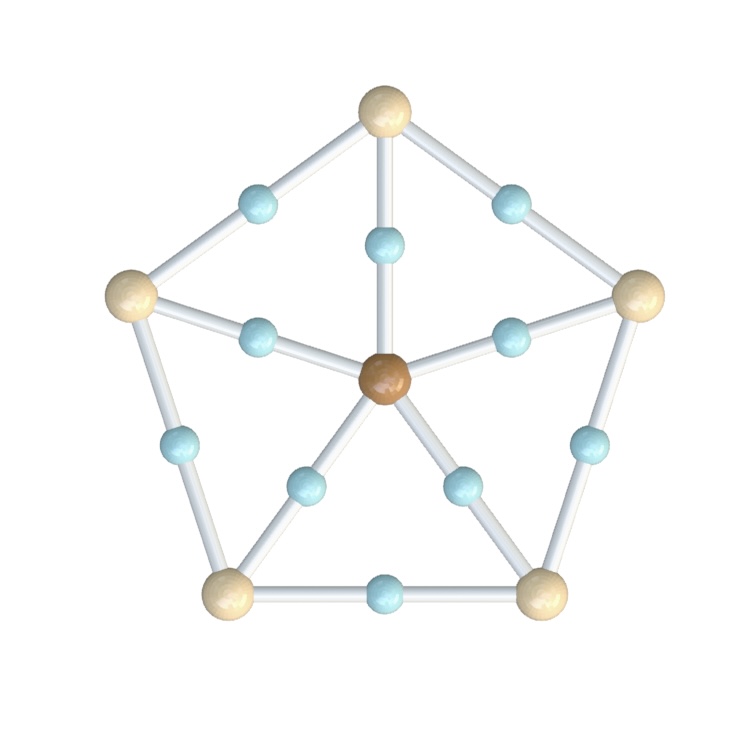}{Five Triangles} & \makecell{\cmark\\2D} & 
\makecell{$\{5,3,3,5\}$\\$k=3$} & \uniformimage{Graphics/MTB/one_edge}{One Edge} & \makecell{\xmark\\1D} 
\\

\midrule

\makecell{$\{5,3,3,3\}$\\$k=2$\\$\chi_P/\chi_L\geq33$\\$(c)$} & \uniformimage{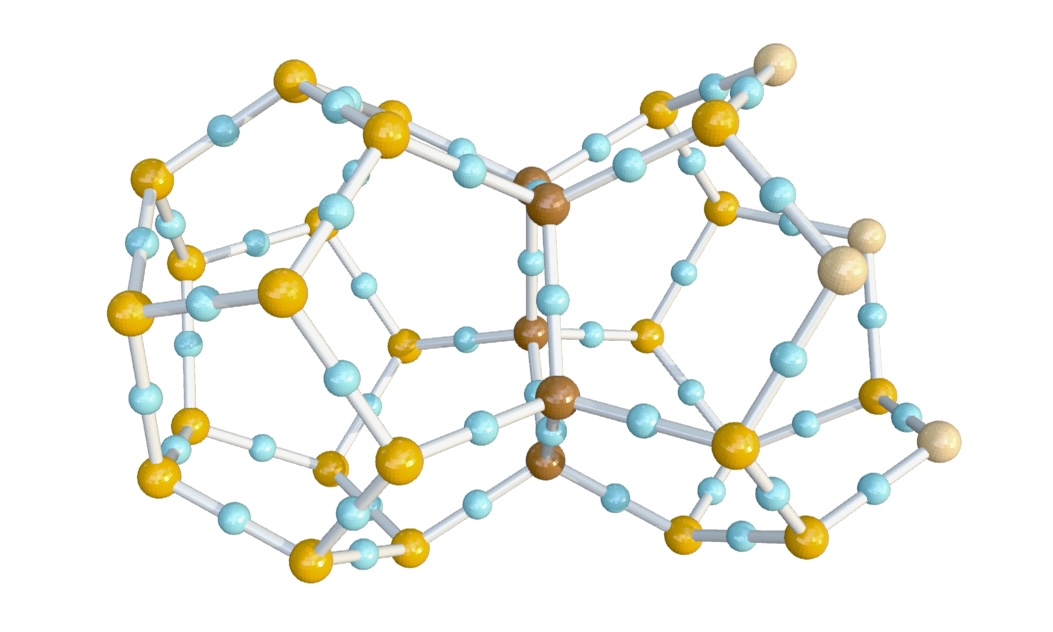}{Defected Two Dodecahedra} & \makecell{\cmark\\3D} &
\makecell{$\{5,3,3,5\}$\\$k=2$} & \uniformimage{Graphics/MTB/one_pentagon}{One Pentagon} & \makecell{\xmark\\2D}  
\\

\midrule

\makecell{$\{5,3,3,3\}$\\$k=1$\\$\chi_P/\chi_L\geq7$\\$(d)$} & \uniformimage{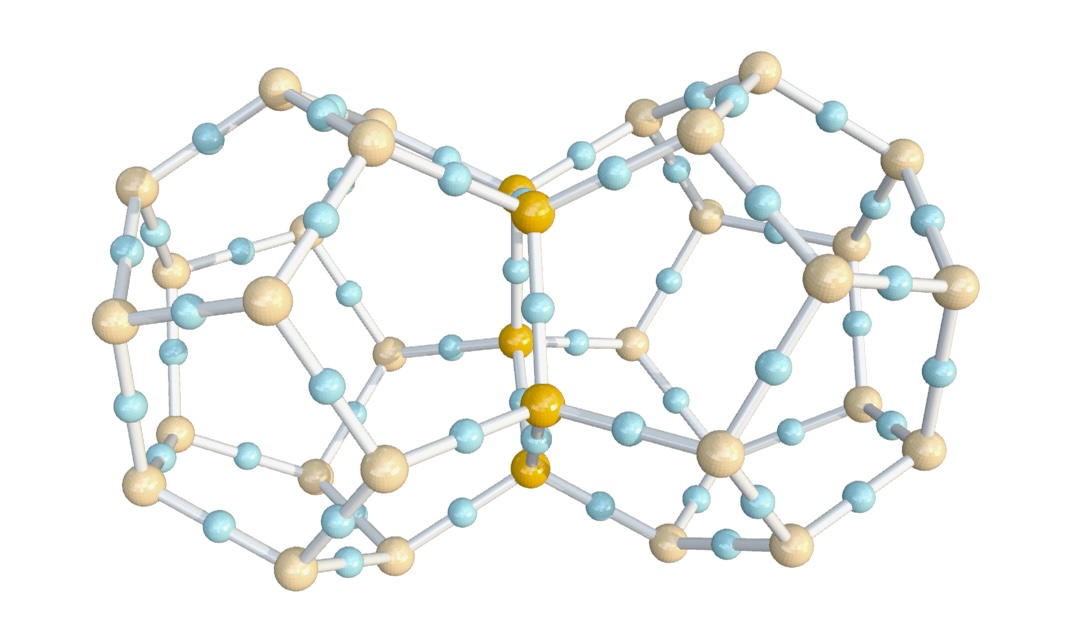}{Two Dodecahedra} & \makecell{\cmark\\3D} &
\makecell{$\{5,3,3,5\}$\\$k=1$} & \uniformimage{Graphics/MTB/one_dodecahedron}{One Dodecahedron} & \makecell{\cmark\\3D}
\\

\midrule

\makecell{$\{5,3,3,4\}$\\$k=3$} & \uniformimage{Graphics/MTB/one_edge}{One Edge} & \makecell{\xmark\\1D} &
\makecell{$\{4,3,3,5\}$\\$k=3$} & \uniformimage{Graphics/MTB/one_edge}{One Edge} & \makecell{\xmark\\2D}\\

\midrule

\makecell{$\{5,3,3,4\}$\\$k=2$} & \uniformimage{Graphics/MTB/one_pentagon}{One Pentagon} & \makecell{\xmark\\2D} &
\makecell{$\{4,3,3,5\}$\\$k=2$} & \uniformimage{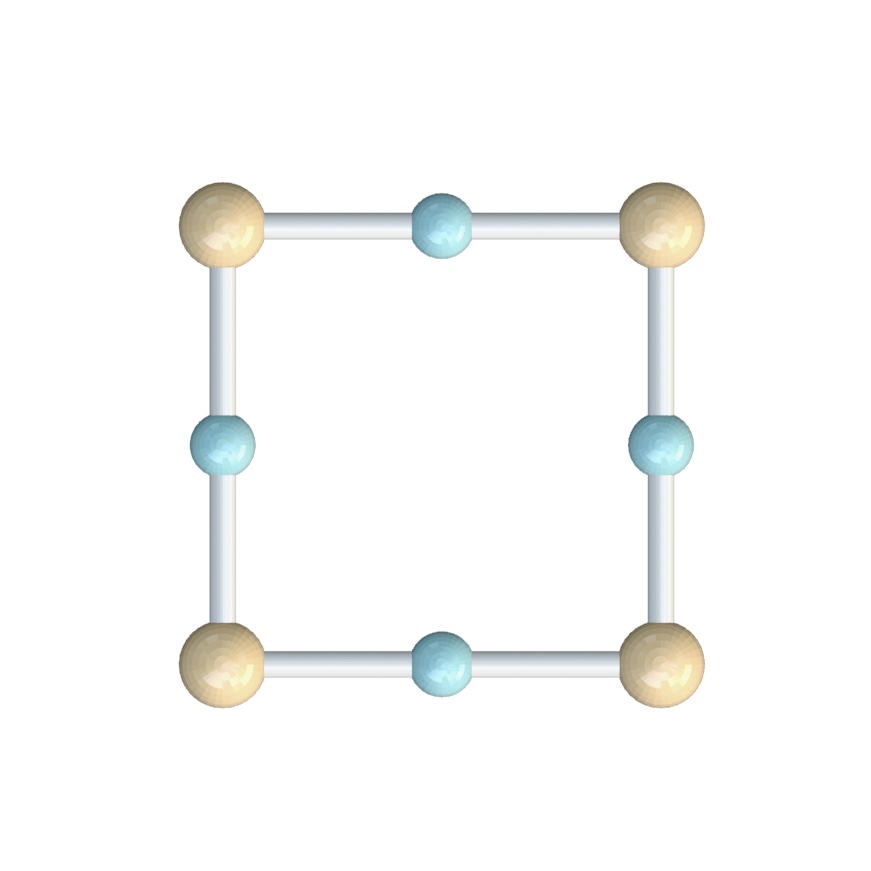}{One Square} & \makecell{\xmark\\3D} \\

\midrule

\makecell{$\{5,3,3,4\}$\\$k=1$} & \uniformimage{Graphics/MTB/one_dodecahedron}{One Dodecahedron} & \makecell{\cmark\\3D} &
\makecell{$\{4,3,3,5\}$\\$k=1$} & \uniformimage{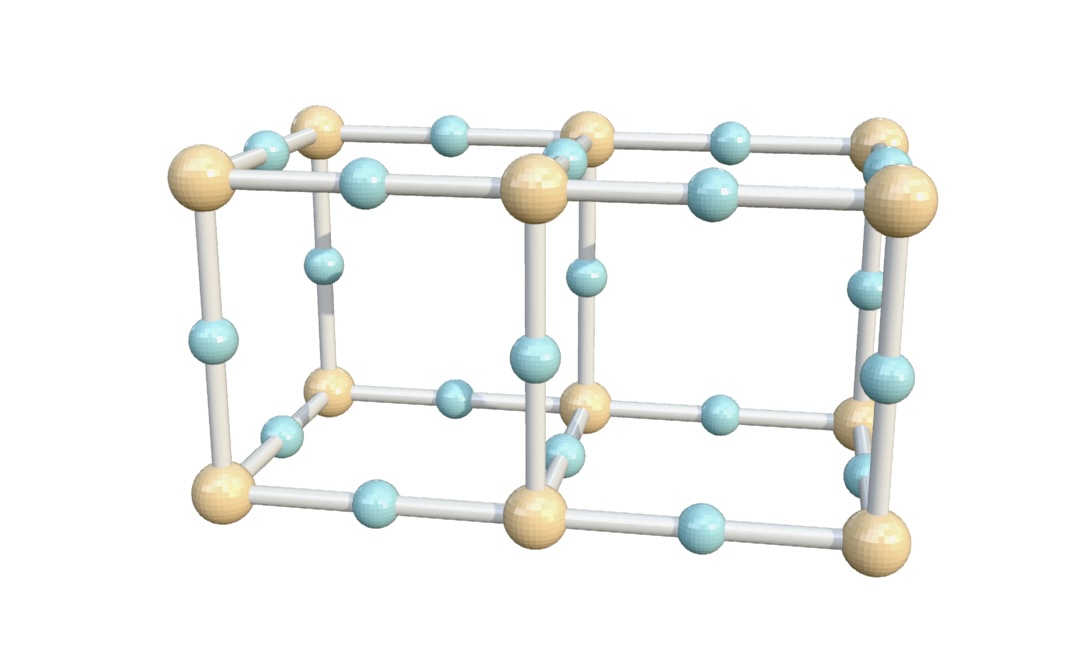}{Two Cubes} & \makecell{\cmark\\3D} 
\\

\midrule

\makecell{$\{3,3,3,5\}$\\$k=12$} & \uniformimage{Graphics/MTB/one_edge}{One Edge} & \makecell{\xmark\\1D} &
\makecell{$\{3,3,3,5\}$\\$k=8$} & \uniformimage{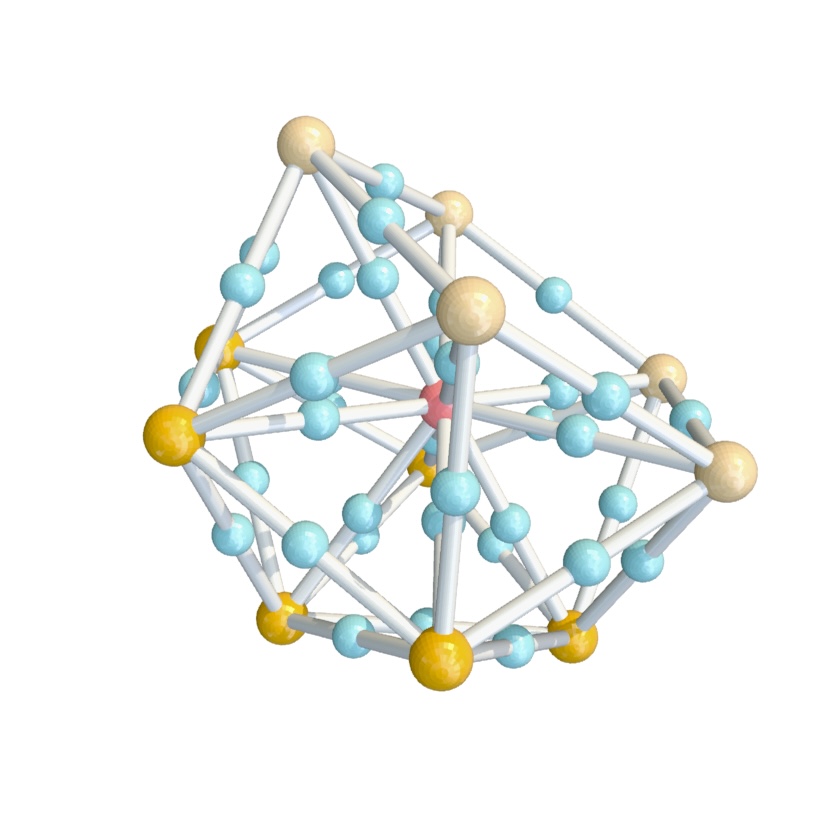}{Nineteen Tetrahedra} & \makecell{\cmark\\3D} \\

\midrule

\makecell{$\{3,3,3,5\}$\\$k=11$} & \uniformimage{Graphics/MTB/one_triangle}{One Triangle} & \makecell{\xmark\\2D} &
\makecell{$\{3,3,3,5\}$\\$k=7$\\$(e)$} & \smalluniformimage{Graphics/MTB/one_dodecahedron}{One Dodecahedron}\smalluniformimage{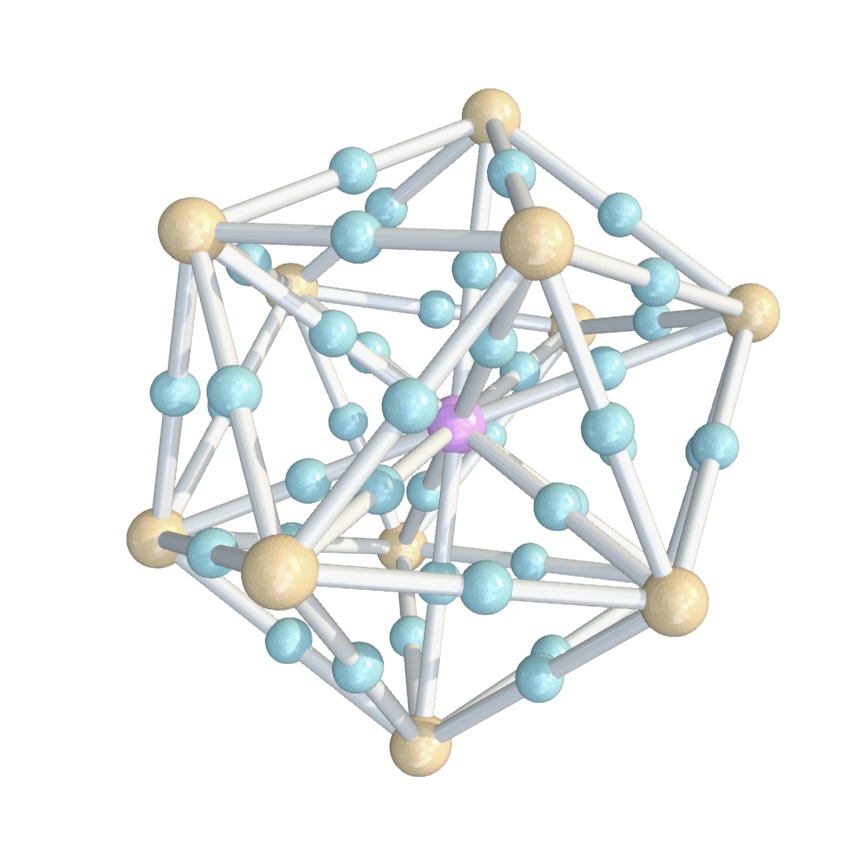}{Twenty Tetrahedra} & \makecell{\cmark\\3D} 
\\

\midrule

\makecell{$\{3,3,3,5\}$\\$k=10$} & \uniformimage{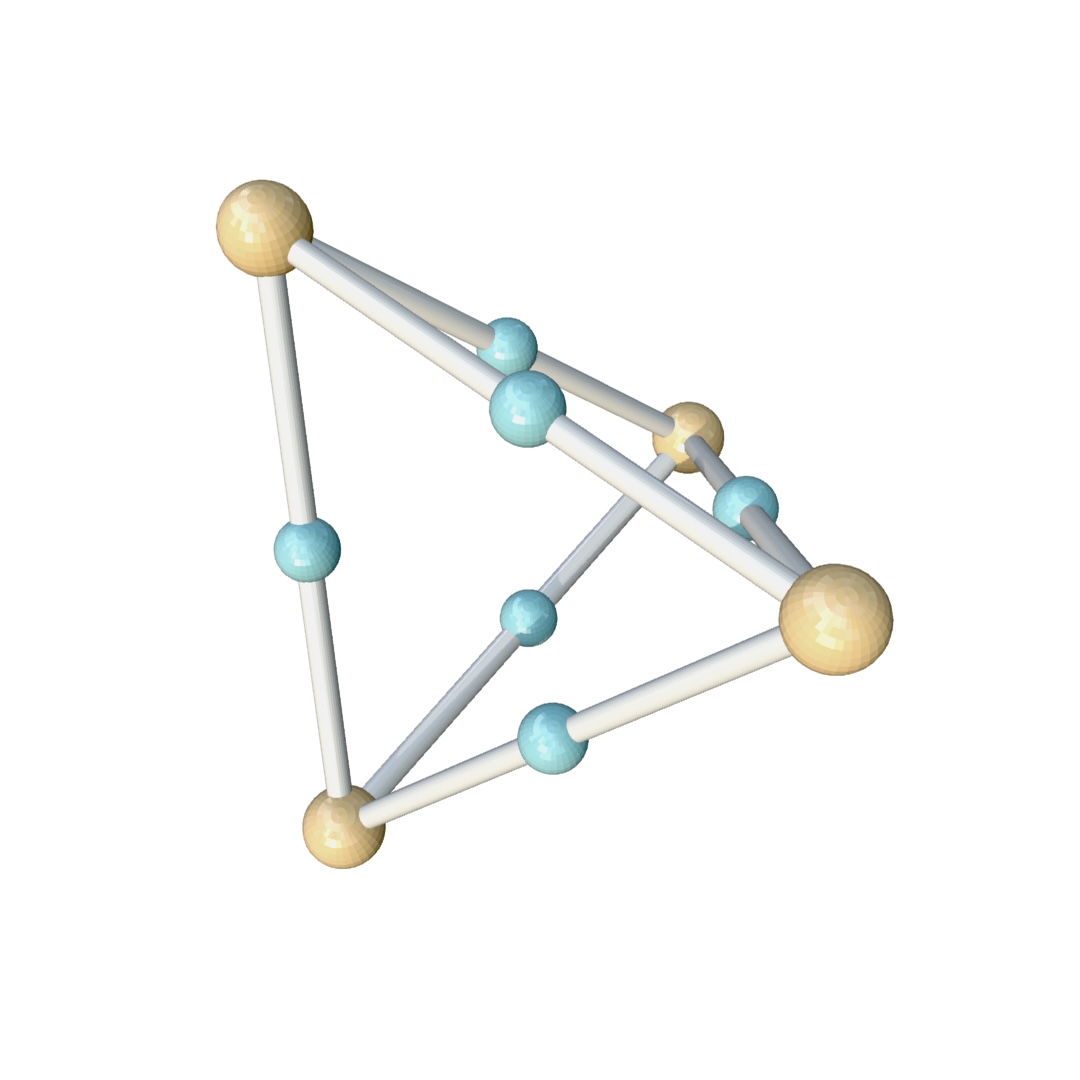}{One Tetrahedron} & \makecell{\cmark\\3D} &
\makecell{$\{3,3,3,5\}$\\$k=6$\\$(f)$} & \uniformimage{Graphics/MTB/two_hundred_two_tetrahedra}{Two Hundred Two Tetrahedra} & \makecell{\cmark\\3D}
\\

\midrule

\makecell{$\{3,3,3,5\}$\\$k=9$} & \uniformimage{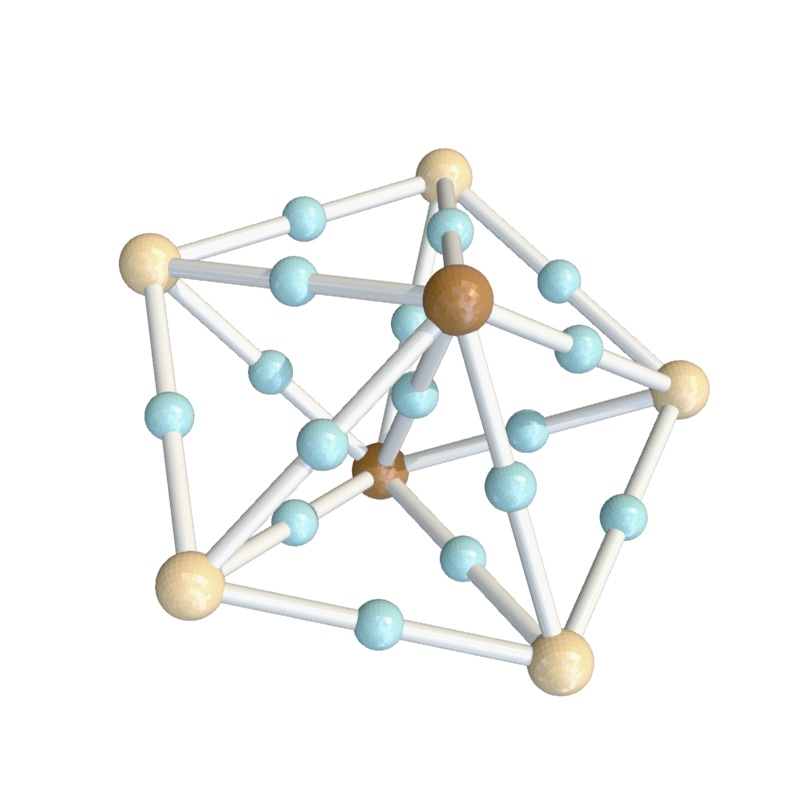}{Five Tetrahedra} & \makecell{\cmark\\3D} &
\makecell{$\{3,3,3,5\}$\\$k=5$\\$(g)$} & \uniformimage{Graphics/MTB/two_hundred_sixty-two_tetrahedra}{Two Hundred Sixty-Two Tetrahedra} & \makecell{\cmark\\3D} 
\\

\bottomrule
\label{tab:summary}
\end{longtable}
}

The lattices in this table are visualized using the standard tensor network representation, where cyan nodes located at the centers of edges denote edge tensors. The vertex tensors are color-coded according to their number of input indices:  
\begin{itemize}
\item \textbf{Yellow}: vertices with \( k \) inputs  
\item \textbf{Orange}: \( k{-}1 \) inputs  
\item \textbf{Brown}: \( k{-}2 \) inputs  
\item \textbf{Red}: 2 inputs  
\item \textbf{Violet}: 1 input
\end{itemize}

If a tensor satisfies multiple criteria, we assign the color corresponding to the highest level (i.e., leftmost in the list). For example, when \( k=3 \), a vertex with two input indices satisfies both the orange and red conditions, so we use orange.

The ``Corr.'' column indicates whether the corresponding code exhibits \emph{nontrivial boundary correlation functions}. A check mark denotes that nontrivial correlations are supported. The ``Residual'' column refers to the shape of the \emph{residual region} in standard complementary recovery, as determined by the minimal entanglement wedge structure.

The following comments clarify specific entries and assumptions in the table:
\begin{enumerate}[label=\textbf{(\alph*)}]
  \item The ``Two Edge A'' type multi-tensor block (MTB) appearing in HTNs on \(\{p,3\}\) tilings was first analyzed in the original HTN paper~\cite{HTN}. This case was omitted in~\cite{HIC-qubit} due to its assumption that the logical bond dimension equals the physical one. Here, we relax this assumption by considering the regime where the ratio \(\chi_P / \chi_L\) is sufficiently large. Unless otherwise stated, all entries in the table assume \(\chi_P / \chi_L \geq 1\).
  
  \item Although the ``One Pentagon'' MTB may appear redundant given the presence of the ``Five Triangles'' MTB---since the former can be viewed as a subregion of the latter---both are retained as separate entries. This is because they involve distinct angular configurations of input indices, as clarified in the vertex figure illustration (Fig.~\ref{fig:one_pentagon and five triangles}).

  \item MTBs are sensitive to the bond dimension ratio \(\chi_P/\chi_L\). In the double-dodecahedron case, any pair of vertices connected by an edge---excluding those on the shared pentagonal interface---can be deleted to define the MTB, demonstrating the freedom of defect placement.

  \item This case is a higher-dimensional analogue of case~(a), where the MTB again depends on a sufficiently large bond dimension ratio \(\chi_P/\chi_L\).

  \item The lattice is a sub-lattice of the 600-cell. Its geometric structure is illustrated in Fig.~\ref{fig:two hundred two tetrahedra}.

  \item This is another sub-lattice of the 600-cell, with detailed construction shown in Fig.~\ref{fig:two hundred sixty-two tetrahedra}.
\end{enumerate}

\begin{figure}
    \includegraphics[width=0.48\linewidth,page=1]{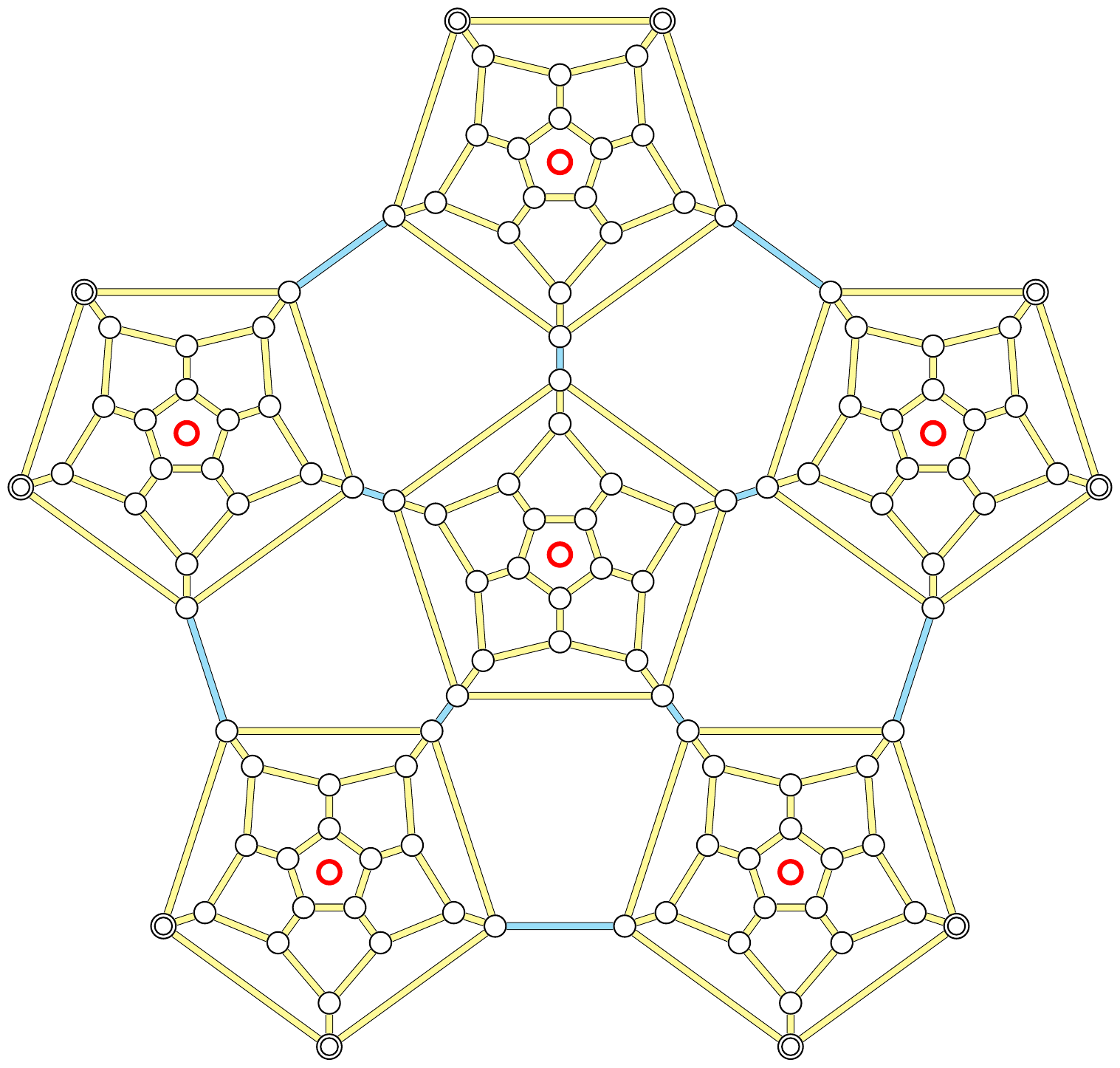}
    \includegraphics[width=0.48\linewidth,page=2]{Graphics/MTB/MTB_3,5,3.pdf}
    \caption{The vertex figure representation of ``One Pentagon" and ``Five Triangles" blocks of honeycomb $\{3,5,3\}$.\label{fig:one_pentagon and five triangles}}
\end{figure}

\begin{figure}
    \includegraphics[width=0.24\linewidth,page=1]{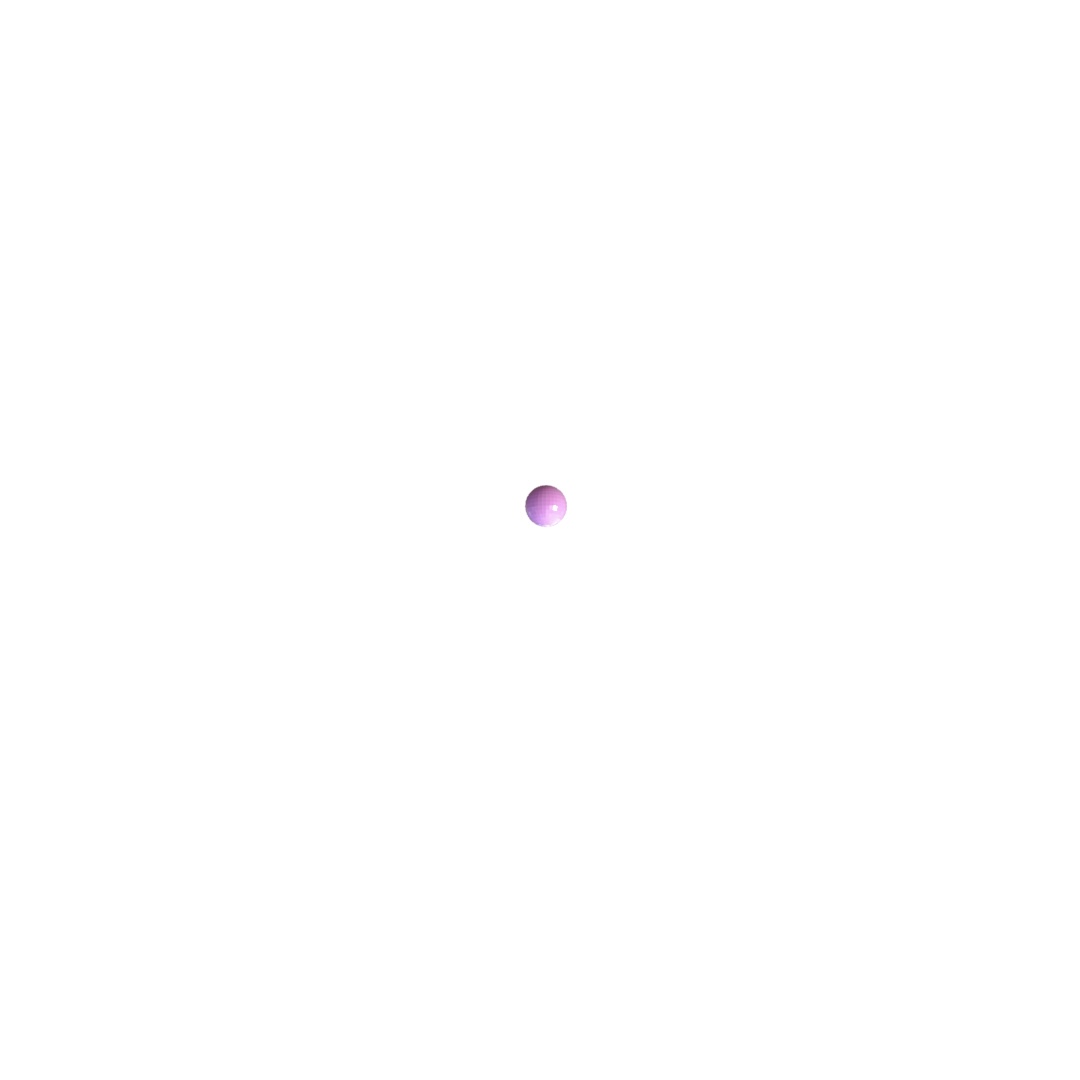}
    \includegraphics[width=0.24\linewidth,page=1]{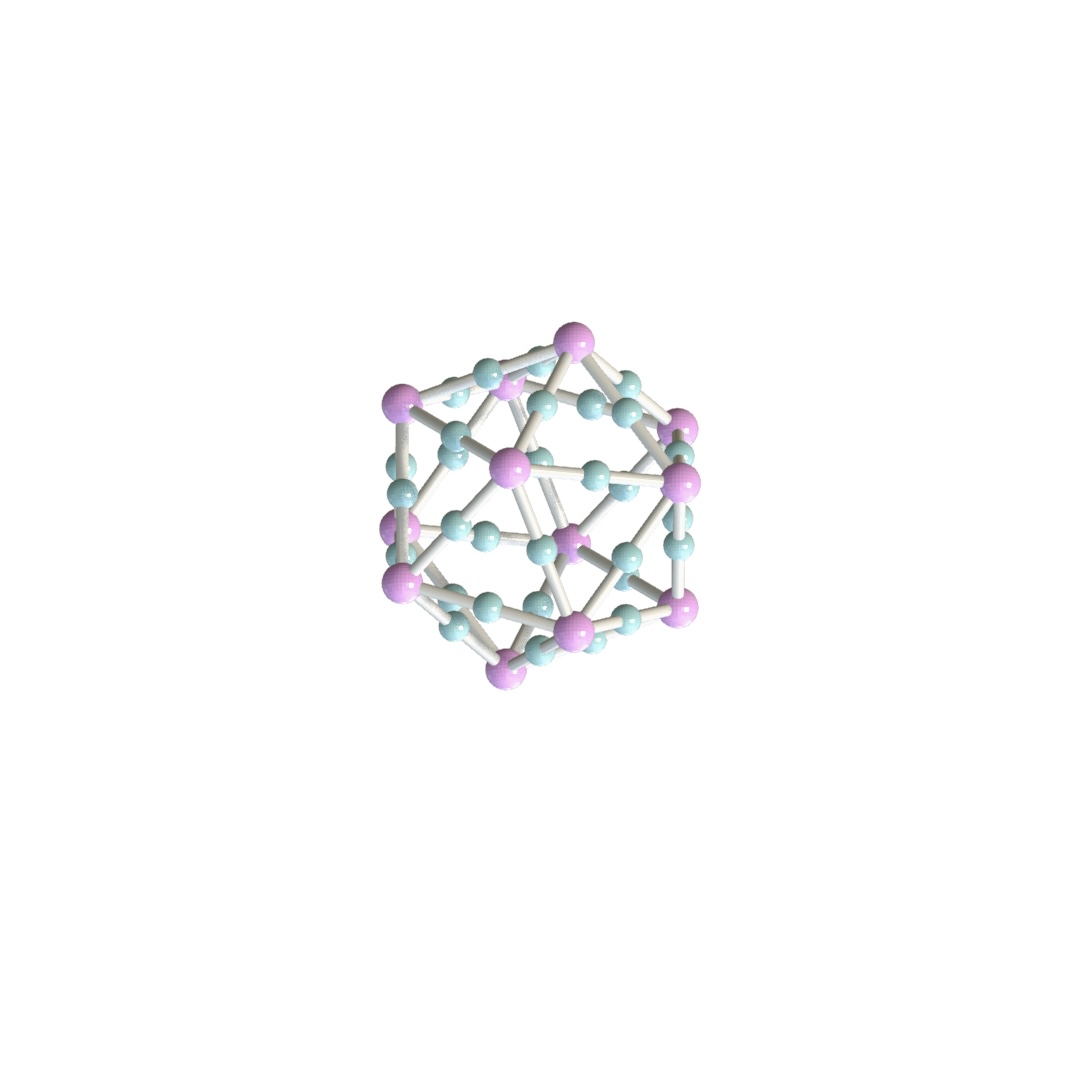}
    \includegraphics[width=0.24\linewidth,page=1]{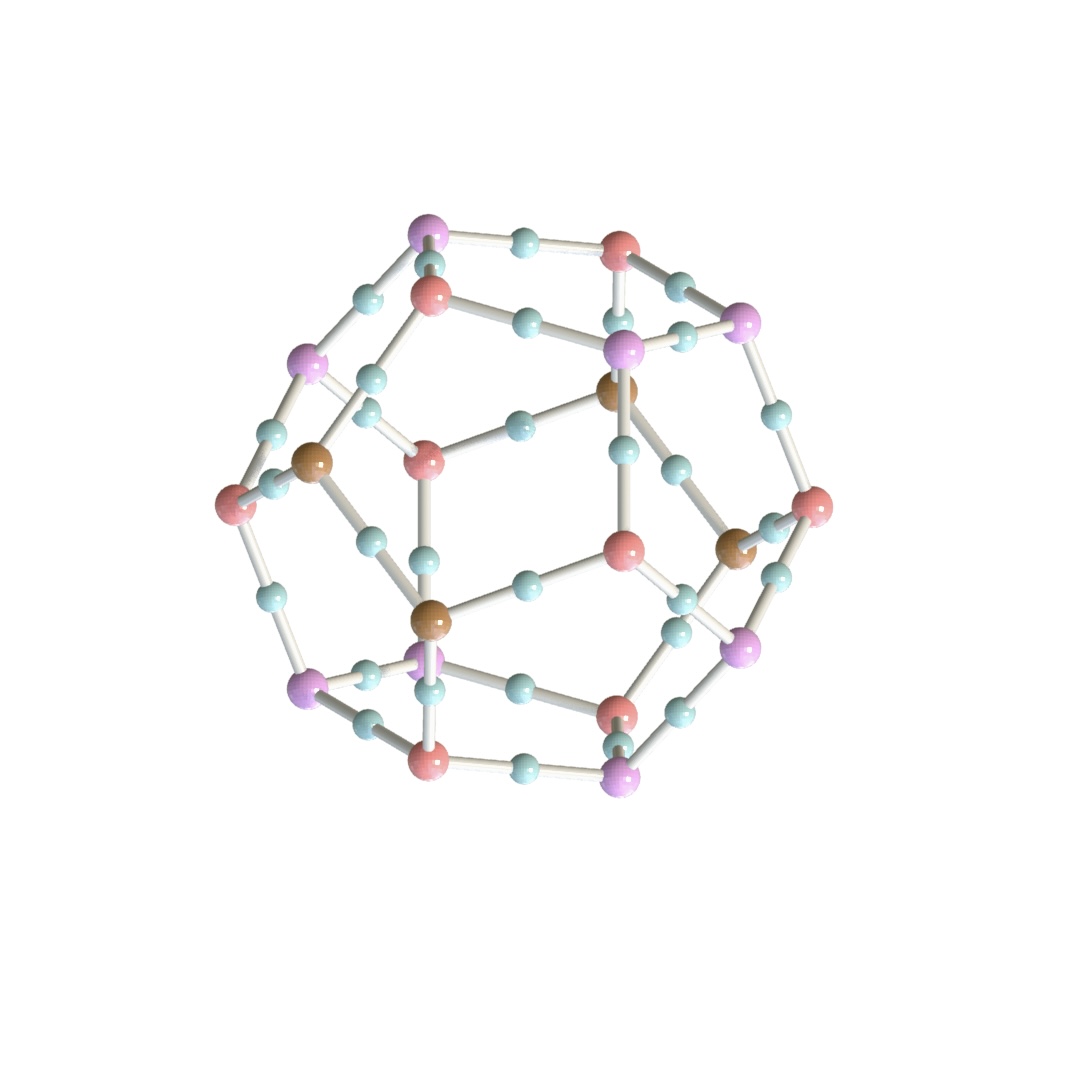}
    \includegraphics[width=0.24\linewidth,page=1]{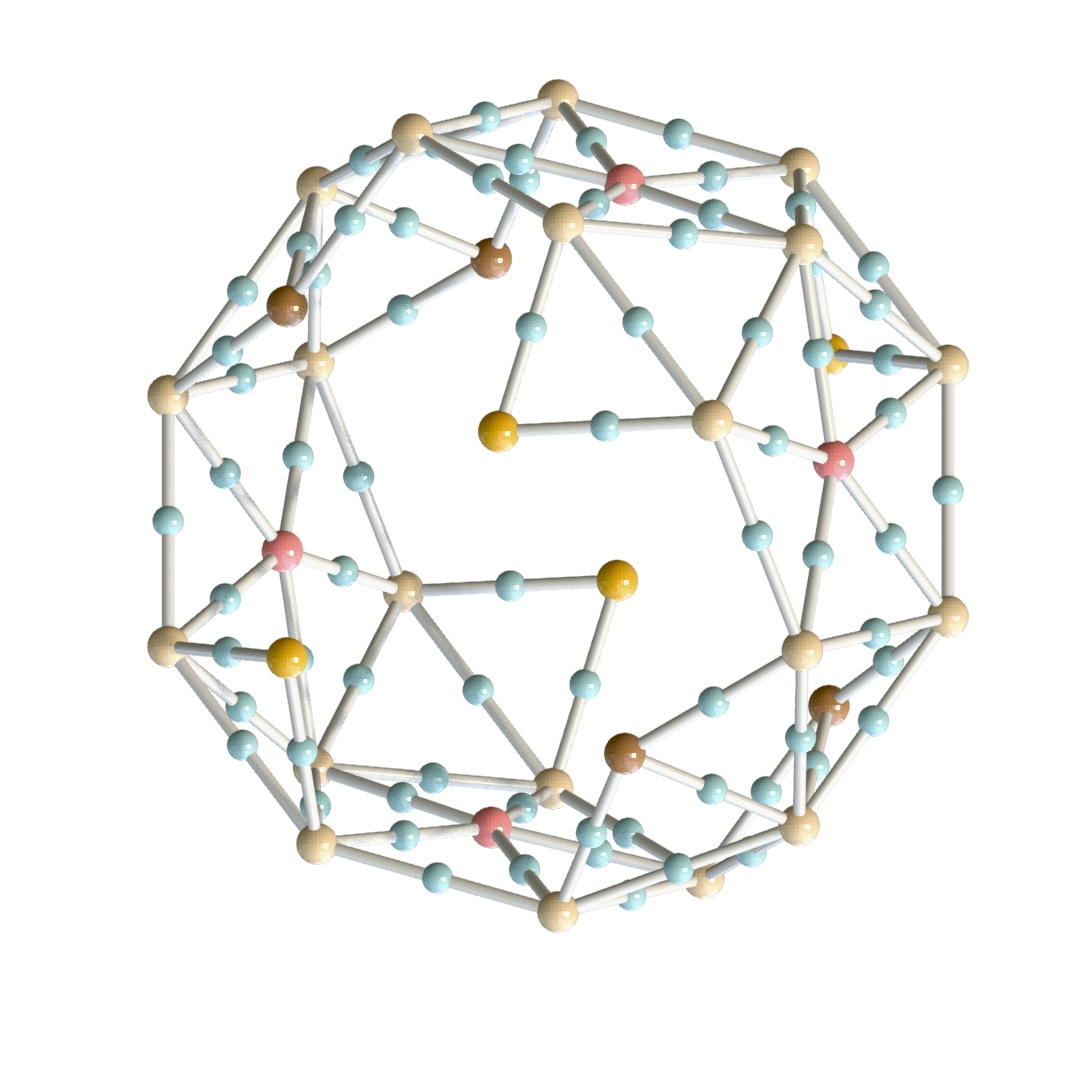}
    \includegraphics[width=0.24\linewidth,page=1]{Graphics/MTB/two_hundred_two_tetrahedra/202-1.jpg}
    \includegraphics[width=0.24\linewidth,page=1]{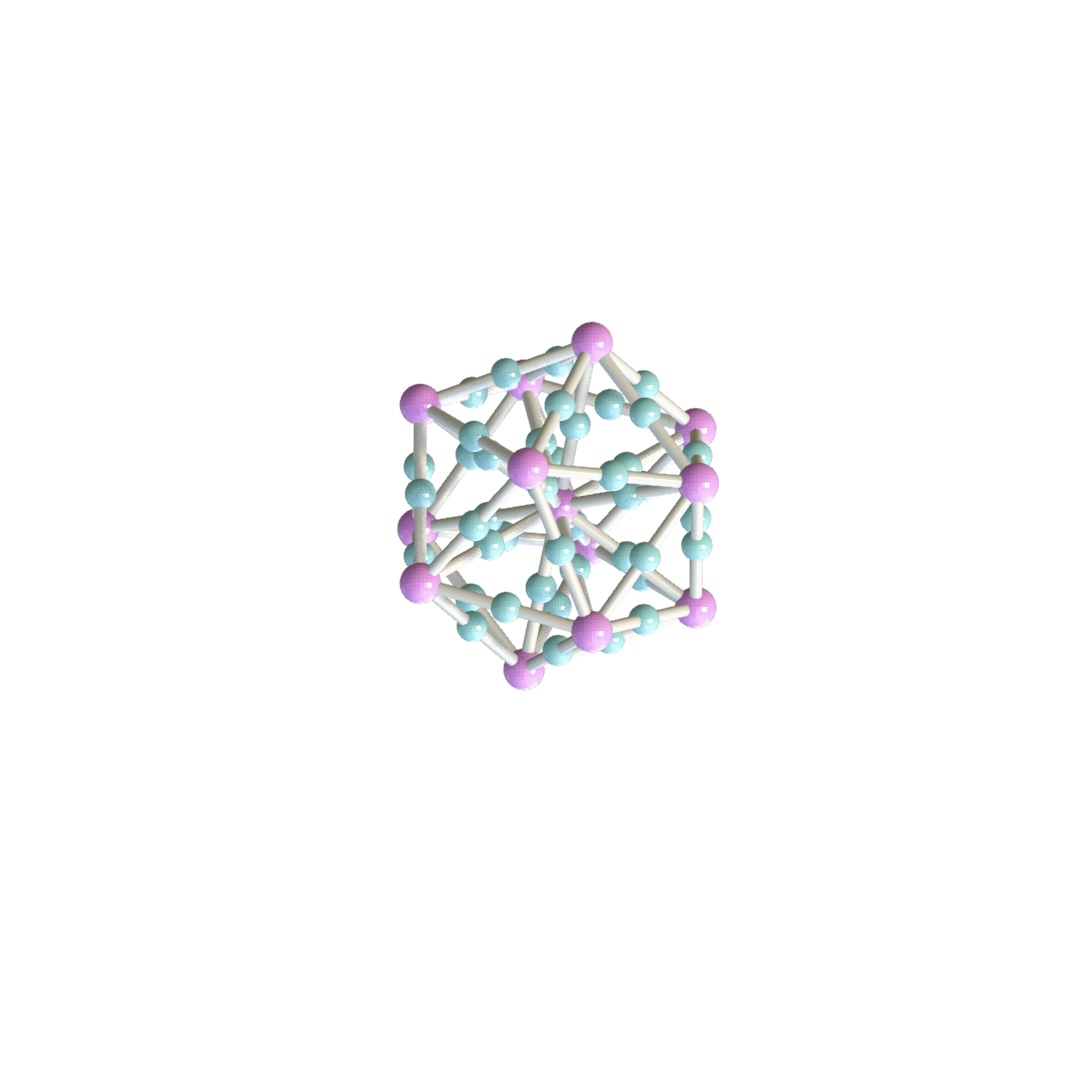}
    \includegraphics[width=0.24\linewidth,page=1]{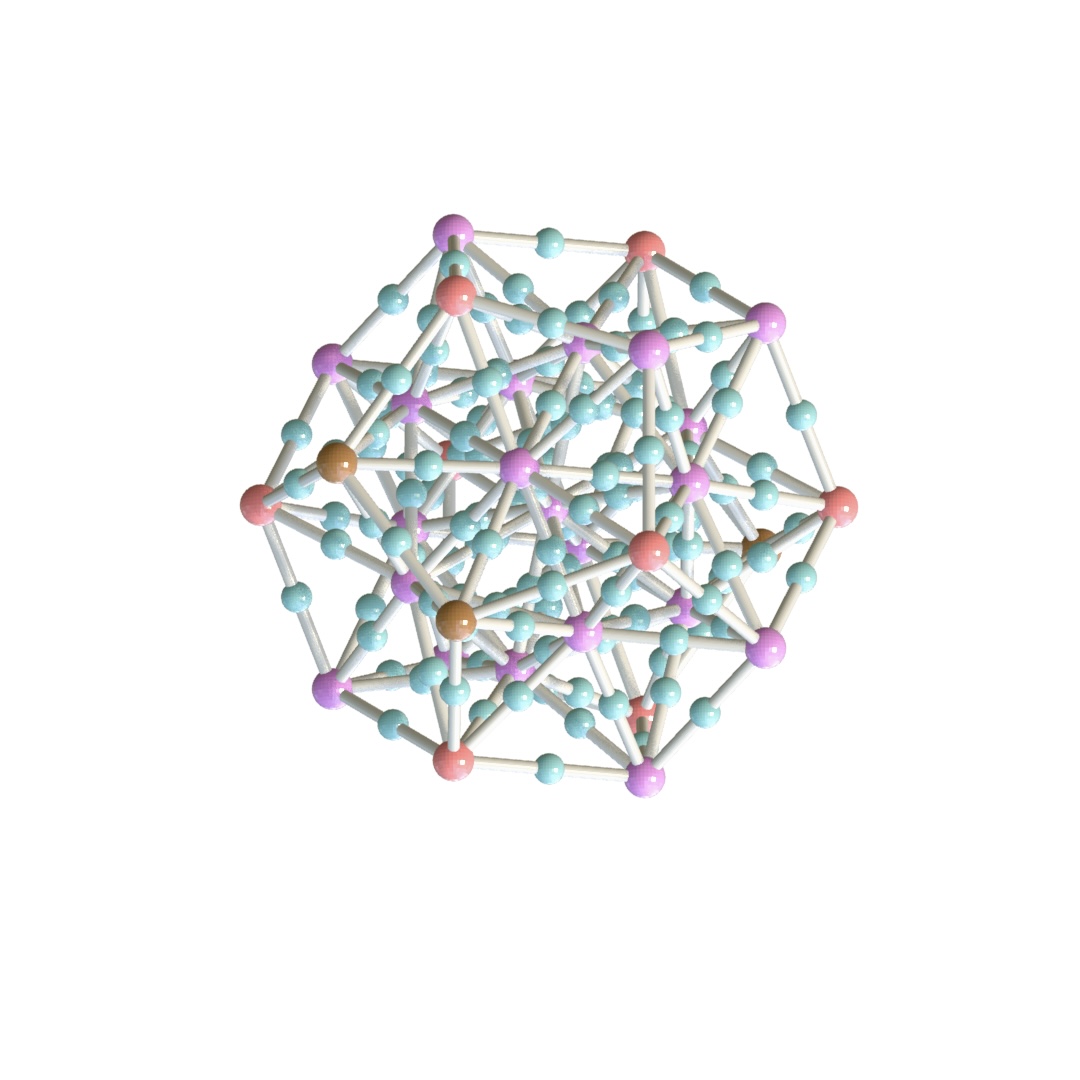}
    \includegraphics[width=0.24\linewidth,page=1]{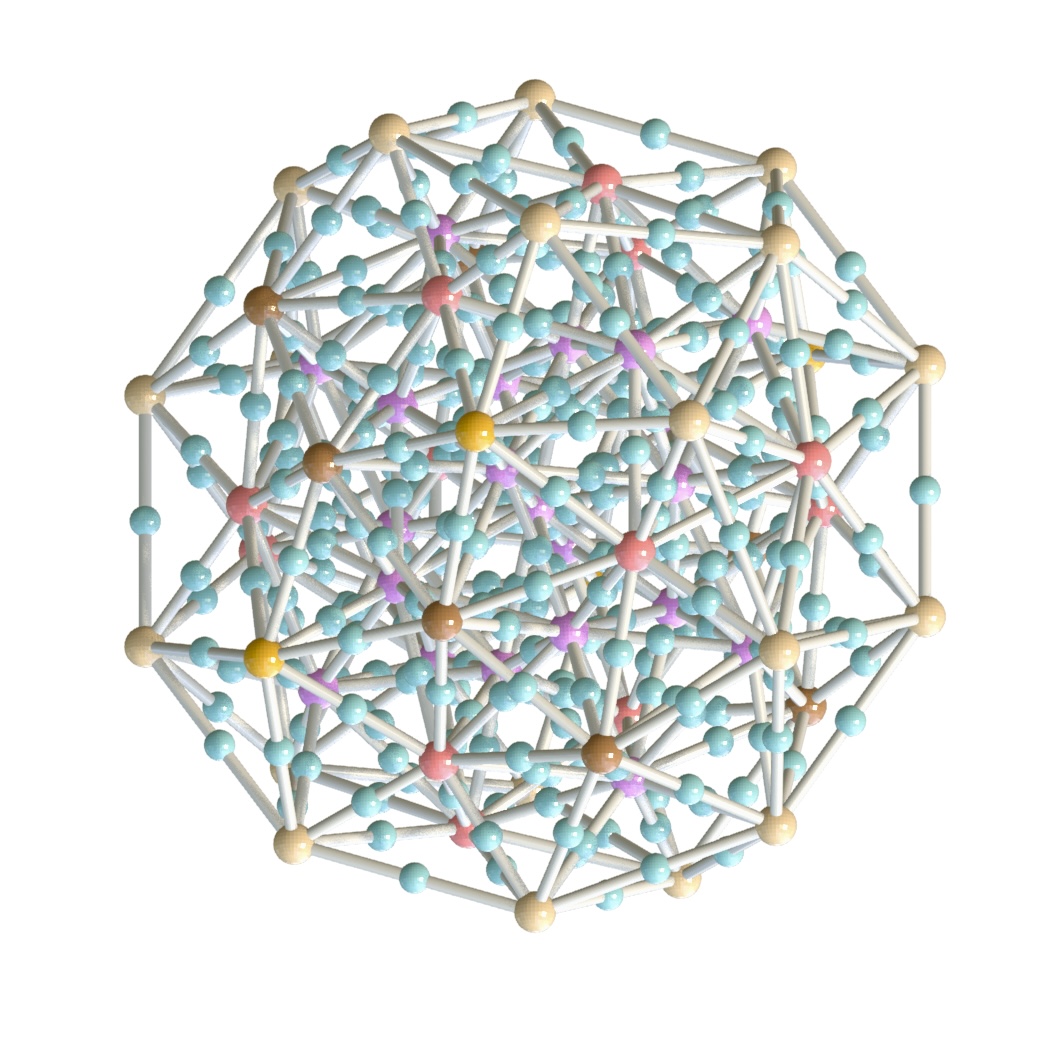}
    \caption{The decomposition of ``Two Hundred Two Tetrahedra". \label{fig:two hundred two tetrahedra}}
\end{figure}

\begin{figure}
    \includegraphics[width=0.24\linewidth,page=1]{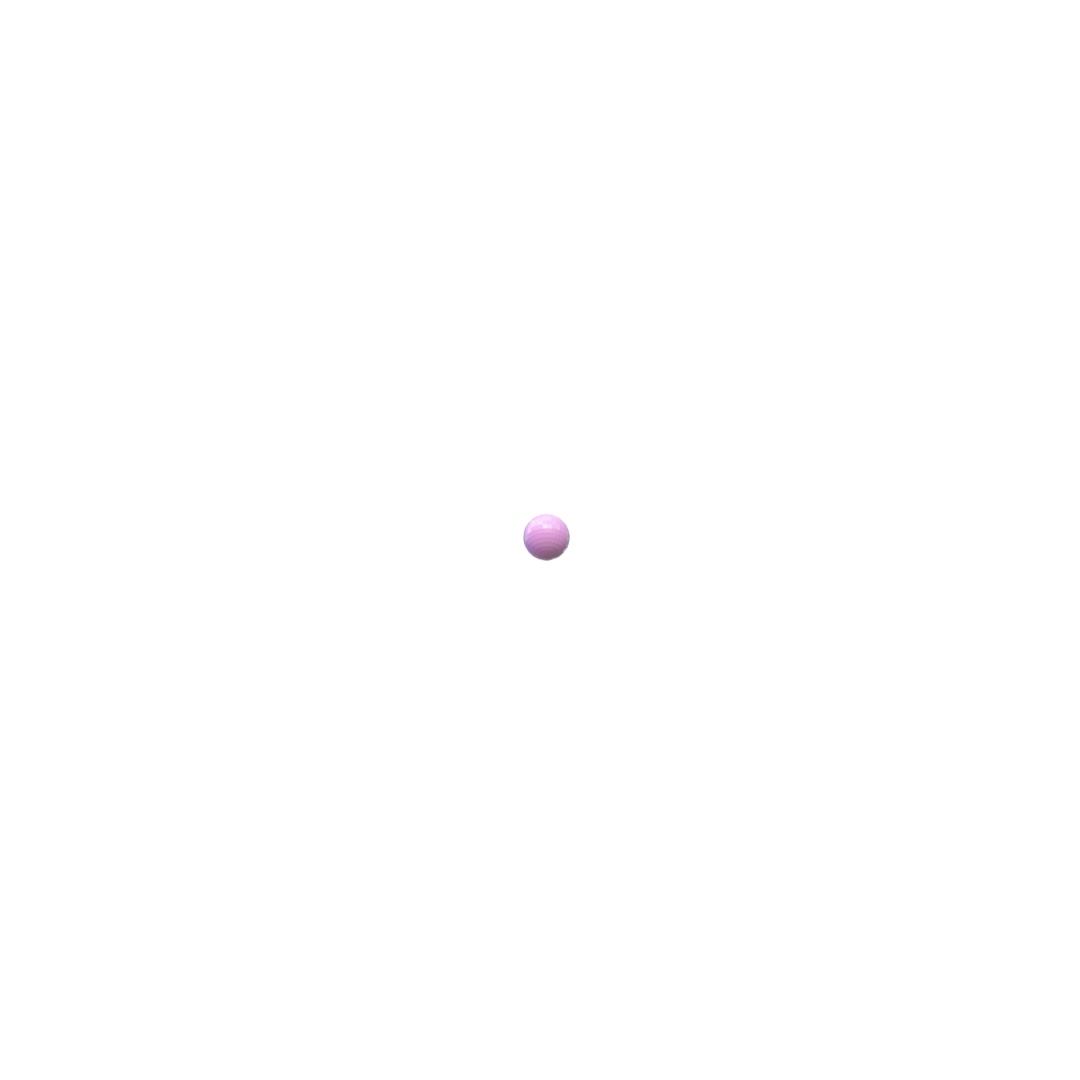}
    \includegraphics[width=0.24\linewidth,page=1]{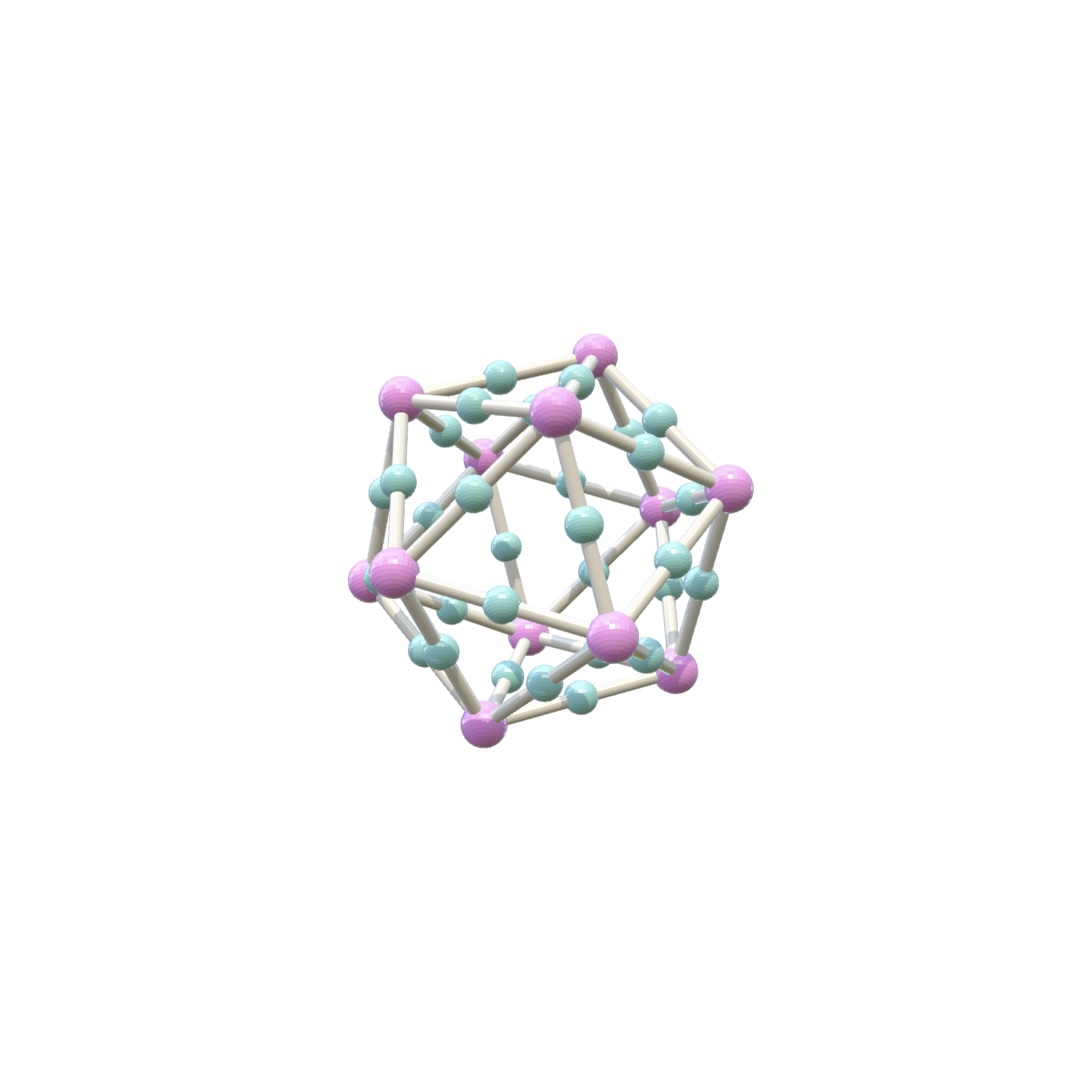}
    \includegraphics[width=0.24\linewidth,page=1]{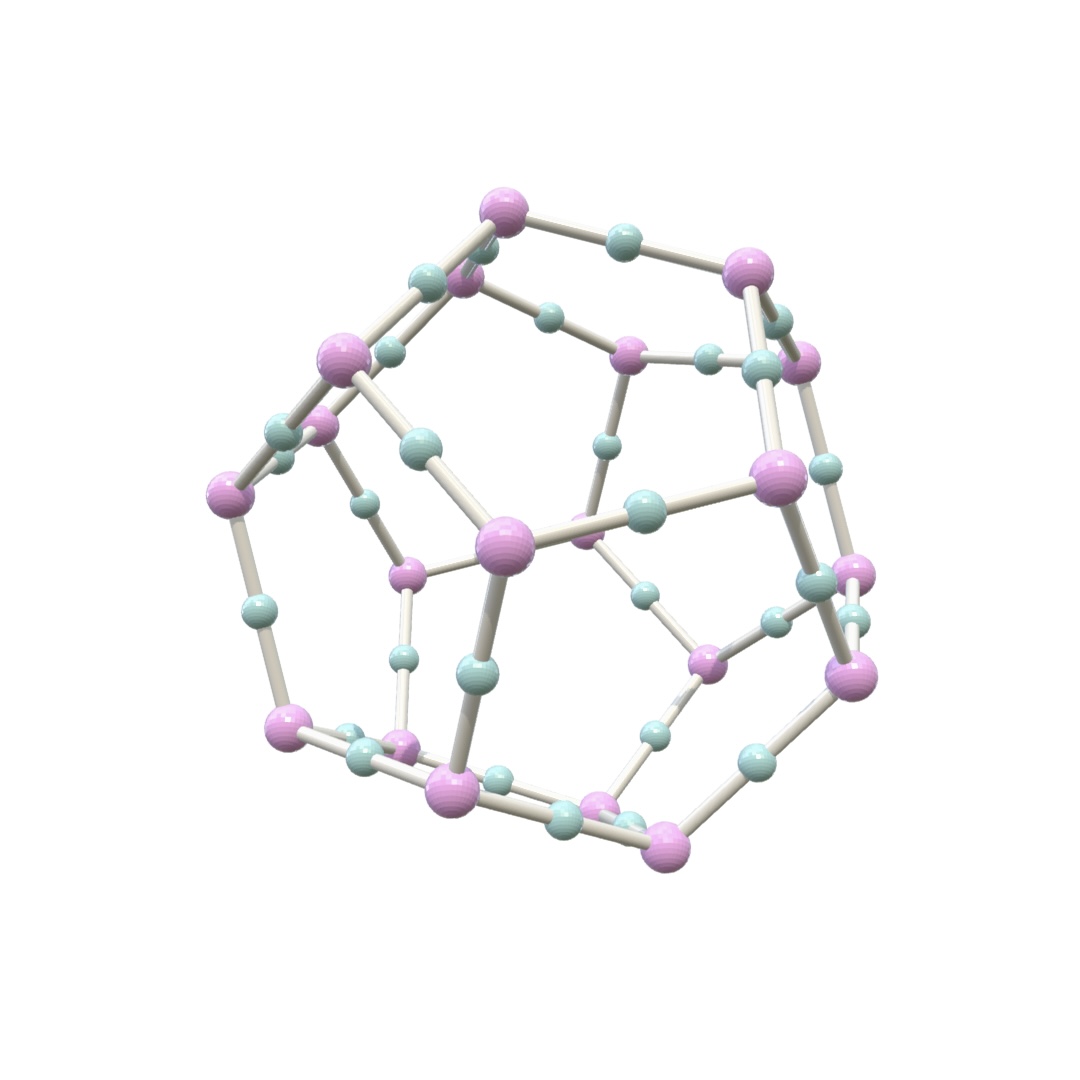}
    \includegraphics[width=0.24\linewidth,page=1]{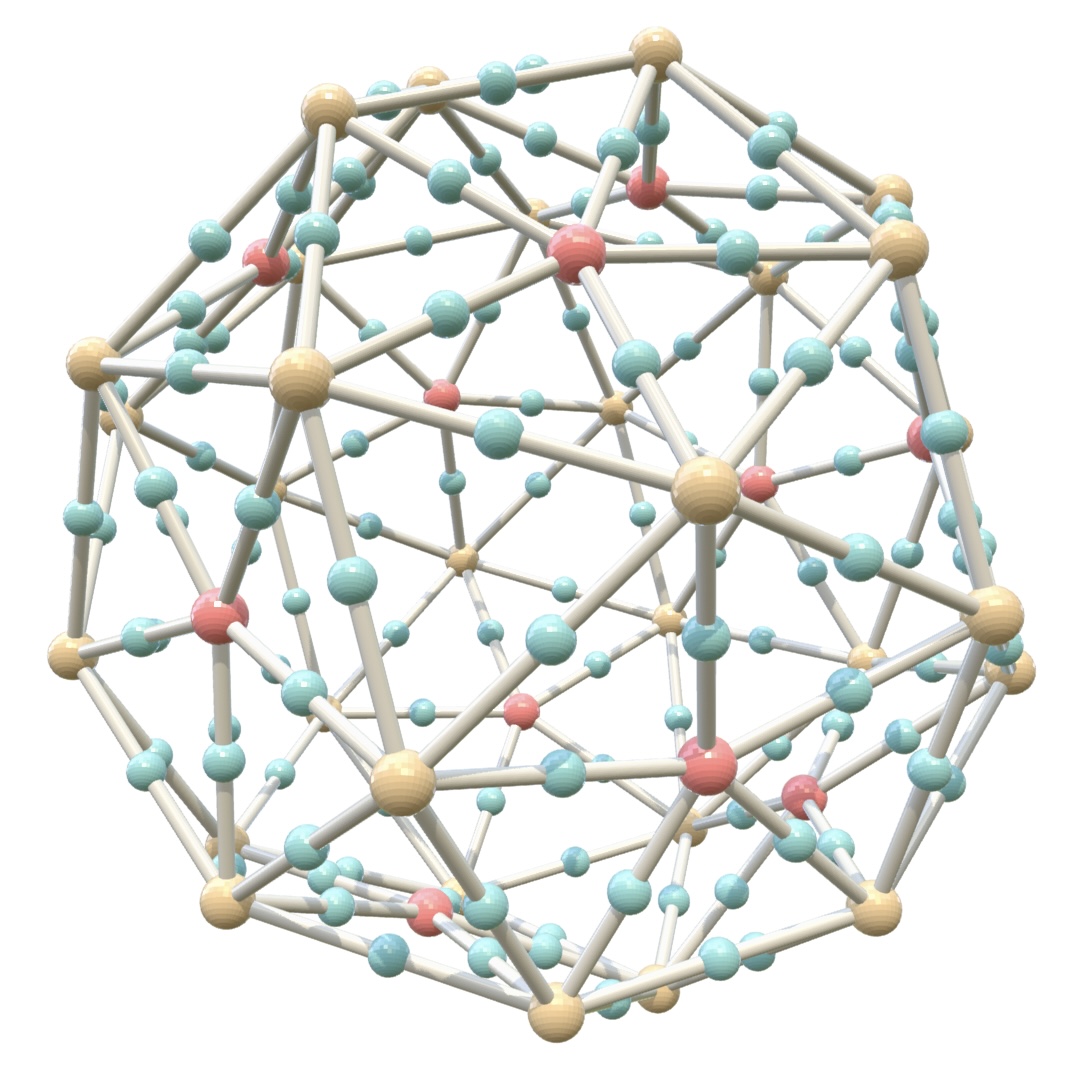}
    \includegraphics[width=0.24\linewidth,page=1]{Graphics/MTB/two_hundred_sixty-two_tetrahedra/262-1.jpg}
    \includegraphics[width=0.24\linewidth,page=1]{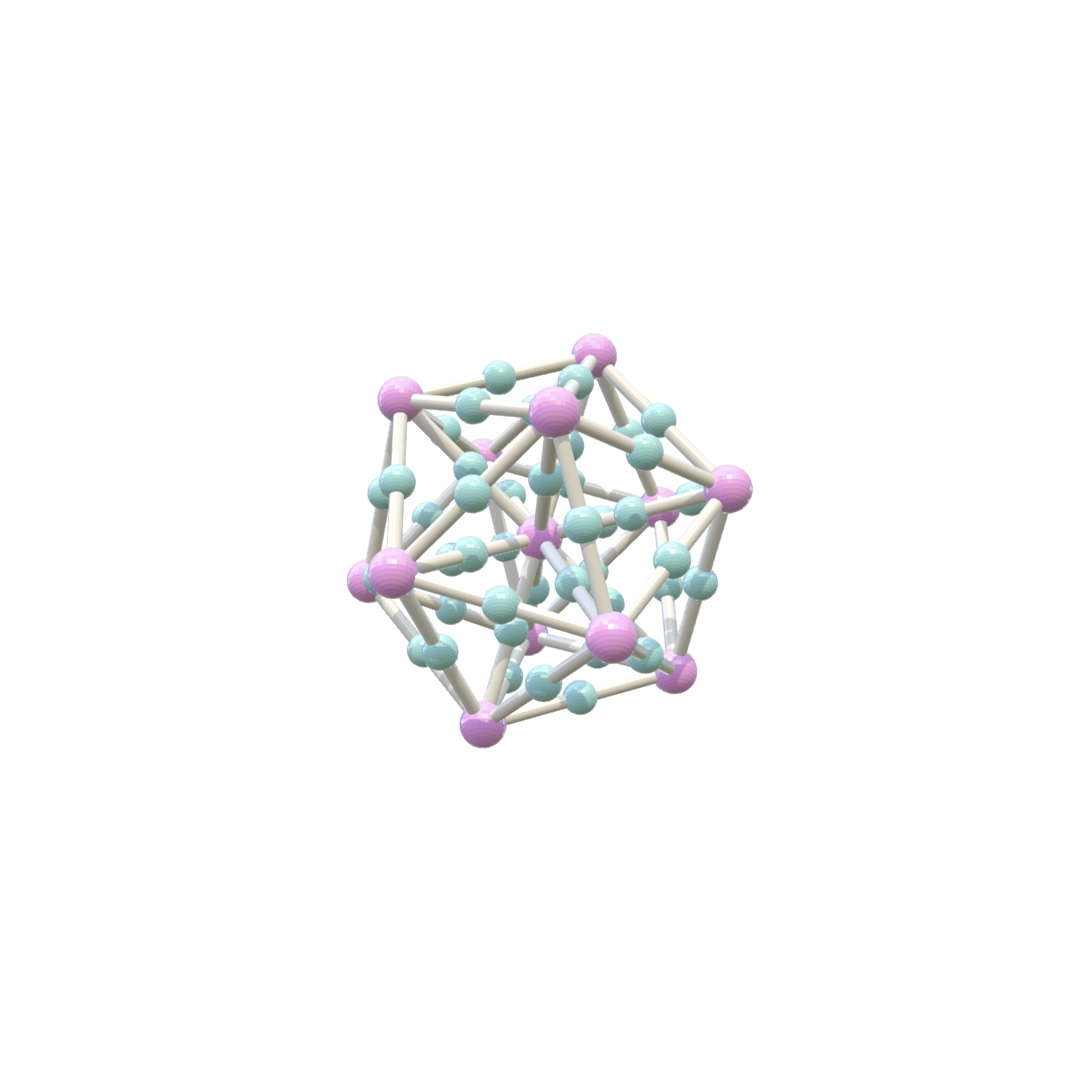}
    \includegraphics[width=0.24\linewidth,page=1]{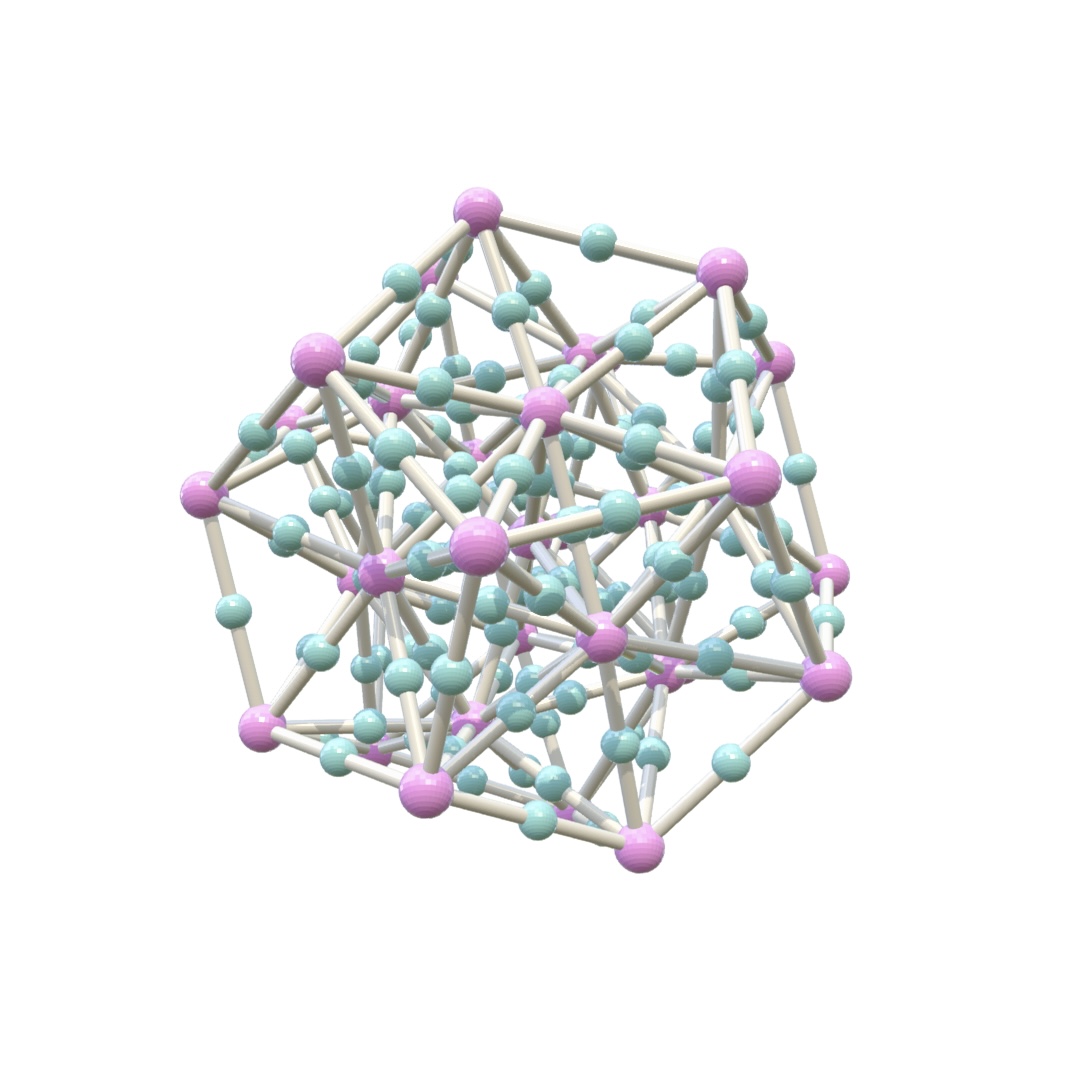}
    \includegraphics[width=0.24\linewidth,page=1]{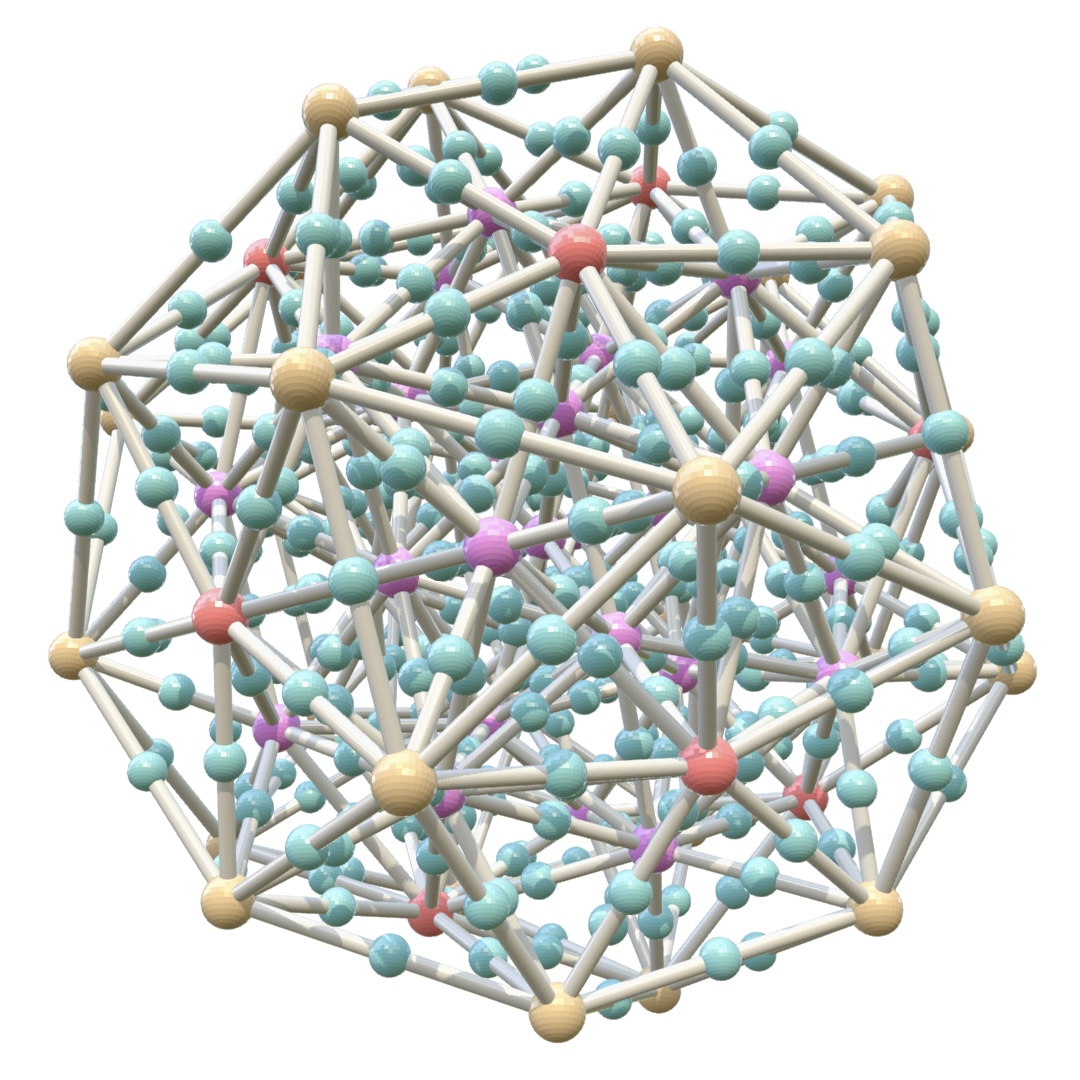}
    \caption{The decomposition of ``Two Hundred Sixty-Two Tetrahedra". \label{fig:two hundred sixty-two tetrahedra}}
\end{figure}

\section{Construction of Angular \(k\)-Uniform, Rotationally Invariant Vertex Codes on Regular Polytopes\label{App.B}}

{\renewcommand{\arraystretch}{1.5}

\begin{table*}[ht]

\centering

\begin{tabularx}{0.9\textwidth}{@{} 
c
c
c
c >{\centering\arraybackslash}
c
c
c
@{}}

\toprule

  \makecell{\textbf{Vertex}\\ \textbf{Figure}}
& \makecell{\textbf{Angular}\\ \textbf{Uniformity}}
& \makecell{\textbf{Truncation}}
& \makecell{\textbf{Bond}\\ \textbf{Dimension}}
& \makecell{\textbf{Type of}\\ \textbf{X-I Code}}
& \makecell{\textbf{Multi-angularity}}
\\

\midrule

  \makecell{$\{p\}$\\$p$ is odd}
& \makecell{$k=1$}
& \makecell{$t\{2,2p\}$}
& \makecell{$\chi_P=4$}
& \makecell{$p$ copies of $\Xi(4)$ on antipodal edges\\each connecting two indices}
& \makecell{\xmark}
\\

\midrule

  \makecell{$\{p\}$\\$p$ is even}
& \makecell{$k=1$}
& \makecell{$\{p\}$}
& \makecell{$\chi_P=1$}
& \makecell{One $\Xi(p)$ on the $p$-gon}
& \makecell{\xmark}
\\

\midrule

  \makecell{$\{3,3\}$}
& \makecell{$k=1$}
& \makecell{$tr\{3,3\}$}
& \makecell{$\chi_P=6$}
& \makecell{Four $\Xi(6)$ on hexagons connecting 3 indices}
& \makecell{\xmark}
\\

\midrule

  \makecell{$\{3,4\}$}
& \makecell{$k=1$}
& \makecell{$t\{3,4\}$}
& \makecell{$\chi_P=4$}
& \makecell{Six $\Xi(4)$ on antipodal edges\\each connecting two indices}
& \makecell{\xmark}
\\

\midrule

  \makecell{$\{3,4\}$}
& \makecell{$k=1$}
& \makecell{$tr\{4,3\}$}
& \makecell{$\chi_P=8$}
& \makecell{Eight $\Xi(6)$ on the hexagon}
& \makecell{\cmark}
\\

\midrule

  \makecell{$\{3,4\}$}
& \makecell{$k=2$}
& \makecell{$\{3,4\}$}
& \makecell{$\chi_P=1$}
& \makecell{One $\Xi(6)$ on the octahedron}
& \makecell{\xmark}
\\

\midrule

  \makecell{$\{3,5\}$}
& \makecell{$k=1$}
& \makecell{$\{3,5\}$}
& \makecell{$\chi_P=1$}
& \makecell{One $\Xi(12)$ on the icosahedron}
& \makecell{\xmark}
\\

\midrule

  \makecell{$\{3,5\}$}
& \makecell{$k=2$}
& \makecell{$rr\{5,3\}$}
& \makecell{$\chi_P=5$}
& \makecell{Ten $\Xi(6)$ on antipodal triangles}
& \makecell{\xmark}
\\

\midrule

  \makecell{$\{4,3\}$}
& \makecell{$k=1$}
& \makecell{$t\{4,3\}$}
& \makecell{$\chi_P=3$}
& \makecell{Six $\Xi(4)$ on antipodal edges\\each connecting two indices}
& \makecell{\xmark}
\\

\midrule

  \makecell{$\{5,3\}$}
& \makecell{$k=1$}
& \makecell{$t\{5,3\}$}
& \makecell{$\chi_P=3$}
& \makecell{Ten $\Xi(4)$ on antipodal edges\\each connecting two indices}
& \makecell{\xmark}
\\

\midrule

  \makecell{$\{5,3\}$}
& \makecell{$k=2$}
& \makecell{$tr\{5,3\}$}
& \makecell{$\chi_P=6$}
& \makecell{Twelve $\Xi(10)$ on decagons}
& \makecell{\cmark}
\\

\midrule

  \makecell{$\{5,3\}$}
& \makecell{$k=4$}
& \makecell{$rr\{5,3\}$}
& \makecell{$\chi_P=3$}
& \makecell{Six $\Xi(10)$ on antipodal pentagons}
& \makecell{\xmark}
\\

\midrule

  \makecell{$\{3,3,3\}$}
& \makecell{$k=1$}
& \makecell{$tr\{3,3,3\}$}
& \makecell{$\chi_P=12$}
& \makecell{Ten $\Xi(6)$ on hexagons connecting three indices}
& \makecell{\xmark}
\\

\midrule

  \makecell{$\{3,3,4\}$}
& \makecell{$k=1$}
& \makecell{$t\{3,3,4\}$}
& \makecell{$\chi_P=5$}
& \makecell{Twelve $\Xi(4)$ on antipodal edges\\each connecting two indices}
& \makecell{\xmark}
\\

\midrule

  \makecell{$\{3,3,5\}$}
& \makecell{$k=1$}
& \makecell{$t\{3,3,5\}$}
& \makecell{$\chi_P=12$}
& \makecell{Three hundred sixty $\Xi(4)$ on antipodal edges\\each connecting two indices}
& \makecell{\xmark}
\\

\midrule

  \makecell{$\{3,4,3\}$}
& \makecell{$k=1$}
& \makecell{$t\{3,4,3\}$}
& \makecell{$\chi_P=6$}
& \makecell{Forty-eight $\Xi(4)$ on antipodal edges\\each connecting two indices}
& \makecell{\xmark}
\\

\midrule

  \makecell{$\{4,3,3\}$}
& \makecell{$k=1$}
& \makecell{$t\{4,3,3\}$}
& \makecell{$\chi_P=4$}
& \makecell{Sixteen $\Xi(4)$ on antipodal edges\\each connecting two indices}
& \makecell{\xmark}
\\

\midrule

  \makecell{$\{5,3,3\}$}
& \makecell{$k=1$}
& \makecell{$t\{5,3,3\}$}
& \makecell{$\chi_P=4$}
& \makecell{Six hundred $\Xi(4)$ on antipodal edges\\each connecting two indices}
& \makecell{\xmark}
\\

\bottomrule

\end{tabularx}

\caption{
Summary of known constructions of angular \(k\)-uniform, rotationally invariant vertex codes with regular vertex figures, built using \(X\)–\(I\) codes. The table classifies each construction according to its vertex figure, angular \(k\)-uniformity, truncation geometry, physical bond dimension \(\chi_P\), and whether it supports multi-angular uniformity. The list encompasses all regular polytopes in two, three, and four dimensions that may serve as building blocks for hyperinvariant codes (HICs) on compact regular hyperbolic honeycombs. Constructions for some \(k\) values and multi-angular uniformity remain open and are left for future exploration.
\label{tab:Codes}}
\end{table*}
}

\end{widetext}

\end{document}